\documentclass{article}

\usepackage{PRIMEarxiv}

\usepackage[utf8]{inputenc} 
\usepackage[T1]{fontenc}    
\usepackage{hyperref}       
\usepackage{url}            
\usepackage{booktabs}       
\usepackage{amsfonts}       
\usepackage{nicefrac}       
\usepackage{microtype}      
\usepackage{lipsum}
\usepackage{natbib}
\usepackage{fancyhdr}       
\usepackage{graphicx}       
\graphicspath{{media/}}     
\usepackage{pgfplots}
\usepackage{amssymb}
\pgfplotsset{width=10cm,compat=1.9}
\usepackage{amsmath}
\usepackage{amsthm}
\usepackage{multirow}

\usepackage{csvsimple}
\usepackage{siunitx}
\usepackage{tikz}
\usetikzlibrary{arrows.meta,calc,positioning,decorations.pathreplacing}
\usepackage{verbatim}
\usepackage{appendix}

\title{Nonlinear Shrinkage for Higher Moments Estimation in Large Portfolios}
\newcommand{\paperkeywords}[1]{Portfolio optimization#1Nonlinear shrinkage#1Higher-order moments#1%
  Cokurtosis#1Tensor decomposition#1Cross-validation#1Higher-Order Average Oracle}
\makeatletter
\hypersetup{
  pdftitle={\@title},
  pdfauthor={Paul Ruelloux, Christian Bongiorno, Damien Challet},
  pdfkeywords={\paperkeywords{, }}
}
\makeatother

\author{
  Paul Ruelloux$^{1,2}$, Christian Bongiorno$^{1,*}$, Damien Challet$^{1}$ \\[0.3em]
  $^{1}$Université Paris-Saclay, CentraleSupélec, \\Mathématiques et Informatique pour la Complexité et les Systèmes, \\
  91190, Gif-sur-Yvette, France\\ \\
  $^{2}$Barclays Bank Ireland PLC\\
  75008, Paris, France\\
  [0.3em]
  $^{*}$Corresponding author\enspace \href{mailto:christian.bongiorno@centralesupelec.fr}{\texttt{christian.bongiorno@centralesupelec.fr}}
}

\begin{document}
\maketitle

\begin{abstract}
Portfolio risk is not limited to the variance of returns. The moments of portfolio returns depend directly on the covariance matrix and higher-order co-dependences (coskewness, cokurtosis, etc.). In practice, however, these objects are extremely noisy, especially for large portfolios. Here we introduce filtering methods based on tensor decompositions that extend cross-validated covariance cleaning and the Average Oracle methodology to even-order co-moment tensors. Using returns on historical Russell 1000 constituents, we find that the proposed filters improve out-of-sample estimation accuracy relative to sample estimators, both for correlation and cokurtosis tensors, as measured by relative squared Frobenius errors. These improvements translate into lower relative errors on portfolio volatility and kurtosis, and into lower out-of-sample volatility and kurtosis when the portfolio objective penalizes both variance and the fourth moment.
\end{abstract}


\section{Introduction}
Understanding the collective fluctuations of financial assets is a central problem in quantitative finance. Their dependence structure determines how individual fluctuations combine at the portfolio level. Covariance provides a natural starting point for describing this structure, while higher-order co-moments capture non-Gaussian features of the joint distribution.

Estimating these quantities becomes increasingly difficult as the number of assets grows. Higher-order co-moments involve a rapidly increasing number of coefficients, whose estimation from limited samples is highly sensitive to sampling noise \cite{ledoit2004well,martellini2010improved}. Extending the estimation window can reduce sampling variability, but in financial markets it may also introduce bias because return distributions evolve over time \cite{stuaricua2005nonstationarities,cont2013running}.

In the high-dimensional regime, where the number of assets is comparable to the available sample size, the sample covariance matrix suffers from large estimation error, a phenomenon referred to as the curse of dimensionality \cite{laloux1999noise, kondor2007noise}. To mitigate estimation noise in high dimensions, a standard strategy is to filter the eigenvalues of the sample covariance matrix \cite{bun2017cleaning}. In this regime, random matrix theory describes how finite sampling distorts the eigenvalue distribution and provides filtering methods known as non-linear shrinkage \cite{ledoit2011eigenvectors}. While these methods are theoretically well-grounded under stylized assumptions on the return-generating process (such as independent and identically distributed observations or stationary dependence structures), those assumptions are only approximately satisfied in financial return series (see however \cite{ledoit2003improved, burda2022cleaning}).

A practical and widely used alternative is cross-validated spectral filtering based on sample splitting \cite{ledoit2012nonlinear}. The basic idea is to use one part of the data to identify the main directions of co-movement in returns (the empirical risk factors), and a different part of the data to reassess the strength of each direction. In this way, the method retains the factor directions learned In-Sample (IS) but replaces their sample variances, which are strongly contaminated by high-dimensional noise, with Out-Of-Sample (OOS) estimates of the variance explained along the same directions. Repeating this split-and-recalibrate step over multiple random partitions and averaging the resulting corrections further reduces sensitivity to any particular split, which yields a stable filtered covariance estimator. When the assumptions underlying random-matrix-based non-linear shrinkage are approximately satisfied, this cross-validated procedure recovers the same asymptotic spectral correction \cite{ledoit2012nonlinear}, while remaining effective even when those assumptions are not strictly met.

Cross-validated spectral filtering may split the data at random or in a time-ordered (walk-forward) fashion; it may also be performed within a fixed asset universe or in an asset-agnostic manner by sampling assets from a larger pool. Because dependences in financial markets change over time, overfitting transient features is a major source of error; enforcing causality through chronological splits and expanding the effective calibration sample by borrowing cross-sectional information both improve robustness. The Average Oracle \citep{bongiorno2023filtering} follows the same core logic as walk-forward cross-validation; it averages the so-called oracle (i.e., optimal given future data) eigenvalues over many past time windows, which yields a fixed vector of filtered eigenvalues that can be applied to the current window and periodically refreshed if needed. This additional averaging acts as a regularizer against regime-specific fluctuations and, in empirical evaluations, systematically delivers more accurate covariance estimates than methods relying solely on the most recent data \cite{bongiorno2024averageoracle}.

Our objective is to filter higher-order co-moments to improve the estimation of portfolio return  moments for large portfolios, where naive sample estimates are dominated by finite-sample noise and further destabilized by non-stationarity. To make higher-order co-moments usable in practice, we extend covariance matrix cleaning methods to higher-order dependences encoded by tensors. Rather than seeking a fully explicit tensor analog of analytical non-linear shrinkage, we generalize two data-driven numerical procedures to the tensor setting, namely cross-validation and the Average Oracle.

We focus on the moments of portfolio returns, which are obtained by contracting the asset co-moment tensors with the portfolio weights. Specifically, we evaluate forecasts of portfolio volatility and kurtosis, as well as portfolio allocations whose objective penalizes both variance and the fourth central moment. More accurate estimates of  portfolio return moments can also benefit broader risk management applications; Cornish--Fisher-type expansions, for instance, map the leading moments of a distribution to its quantiles, and thus to tail-risk measures such as Value-at-Risk \cite{CornishFisher1937,Withers1984,MaillardCornishFisher}.

A first conceptual difficulty is that tensors generally do not possess a spectral decomposition analogous to that of matrices. For a co-moment tensor of order $k\geq 3$ (\textit{e.g.}, cokurtosis for $k=4$), there is no unique notion of tensor diagonalization by an orthogonal basis of vectors. A natural substitute is to rely on multilinear factorizations such as the Higher-Order Singular Value Decomposition (HOSVD), a constrained variant of the Tucker decomposition that represents a tensor as a core tensor rotated along each tensor mode (i.e., index axis) \cite{de2000multilinear}. This mirrors the role of eigenvectors in the matrix setting by providing interpretable co-movement directions, while also making clear why a direct spectral-style shrinkage is not immediately available. This framework remains consistent with the classical case, since the HOSVD of a matrix reduces to the standard matrix singular value decomposition \cite{de2000multilinear, cichocki2015tensor}. While other tensor decompositions exist (such as the orthogonally decomposable tensor model of Ref.~\cite{anandkumar2014tensor}), their structural assumptions are too restrictive for the empirical co-moment tensors considered here.

The remainder of the paper is organized as follows. Section~\ref{sec:theory} introduces co-moment tensors, and their relationships with portfolio return moments, tensor decompositions, and the oracle construction. Section~\ref{sec:comoment_shrinkage} introduces the Cross-Validation and Higher-Order Average Oracle estimators, and discusses the joint distributional consistency of the filtered covariance and cokurtosis. Section~\ref{sec:realdataapp} reports empirical results on U.S. equity returns. We assess the estimation accuracy at the tensor and portfolio-moment levels, and report the out-of-sample volatility and kurtosis of portfolios constructed from filtered moments. Section~\ref{sec:conclusion} concludes. The Supplementary Information (SI) part contains the mathematical proofs, details on the dataset and experimental protocol, the statistical inference methodology, an in-depth study of joint distributional consistency, controlled synthetic experiments, and additional empirical results.

\section{Theoretical Background}
\label{sec:theory}

\subsection{Co-moment tensors}
\label{sec:comoment_def}

Let $\boldsymbol{x}=(x_1,\ldots,x_n)^\top\in\mathbb{R}^n$ be a random vector of asset returns with mean $\boldsymbol{\mu}:=\mathbb{E}[\boldsymbol{x}]$. Beyond the covariance matrix
\begin{equation}
\label{eq:co-moment_tensor}
\boldsymbol{\Sigma} := \mathbb{E}\!\left[(\boldsymbol{x} - \boldsymbol{\mu})(\boldsymbol{x} - \boldsymbol{\mu})^\top\right] \in \mathbb{R}^{n\times n},
\qquad
\Sigma_{ij} = \mathbb{E}\!\left[(x_i-\mu_i)(x_j-\mu_j)\right],
\end{equation}
higher-order central co-moments capture non-Gaussian features of the joint return distribution, such as asymmetry and heavy tails. In particular, the third- and fourth-order co-moments are encoded by the coskewness tensor $\boldsymbol{\mathcal{S}}\in\mathbb{R}^{n\times n\times n}$ and the cokurtosis tensor $\boldsymbol{\mathcal{K}}\in\mathbb{R}^{n\times n\times n\times n}$, whose entries are
\begin{equation}
\label{eq:comomententries}
\mathcal{S}_{ijk} = \mathbb{E}\!\left[(x_i-\mu_i)(x_j-\mu_j)(x_k-\mu_k)\right],
\qquad
\mathcal{K}_{ijkl} = \mathbb{E}\!\left[(x_i-\mu_i)(x_j-\mu_j)(x_k-\mu_k)(x_l-\mu_l)\right].
\end{equation}
More generally, the order-$k$ central co-moment tensor is defined as $\boldsymbol{\mathcal{M}}^{(k)} := \mathbb{E}[(\boldsymbol{x}-\boldsymbol{\mu})^{\otimes k}] \in (\mathbb{R}^n)^{\otimes k}$ (with $\boldsymbol{\Sigma}=\boldsymbol{\mathcal{M}}^{(2)}$, $\boldsymbol{\mathcal{S}}=\boldsymbol{\mathcal{M}}^{(3)}$, and $\boldsymbol{\mathcal{K}}=\boldsymbol{\mathcal{M}}^{(4)}$). Because the products in Eq.~\eqref{eq:comomententries} do not depend on the ordering of the factors, co-moment tensors are supersymmetric, that is, their entries are invariant under any permutation of their indices (for instance, $\mathcal{K}_{ijkl} = \mathcal{K}_{jikl} = \mathcal{K}_{klij}$). We present our framework for fourth-order cokurtosis ($k=4$) alongside covariance ($k=2$) in the main text, and defer the general order-$k$ tensor formalism and proofs to SI, Sec.~\ref{sec:appmodematrices}--\ref{sec:appevenprojection}.

In practice, co-moments must be estimated from finite samples. Given $\Delta t$ observations $\boldsymbol{X}=\{\boldsymbol{x}_{1},\ldots,\boldsymbol{x}_{\Delta t}\}\in\mathbb{R}^{n\times\Delta t}$, the sample mean, covariance, and cokurtosis estimators are
\begin{equation}
\label{eq:sample_estimators}
\hat{\boldsymbol{\mu}} = \frac{1}{\Delta t} \sum_{t=1}^{\Delta t} \boldsymbol{x}_{t},
\qquad
\widehat{\boldsymbol{\Sigma}} = \frac{1}{\Delta t} \sum_{t=1}^{\Delta t}
(\boldsymbol{x}_{t} - \hat{\boldsymbol{\mu}})(\boldsymbol{x}_{t} - \hat{\boldsymbol{\mu}})^\top,
\qquad
\widehat{\mathcal{K}}_{ijkl} = \frac{1}{\Delta t} \sum_{t=1}^{\Delta t}
(x_{it}-\hat{\mu}_i)(x_{jt}-\hat{\mu}_j)(x_{kt}-\hat{\mu}_k)(x_{lt}-\hat{\mu}_l),
\end{equation}
which are the moments of the empirical distribution. No parametric distribution is assumed, so we refer to them as sample estimators (Sample). At fixed $n$, they are asymptotically consistent as $\Delta t\to\infty$, albeit biased in finite samples \cite{joanes1998comparing,afendras2020limiting}; since we do not assume a specific parametric distribution for returns or rely on maximum-likelihood estimation, we adopt the empirical-moment convention and use the common $1/\Delta t$ normalization for both $\widehat{\boldsymbol{\Sigma}}$ and $\widehat{\boldsymbol{\mathcal{K}}}$. As the number of assets $n$ grows, the number of cokurtosis entries $n^4$ increases rapidly (even accounting for supersymmetry, there are $\binom{n+3}{4} = O(n^4)$ distinct coefficients), so that sample estimates are heavily contaminated by estimation noise.

\subsection{Portfolio moments}
\label{subsec:portfolio_moments}

For a portfolio with weight vector $\boldsymbol{w}\in\mathbb{R}^n$ and return $x_p = \sum_{i=1}^n w_i x_i$, the univariate moments of portfolio returns are obtained by contracting the asset co-moments with $\boldsymbol{w}$ along all indices. Given estimates $\hat{\mu}_i$, $\widehat{\Sigma}_{ij}$, $\widehat{\mathcal{S}}_{ijk}$, and $\widehat{\mathcal{K}}_{ijkl}$, the estimated portfolio mean, variance, standardized skewness, and standardized kurtosis read
\begin{equation}
\label{eq:contraction}
\hat{\mu}_p = \sum_{i=1}^n\hat{\mu}_i w_i,\qquad
\hat{\sigma} ^{2}= \sum_{i,j=1}^n \widehat{\Sigma}_{ij}\,w_i w_j,\qquad
\hat{\mathfrak{s}} = \sum_{i,j,k=1}^n \frac{\widehat{\mathcal{S}}_{ijk}\,w_i w_j w_k}{\hat{\sigma}^3},\qquad
\hat{\kappa} =\sum_{i,j,k,l = 1}^n \frac{\widehat{\mathcal{K}}_{ijkl}\,w_i w_j w_k w_l}{\hat{\sigma}^{4}}.
\end{equation}
Because higher-order sample co-moments react strongly to outliers and accumulate estimation errors across $O(n^4)$ terms, plug-in estimates of portfolio kurtosis and fourth-moment-penalized allocations deteriorate even at moderate portfolio dimensions unless the cokurtosis tensor is regularized.

\subsection{Symmetric HOSVD of co-moment tensors}
\label{sec:decomposition}

For a covariance matrix $\boldsymbol{\Sigma}$, spectral filtering relies on the eigenvalue decomposition $\boldsymbol{\Sigma} = \boldsymbol{U}\boldsymbol{\Lambda}\boldsymbol{U}^\top$, where $\boldsymbol{U}\in O(n)$ is the orthogonal matrix of eigenvectors and $\boldsymbol{\Lambda}=\mathrm{diag}(\lambda_1,\ldots,\lambda_n)$ contains the eigenvalues. Higher-order tensors do not admit a direct analog of spectral diagonalization, since rotating a fourth-order tensor by a single orthogonal basis $\boldsymbol{U}$ cannot, in general, reduce it to a diagonal array of $n$ numbers \cite{kolda2009tensor}.

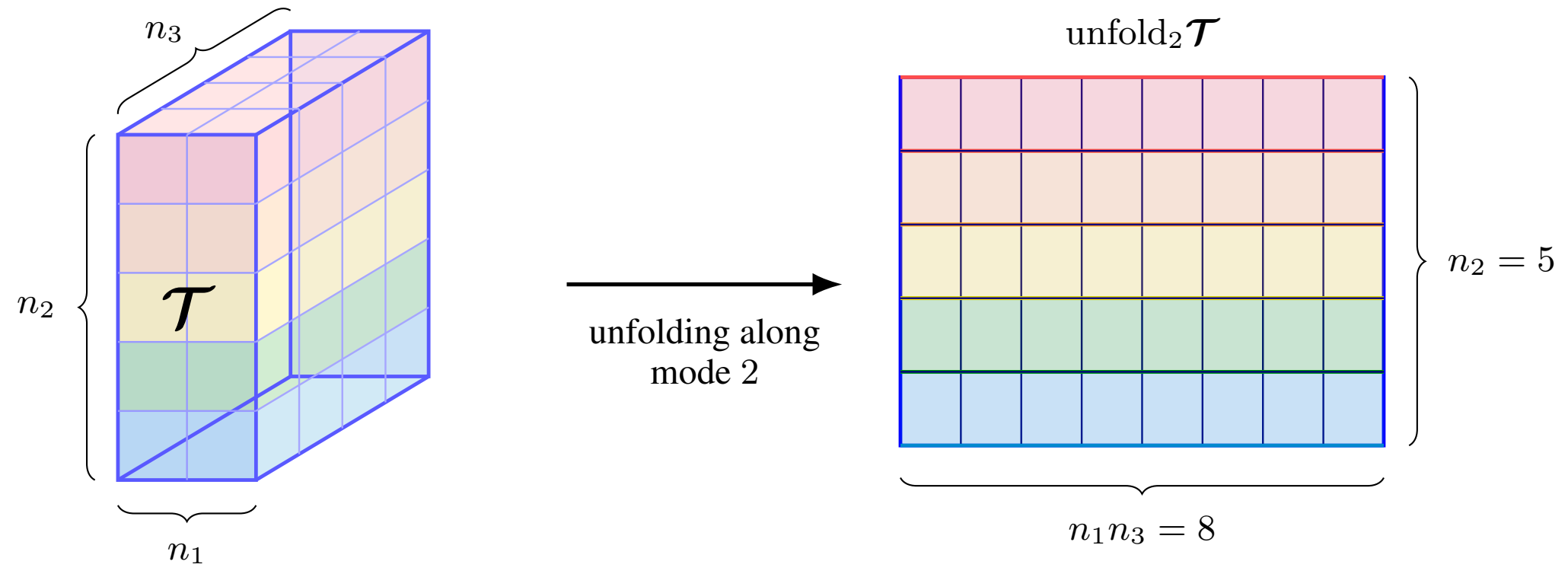
\begin{figure}[htbp]
\centering
\begin{tikzpicture}[
    >=Latex,
    font=\small,
    every node/.style={font=\small}
]

\coordinate (A) at (1.5,0);
\coordinate (B) at (2.7,0);
\coordinate (C) at (2.7,3.0);
\coordinate (D) at (1.5,3.0);

\coordinate (shift) at (1.5,0.9);

\coordinate (A2) at ($(A)+(shift)$);
\coordinate (B2) at ($(B)+(shift)$);
\coordinate (C2) at ($(C)+(shift)$);
\coordinate (D2) at ($(D)+(shift)$);

\fill[blue!4] (A2)--(B2)--(C2)--(D2)--cycle;
\fill[blue!7] (A)--(B)--(C)--(D)--cycle;

\foreach \tlow/\thigh/\col in {
    0.00/0.20/cyan!70!blue,
    0.20/0.40/green!60!black,
    0.40/0.60/yellow!80!orange,
    0.60/0.80/orange!85,
    0.80/1.00/red!70
}{
    
    \fill[\col, opacity=0.22]
        ($(A)!\tlow!(D)$) -- ($(B)!\tlow!(C)$) --
        ($(B)!\thigh!(C)$) -- ($(A)!\thigh!(D)$) -- cycle;

    \fill[\col, opacity=0.18]
        ($(B)!\tlow!(C)$) -- ($(B2)!\tlow!(C2)$) --
        ($(B2)!\thigh!(C2)$) -- ($(B)!\thigh!(C)$) -- cycle;
}

\fill[red!70, opacity=0.14] (D)--(C)--(C2)--(D2)--cycle;

\draw[blue!65, line width=0.9pt] (A)--(B)--(C)--(D)--cycle;
\draw[blue!65, line width=0.9pt] (A2)--(B2)--(C2)--(D2)--cycle;
\draw[blue!65, line width=0.9pt] (A)--(A2);
\draw[blue!65, line width=0.9pt] (B)--(B2);
\draw[blue!65, line width=0.9pt] (C)--(C2);
\draw[blue!65, line width=0.9pt] (D)--(D2);

\draw[blue!40, line width=0.45pt] ($(A)!0.5!(B)$)--($(D)!0.5!(C)$);

\foreach \t in {0.2,0.4,0.6,0.8}{
    \draw[blue!40, line width=0.45pt] ($(A)!\t!(D)$)--($(B)!\t!(C)$);
}

\draw[blue!35, line width=0.45pt] ($(D)!0.5!(C)$)--($(D2)!0.5!(C2)$);

\foreach \t in {0.25,0.5,0.75}{
    \draw[blue!35, line width=0.45pt] ($(D)!\t!(D2)$)--($(C)!\t!(C2)$);
}

\foreach \t in {0.2,0.4,0.6,0.8}{
    \draw[blue!35, line width=0.45pt] ($(B)!\t!(C)$)--($(B2)!\t!(C2)$);
}

\foreach \t in {0.25,0.5,0.75}{
    \draw[blue!35, line width=0.45pt] ($(B)!\t!(B2)$)--($(C)!\t!(C2)$);
}

\node at ($(A)!0.5!(C)$) {\Large$\boldsymbol{\mathcal T}$};

\draw[decorate,decoration={brace,mirror,amplitude=4pt}]
($(A)+(0,-0.22)$) -- ($(B)+(0,-0.22)$)
node[midway,below=6pt] {\(n_1\)};

\draw[decorate,decoration={brace,amplitude=4pt}]
($(A)-(0.20,0)$) -- ($(D)+(-0.20,0)$)
node[midway,left=6pt] {\(n_2\)};

\draw[decorate,decoration={brace,amplitude=4pt}]
($(D)+(0,0.18)$) -- ($(D2)+(0,0.18)$)
node[midway,above left=2pt] {\(n_3\)};

\draw[->, line width=1.0pt] (5.4,1.7) -- (7.8,1.7);
\node[align=center] at (6.6,1.12) {unfolding along\\mode \(2\)};

\coordinate (M1) at (8.3,0.30);
\coordinate (M2) at (12.5,0.30);
\coordinate (M3) at (12.5,3.50);
\coordinate (M4) at (8.3,3.50);

\fill[blue!4] (M1)--(M2)--(M3)--(M4)--cycle;
\draw[blue!50!blue, line width=0.9pt] (M1)--(M2)--(M3)--(M4)--cycle;

\node at ($(M4)!0.5!(M3)+(0,0.38)$)
{\(\mathrm{unfold}_2\boldsymbol{\mathcal T}\)};

\foreach \x in {8.825,9.35,9.875,10.40,10.925,11.45,11.975}{
    \draw[blue!50!black, line width=0.45pt] (\x,0.30)--(\x,3.50);
}

\foreach \y/\col in {
    0.30/cyan!70!blue,
    0.94/green!60!black,
    1.58/yellow!80!orange,
    2.22/orange!85,
    2.86/red!70
}{
    \fill[\col, opacity=0.18] (8.3,\y) rectangle (12.5,\y+0.64);
    \draw[\col, line width=0.9pt] (8.3,\y)--(12.5,\y);
    \draw[\col, line width=0.9pt] (8.3,\y+0.64)--(12.5,\y+0.64);
}

\foreach \y in {0.94,1.58,2.22,2.86}{
    \draw[blue!50!black, line width=0.45pt] (8.3,\y)--(12.5,\y);
}

\draw[decorate,decoration={brace,mirror,amplitude=4pt}]
(12.72,0.30) -- (12.72,3.50)
node[midway,right=6pt] {\(n_2=5\)};

\draw[decorate,decoration={brace,mirror,amplitude=4pt}]
(8.3,0.02) -- (12.5,0.02)
node[midway,below=6pt] {\(n_1n_3=8\)};
\end{tikzpicture}
\caption{Mode-\(2\) unfolding of a tensor \(\boldsymbol{\mathcal T}\in\mathbb{R}^{2\times 5\times 4}\).}
\label{fig:mode_2_unfolding}
\end{figure}

A natural multilinear generalization is the Higher-Order Singular Value Decomposition (HOSVD) \cite{de2000multilinear}. To extract an orthogonal basis $\boldsymbol{U}\in O(n)$ from a supersymmetric tensor such as the cokurtosis $\boldsymbol{\mathcal{K}}\in\mathbb{R}^{n\times n\times n\times n}$, one unfolds (or matricizes) the tensor along its first mode into a rectangular matrix $\boldsymbol{K}_{(1)} \in \mathbb{R}^{n \times n^3}$, whose $i$-th row collects all entries $\mathcal{K}_{ijkl}$ as $(j,k,l)$ varies (Fig.~\ref{fig:mode_2_unfolding} illustrates unfolding on a third-order tensor; by supersymmetry, unfolding along any mode yields the exact same matrix). Computing the standard matrix singular value decomposition $\boldsymbol{K}_{(1)} = \boldsymbol{U}\boldsymbol{\Lambda}\boldsymbol{V}^\top$ (or equivalently diagonalizing the $n\times n$ matrix $\boldsymbol{K}_{(1)}\boldsymbol{K}_{(1)}^\top$) provides the orthogonal mode matrix $\boldsymbol{U}\in O(n)$ of left singular vectors.

Using this shared orthogonal basis $\boldsymbol{U}$ across all four modes defines the symmetric HOSVD of $\boldsymbol{\mathcal{K}}$, which expresses the tensor in terms of a rotated core tensor $\boldsymbol{\mathcal{G}}\in\mathbb{R}^{n\times n\times n\times n}$:
\begin{equation}
\label{eq:symhosvd_core}
\boldsymbol{\mathcal{K}} = \boldsymbol{\mathcal{G}} \times_1 \boldsymbol{U} \times_2 \boldsymbol{U} \times_3 \boldsymbol{U} \times_4 \boldsymbol{U},
\qquad
\mathcal{K}_{ijkl} = \sum_{a,b,c,d=1}^n U_{ia} U_{jb} U_{kc} U_{ld}\, \mathcal{G}_{abcd}.
\end{equation}
Conversely, rotating the centered returns into the basis $\boldsymbol{U}$ defines $\boldsymbol{y} := \boldsymbol{U}^\top(\boldsymbol{x}-\boldsymbol{\mu})$ and shows that the core tensor $\boldsymbol{\mathcal{G}}$ is simply the fourth-order co-moment tensor of the rotated coordinates $\boldsymbol{y}$:
\begin{equation}
\label{eq:core_as_rotated_moments}
\boldsymbol{\mathcal{G}} = \boldsymbol{\mathcal{K}} \times_1 \boldsymbol{U}^\top \times_2 \boldsymbol{U}^\top \times_3 \boldsymbol{U}^\top \times_4 \boldsymbol{U}^\top,
\qquad
\mathcal{G}_{abcd} = \mathbb{E}\!\left[y_a y_b y_c y_d\right].
\end{equation}
By construction of the SVD of $\boldsymbol{K}_{(1)}$, any two distinct slices of $\boldsymbol{\mathcal{G}}$ along the same mode are orthogonal with respect to the Frobenius inner product:
\begin{equation}
\label{eq:all_orthogonality_def}
\sum_{b,c,d=1}^n \mathcal{G}_{abcd}\,\mathcal{G}_{a'bcd} = 0
\qquad \forall\, a \neq a'.
\end{equation}
We denote by $\mathbb{A}_4$ (and more generally $\mathbb{A}_k$; see SI, Sec.~\ref{sec:appmodematrices}) the set of supersymmetric fourth-order tensors satisfying this mutual slice-orthogonality condition~\eqref{eq:all_orthogonality_def}, which are called all-orthogonal core tensors (or cores for short). For $k=2$, all-orthogonality reduces to the requirement that the eigenvalue core $\boldsymbol{\Lambda} = \boldsymbol{U}^\top\boldsymbol{\Sigma}\boldsymbol{U}$ be diagonal.

\subsection{Oracle core, sign invariance, and the even-core matrix}
\label{sec:oracle_comoment}

By analogy with spectral covariance filtering, we clean a sample co-moment tensor by keeping its empirical basis $\widehat{\boldsymbol{U}}\in O(n)$ fixed and regularizing only its core tensor. For a candidate core $\boldsymbol{\mathcal{G}}$, let
\begin{equation}
\label{eq:oracle_reconstruction_map}
\widehat{\boldsymbol{\mathcal{K}}}(\boldsymbol{\mathcal{G}}) := \boldsymbol{\mathcal{G}} \times_1 \widehat{\boldsymbol{U}} \times_2 \widehat{\boldsymbol{U}} \times_3 \widehat{\boldsymbol{U}} \times_4 \widehat{\boldsymbol{U}}
\end{equation}
denote the reconstructed cokurtosis tensor. If the true population tensor $\boldsymbol{\mathcal{K}}$ were known, the ideal constrained oracle core would minimize the squared Frobenius error $\|\widehat{\boldsymbol{\mathcal{K}}}(\boldsymbol{\mathcal{G}}) - \boldsymbol{\mathcal{K}}\|_F^2$ over $\boldsymbol{\mathcal{G}}\in\mathbb{A}_4$:
\begin{equation}
\label{eq:true_oracle}
\boldsymbol{\mathcal{G}}^{\star}
:=
\arg\min_{\boldsymbol{\mathcal{G}}\in\mathbb{A}_4}
\left\|
\widehat{\boldsymbol{\mathcal{K}}}(\boldsymbol{\mathcal{G}})
-
\boldsymbol{\mathcal{K}}
\right\|_F^2.
\end{equation}
Unlike the matrix case ($k=2$), where projecting onto $\mathbb{A}_2$ simply takes the diagonal of $\widehat{\boldsymbol{U}}^\top\boldsymbol{\Sigma}\widehat{\boldsymbol{U}}$, Eq.~\eqref{eq:true_oracle} has no closed-form solution for $k=4$. However, if we relax the constraint $\boldsymbol{\mathcal{G}}\in\mathbb{A}_4$ and allow $\boldsymbol{\mathcal{G}}$ to be an arbitrary tensor in $\mathbb{R}^{n\times n\times n\times n}$, the orthogonal invariance of the Frobenius norm yields the explicit unconstrained minimizer (SI, Sec.~\ref{app:proofoptimality})
\begin{equation}
\label{eq:unconstrained_oracle}
\widetilde{\boldsymbol{\mathcal{G}}}
:=
\arg\min_{\boldsymbol{\mathcal{G}}\in\mathbb{R}^{n\times n\times n\times n}}
\left\|
\widehat{\boldsymbol{\mathcal{K}}}(\boldsymbol{\mathcal{G}})
-
\boldsymbol{\mathcal{K}}
\right\|_F^2
=
\boldsymbol{\mathcal{K}} \times_1 \widehat{\boldsymbol{U}}^\top \times_2 \widehat{\boldsymbol{U}}^\top \times_3 \widehat{\boldsymbol{U}}^\top \times_4 \widehat{\boldsymbol{U}}^\top,
\qquad
\widetilde{\mathcal{G}}_{abcd} = \mathbb{E}\!\left[y_a y_b y_c y_d\,\middle|\,\widehat{\boldsymbol{U}}\right],
\end{equation}
where $\boldsymbol{y} = \widehat{\boldsymbol{U}}^\top(\boldsymbol{x}-\boldsymbol{\mu})$ denotes an out-of-sample return vector projected onto the fixed sample basis $\widehat{\boldsymbol{U}}$. For a single realization of $\widehat{\boldsymbol{U}}$, $\widetilde{\boldsymbol{\mathcal{G}}}$ is a dense supersymmetric tensor that does not belong to $\mathbb{A}_4$, just as $\widehat{\boldsymbol{U}}^\top\boldsymbol{\Sigma}\widehat{\boldsymbol{U}}$ is not diagonal on a single sample.

In any filtering procedure based on sample splitting or historical calibration, the core is estimated by averaging $\widetilde{\boldsymbol{\mathcal{G}}}$ over multiple realizations of the sample basis $\widehat{\boldsymbol{U}}$ obtained from training subsamples in cross-validation or from historical training windows in calibration:
\begin{equation}
\label{eq:generic_core_estimator}
\mathbb{E}\!\left[\widetilde{\mathcal{G}}_{abcd}\right]
=
\mathbb{E}\!\left[y_a y_b y_c y_d\right].
\end{equation}
Because each singular vector $\widehat{\boldsymbol{u}}_a$ in $\widehat{\boldsymbol{U}}$ is defined only up to an arbitrary sign $\eta_a \in \{-1, +1\}$, flipping the orientation of the $a$-th basis vector ($\widehat{\boldsymbol{u}}_a \mapsto \eta_a \widehat{\boldsymbol{u}}_a$) flips the sign of the rotated coordinate ($y_a \mapsto \eta_a y_a$). Under independent symmetric signs $\eta_a \sim \mathcal{U}(\{-1, +1\})$, each core entry is multiplied by the product of the corresponding signs:
\begin{equation}
\label{eq:sign_expectation}
\mathbb{E}\!\left[\widetilde{\mathcal{G}}_{abcd}\right] \propto \mathbb{E}\!\left[\eta_a \eta_b \eta_c \eta_d\right].
\end{equation}
This sign symmetry has two direct consequences. First, it causes coskewness and all odd-order co-moments to vanish. For a third-order tensor ($k=3$), the analogous expectation involves three signs, $\mathbb{E}[\eta_a \eta_b \eta_c]$. Because three indices cannot all appear an even number of times, at least one index in $(a,b,c)$ has odd multiplicity (1 or 3), so $\mathbb{E}[\eta_a \eta_b \eta_c] = 0$ for every triplet $(a,b,c)$. The same holds for any odd order $k=2m+1$ (SI, Sec.~\ref{sec:appevenprojection}). Geometrically, any filtering procedure that averages cores without fixing singular vector orientations is invariant under axis reflections ($\boldsymbol{y}\overset{d}{=}-\boldsymbol{y}$) and therefore forces all odd-order central co-moments, including coskewness, to vanish asymptotically.

Second, for cokurtosis ($k=4$), sign averaging automatically restores $\mathbb{A}_4$ and reduces the core to an $n\times n$ matrix. Indeed, $\mathbb{E}[\eta_a \eta_b \eta_c \eta_d] = 1$ if and only if every distinct index in $(a,b,c,d)$ appears an even number of times (2 or 4), and is zero otherwise. Thus, all entries of the form $\mathcal{G}_{iiij}$, $\mathcal{G}_{iijk}$, and $\mathcal{G}_{ijkl}$ with distinct indices vanish in expectation. The only surviving entries are the diagonal fourth powers $\mathcal{G}_{iiii}$ and the pairwise terms $\mathcal{G}_{iijj}$ (along with their six index permutations $\mathcal{G}_{ijij}, \mathcal{G}_{ijji}, \ldots$, which all share the same value by supersymmetry).

We therefore define the sign-invariant even projector $\Pi_{\mathrm{even}}$ on fourth-order cores by retaining only even-multiplicity index tuples:
\begin{equation}
\label{eq:even_projection}
\Pi_{\mathrm{even}}\!\left(\boldsymbol{\mathcal{G}}\right)_{abcd}
:=
\begin{cases}
\mathcal{G}_{abcd}, & \text{if } (a,b,c,d) \text{ is of the form } (i,i,i,i) \text{ or a permutation of } (i,i,j,j),\\
0, & \text{otherwise}.
\end{cases}
\end{equation}
As proved in SI, Sec.~\ref{sec:appevenprojection} (for any even order $k$), zeroing out odd-multiplicity tuples automatically restores the all-orthogonality constraint:
\begin{equation}
\label{eq:even_projection_allorthogonal}
\forall~\boldsymbol{\mathcal{G}}\in\mathbb{R}^{n\times n\times n\times n},
\qquad
\Pi_{\mathrm{even}}\!\left(\boldsymbol{\mathcal{G}}\right) \in \mathbb{A}_4,
\end{equation}
because taking the inner product of two distinct mode slices $a\neq a'$ always pairs an even index multiplicity with an odd one. Applying $\Pi_{\mathrm{even}}$ to Eq.~\eqref{eq:unconstrained_oracle} defines the fourth-order Oracle core:
\begin{equation}
\label{eq:oracle_solution}
\boldsymbol{\mathcal{G}}_{\mathrm{Oracle}} := \Pi_{\mathrm{even}}\!\left(\widetilde{\boldsymbol{\mathcal{G}}}\right) \in \mathbb{A}_4.
\end{equation}
Because all six permutations of $(i,i,j,j)$ are identical, any even fourth-order core $\boldsymbol{\mathcal{G}}_{\mathrm{even}} = \Pi_{\mathrm{even}}(\boldsymbol{\mathcal{G}})$ is completely parameterized by a single $n\times n$ symmetric matrix $\boldsymbol{B}$ with entries
\begin{equation}
\label{eq:B_matrix_def}
B_{ij} := \mathcal{G}_{iijj},
\qquad
B_{ii} = \mathcal{G}_{iiii}.
\end{equation}
For the oracle core in Eq.~\eqref{eq:oracle_solution}, this matrix is simply the matrix of cross-moments of squared rotated returns, $B_{\mathrm{Oracle}, ij} = \mathbb{E}[y_i^2 y_j^2 | \widehat{\boldsymbol{U}}]$. Consequently, we never need to store or manipulate an $n^4$ core tensor. For any portfolio weight vector $\boldsymbol{w}$ and rotated weights $\boldsymbol{v} = \widehat{\boldsymbol{U}}^\top \boldsymbol{w}$, contracting the reconstructed cokurtosis tensor $\widehat{\boldsymbol{\mathcal{K}}}(\boldsymbol{\mathcal{G}}_{\mathrm{even}})$ with $\boldsymbol{w}$ reduces to an $O(n^2)$ quadratic form in the squared rotated weights (SI, Sec.~\ref{app:evencore}):
\begin{equation}
\label{eq:m4even}
\hat{m}_4(\boldsymbol{w})
= \sum_{i=1}^n B_{ii} v_i^4 + 3\sum_{i\neq j} B_{ij} v_i^2 v_j^2
= 3\,(\boldsymbol{v}\circ\boldsymbol{v})^\top \boldsymbol{B}\,(\boldsymbol{v}\circ\boldsymbol{v}) - 2\sum_{i=1}^n B_{ii} v_i^4,
\end{equation}
where $\circ$ denotes the elementwise (Hadamard) product.

\section{Co-Moment Shrinkage}
\label{sec:comoment_shrinkage}

We now construct practical estimators that approximate the oracle core by estimating the orthogonal basis $\widehat{\boldsymbol{U}}$ on a training sample and calibrating the $n\times n$ even-core matrix $\boldsymbol{B}$ (along with the filtered eigenvalues for $k=2$) on an independent testing sample, in direct analogy with cross-validated covariance filtering \cite{bartz2016cross}.

\subsection{Cross-Validation estimator}
\label{sec:CVestimator}

Let the in-sample (IS) data be partitioned into train/test folds indexed by $b\in\mathcal{F}$. For fold $b$, let $\widehat{\boldsymbol{\mathcal{K}}}_{\mathrm{train}(b)}$ and $\widehat{\boldsymbol{\mathcal{K}}}_{\mathrm{test}(b)}$ denote the sample cokurtosis tensors computed on the training and testing subsamples, respectively. We first extract the orthogonal mode basis $\widehat{\boldsymbol{U}}_{\mathrm{train}(b)} \in O(n)$ from the symmetric HOSVD of $\widehat{\boldsymbol{\mathcal{K}}}_{\mathrm{train}(b)}$:
\begin{equation}
\label{eq:cv_train_hosvd}
\widehat{\boldsymbol{\mathcal{K}}}_{\mathrm{train}(b)}
=
\widehat{\boldsymbol{\mathcal{G}}}_{\mathrm{train}(b)}
\times_1 \widehat{\boldsymbol{U}}_{\mathrm{train}(b)}
\times_2 \widehat{\boldsymbol{U}}_{\mathrm{train}(b)}
\times_3 \widehat{\boldsymbol{U}}_{\mathrm{train}(b)}
\times_4 \widehat{\boldsymbol{U}}_{\mathrm{train}(b)}.
\end{equation}
Projecting the testing tensor onto $\widehat{\boldsymbol{U}}_{\mathrm{train}(b)}$ and applying $\Pi_{\mathrm{even}}$ yields the fold-$b$ even core
\begin{equation}
\label{eq:cv_test_core}
\widehat{\boldsymbol{\mathcal{G}}}_{(b)}
:=
\Pi_{\mathrm{even}}\!\left(
\widehat{\boldsymbol{\mathcal{K}}}_{\mathrm{test}(b)}
\times_1 \widehat{\boldsymbol{U}}_{\mathrm{train}(b)}^\top
\times_2 \widehat{\boldsymbol{U}}_{\mathrm{train}(b)}^\top
\times_3 \widehat{\boldsymbol{U}}_{\mathrm{train}(b)}^\top
\times_4 \widehat{\boldsymbol{U}}_{\mathrm{train}(b)}^\top
\right).
\end{equation}
In terms of the $n\times n$ matrix representation of Eq.~\eqref{eq:B_matrix_def}, if $\boldsymbol{y}_t = \widehat{\boldsymbol{U}}_{\mathrm{train}(b)}^\top(\boldsymbol{x}_t - \hat{\boldsymbol{\mu}}_{\mathrm{test}(b)})$ denotes the centered testing observations rotated into the training basis, Eq.~\eqref{eq:cv_test_core} amounts to computing the $n\times n$ matrix
\begin{equation}
\label{eq:Bempirical}
\widehat{B}_{(b), ij} = \frac{1}{\Delta t_{\mathrm{test}}} \sum_{t\in\mathrm{test}(b)} y_{it}^2\, y_{jt}^2.
\end{equation}
Averaging across all folds $b\in\mathcal{F}$ yields the cross-validated even core $\widehat{\boldsymbol{\mathcal{G}}}_{\mathrm{CV}}$ (parameterized by $\widehat{\boldsymbol{B}}_{\mathrm{CV}}$):
\begin{equation}
\label{eq:raw_cv_core}
\widehat{\boldsymbol{\mathcal{G}}}_{\mathrm{CV}} := \frac{1}{|\mathcal{F}|} \sum_{b\in\mathcal{F}} \widehat{\boldsymbol{\mathcal{G}}}_{(b)},
\qquad
\widehat{\boldsymbol{B}}_{\mathrm{CV}} = \frac{1}{|\mathcal{F}|} \sum_{b\in\mathcal{F}} \widehat{\boldsymbol{B}}_{(b)}.
\end{equation}
Finally, denoting by $\widehat{\boldsymbol{U}}$ the HOSVD basis of the full IS cokurtosis tensor, the Cross-Validation (CV) cokurtosis estimator is reconstructed as
\begin{equation}
\label{eq:cv_estimator}
\widehat{\boldsymbol{\mathcal{K}}}_{\mathrm{CV}}
:=
\widehat{\boldsymbol{\mathcal{G}}}_{\mathrm{CV}}
\times_1 \widehat{\boldsymbol{U}}
\times_2 \widehat{\boldsymbol{U}}
\times_3 \widehat{\boldsymbol{U}}
\times_4 \widehat{\boldsymbol{U}}.
\end{equation}
The exact same procedure applied at order $k=2$ (where $\widehat{\boldsymbol{B}}$ is replaced by the diagonal matrix of OOS variances $\frac{1}{\Delta t_{\mathrm{test}}}\sum_t y_{it}^2$ along the training eigenvectors) yields the cross-validated covariance estimator.

\subsection{Higher-Order Average Oracle estimator}
\label{sec:HOAOestimator}

Random cross-validation splits shuffle observations within the IS window, so that they mix data from different points in time and measure persistence within the same regime rather than forward transfer across time. To capture chronological persistence, the Higher-Order Average Oracle (HOAO) replaces random internal splits with historical pairs of chronological past-training and subsequent future-testing windows of lengths $\Delta t_{\mathrm{train}}$ and $\Delta t_{\mathrm{test}}$.

For each historical calibration date $\tau \in \mathcal{T}_{\mathrm{cal}}$, let $\widehat{\boldsymbol{\mathcal{K}}}_{\mathrm{train}(\tau)}$ and $\widehat{\boldsymbol{\mathcal{K}}}_{\mathrm{test}(\tau)}$ denote the sample cokurtosis tensors on the training window ending at $\tau$ and on the subsequent testing window. We extract the training basis $\widehat{\boldsymbol{U}}_{\mathrm{train}(\tau)}$ via the symmetric HOSVD
\begin{equation}
\label{eq:hoao_train_hosvd}
\widehat{\boldsymbol{\mathcal{K}}}_{\mathrm{train}(\tau)}
=
\widehat{\boldsymbol{\mathcal{G}}}_{\mathrm{train}(\tau)}
\times_1 \widehat{\boldsymbol{U}}_{\mathrm{train}(\tau)}
\times_2 \widehat{\boldsymbol{U}}_{\mathrm{train}(\tau)}
\times_3 \widehat{\boldsymbol{U}}_{\mathrm{train}(\tau)}
\times_4 \widehat{\boldsymbol{U}}_{\mathrm{train}(\tau)},
\end{equation}
and project the subsequent testing tensor onto $\widehat{\boldsymbol{U}}_{\mathrm{train}(\tau)}$ to obtain the realized oracle even core of window $\tau$:
\begin{equation}
\label{eq:hoao_oracle_core_tau}
\widehat{\boldsymbol{\mathcal{G}}}_{(\tau)}
:=
\Pi_{\mathrm{even}}\!\left(
\widehat{\boldsymbol{\mathcal{K}}}_{\mathrm{test}(\tau)}
\times_1 \widehat{\boldsymbol{U}}_{\mathrm{train}(\tau)}^\top
\times_2 \widehat{\boldsymbol{U}}_{\mathrm{train}(\tau)}^\top
\times_3 \widehat{\boldsymbol{U}}_{\mathrm{train}(\tau)}^\top
\times_4 \widehat{\boldsymbol{U}}_{\mathrm{train}(\tau)}^\top
\right),
\quad
\widehat{B}_{(\tau), ij} =\frac{1}{\Delta t_{\mathrm{test}}} \sum_{t\in\mathrm{test}(\tau)} y_{it}^2\, y_{jt}^2,
\end{equation}
where $\boldsymbol{y}_t = \widehat{\boldsymbol{U}}_{\mathrm{train}(\tau)}^\top(\boldsymbol{x}_t - \hat{\boldsymbol{\mu}}_{\mathrm{test}(\tau)})$. Averaging these realized cores across all historical calibration dates $\tau \in \mathcal{T}_{\mathrm{cal}}$ produces the time-independent HOAO core:
\begin{equation}
\label{eq:hoao_raw_core}
\widehat{\boldsymbol{\mathcal{G}}}_{\mathrm{HOAO}} := \frac{1}{|\mathcal{T}_{\mathrm{cal}}|} \sum_{\tau\in\mathcal{T}_{\mathrm{cal}}} \widehat{\boldsymbol{\mathcal{G}}}_{(\tau)},
\qquad
\widehat{\boldsymbol{B}}_{\mathrm{HOAO}} = \frac{1}{|\mathcal{T}_{\mathrm{cal}}|} \sum_{\tau\in\mathcal{T}_{\mathrm{cal}}} \widehat{\boldsymbol{B}}_{(\tau)}.
\end{equation}
Combining this pre-calibrated core with the HOSVD basis $\widehat{\boldsymbol{U}}$ of the current IS tensor yields the HOAO cokurtosis estimator:
\begin{equation}
\label{eq:HOAOest}
\widehat{\boldsymbol{\mathcal{K}}}_{\mathrm{HOAO}}
:=
\widehat{\boldsymbol{\mathcal{G}}}_{\mathrm{HOAO}}
\times_1 \widehat{\boldsymbol{U}}
\times_2 \widehat{\boldsymbol{U}}
\times_3 \widehat{\boldsymbol{U}}
\times_4 \widehat{\boldsymbol{U}}.
\end{equation}
At second order ($k=2$), the same historical averaging of realized OOS variances along training eigenvectors recovers the Average Oracle covariance estimator of Ref.~\cite{bongiorno2023filtering}.

Applying CV and HOAO to equity returns involves three further practical steps, namely separating asset volatilities from the dependence structure (Sec.~\ref{sec:standardization}), isolating the effect of $\Pi_{\mathrm{even}}$ without calibration (Sec.~\ref{sec:sample_even}), and addressing the joint consistency of separately filtered second and fourth moments (Sec.~\ref{sec:compatibility}).

\subsection{Standardization}
\label{sec:standardization}

Co-moment tensors reflect both individual asset volatilities, which set the scale of each return series, and the cross-sectional dependence structure. Because asset volatilities vary substantially over time, calibrating a core on raw historical returns would carry past volatility levels into the present. We therefore apply our filters to standardized returns and scale back the filtered tensors using the current IS sample volatilities.

Let $\widehat{\boldsymbol{D}}=\mathrm{diag}(\hat\sigma_1,\ldots,\hat\sigma_n)$ be the diagonal matrix of current IS sample volatilities, and define the standardized returns $\boldsymbol{z}_t=\widehat{\boldsymbol{D}}^{-1}(\boldsymbol{x}_t-\hat{\boldsymbol{\mu}})$. At order $k=2$, we filter the sample correlation matrix and restore its diagonal entries to unity to obtain $\widehat{\boldsymbol{C}}$. At order $k=4$, we filter the standardized cokurtosis tensor to obtain $\widehat{\boldsymbol{\mathcal{K}}}_z$. The two orders are filtered in their respective HOSVD bases and restored to the scale of raw returns via
\begin{equation}
\label{eq:destandardization}
\widehat{\boldsymbol{\Sigma}}=\widehat{\boldsymbol{D}}\,\widehat{\boldsymbol{C}}\,\widehat{\boldsymbol{D}},
\qquad
\widehat{\boldsymbol{\mathcal{K}}}=\widehat{\boldsymbol{\mathcal{K}}}_z\times_1\widehat{\boldsymbol{D}}\times_2\widehat{\boldsymbol{D}}\times_3\widehat{\boldsymbol{D}}\times_4\widehat{\boldsymbol{D}},
\end{equation}
so that the fourth central moment of a portfolio $\boldsymbol{w}$ is evaluated via Eq.~\eqref{eq:m4even} using the volatility-scaled rotated weights $\boldsymbol{v} = \widehat{\boldsymbol{U}}^\top\widehat{\boldsymbol{D}}\boldsymbol{w}$. During HOAO calibration, each training and testing window is standardized using its own sample mean and volatility vector. Following Ref.~\cite{bongiorno2023filtering}, calibration windows are drawn at random dates and for random subsets of $n$ assets, so that $\widehat{\boldsymbol{B}}_{\mathrm{HOAO}}$ captures a universal, asset-agnostic dependence structure.

\subsection{Isolating the effect of even projection without calibration}
\label{sec:sample_even}

Both CV and HOAO modify the sample cokurtosis tensor by combining two operations: first, they discard all odd-multiplicity core entries via $\Pi_{\mathrm{even}}$; second, they recalibrate the remaining $n\times n$ matrix $\boldsymbol{B}$ of even-multiplicity entries on out-of-sample validation or historical testing data. Because the projection alone removes $n^4 - 3n^2 + 2n$ noisy coefficients from the sample core, it could conceivably account for part of the error reduction. To separate the effect of the even projection from that of core calibration, we compare CV and HOAO not only with the standard plug-in sample estimator of Eq.~\eqref{eq:sample_estimators} (labeled Sample), but also with the uncalibrated even-projected sample estimator Sample$_{\mathrm{even}}$:
\begin{equation}
\label{eq:sample_even}
\widehat{\boldsymbol{\mathcal{K}}}_{\mathrm{even}}
:=
\Pi_{\mathrm{even}}\big(\widehat{\boldsymbol{\mathcal{G}}}\big)
\times_1 \widehat{\boldsymbol{U}}
\times_2 \widehat{\boldsymbol{U}}
\times_3 \widehat{\boldsymbol{U}}
\times_4 \widehat{\boldsymbol{U}},
\end{equation}
where $(\widehat{\boldsymbol{\mathcal{G}}},\widehat{\boldsymbol{U}})$ is the symmetric HOSVD of the current IS sample cokurtosis tensor (corresponding to $\widehat{B}_{\mathrm{IS}, ij} = \frac{1}{\Delta t_{\mathrm{IS}}}\sum_{t\in\mathrm{IS}} y_{it}^2 y_{jt}^2$). For $k=2$, the sample core is already diagonal, so Sample$_{\mathrm{even}}$ coincides with the standard sample estimator (Sample). For $k=4$, Sample$_{\mathrm{even}}$, CV, and HOAO share the exact same orthogonal basis $\widehat{\boldsymbol{U}}$ and differ only in their $n\times n$ even-core matrices ($\widehat{\boldsymbol{B}}_{\mathrm{IS}}$, $\widehat{\boldsymbol{B}}_{\mathrm{CV}}$, and $\widehat{\boldsymbol{B}}_{\mathrm{HOAO}}$).

\subsection{Distributional consistency of the filtered moments}
\label{sec:compatibility}
Portfolio applications rely on covariance and cokurtosis jointly, since portfolio kurtosis is a ratio of the two and the portfolio allocation problem of Sec.~\ref{sec:allocation} combines both in its objective. Because $\widehat{\boldsymbol{\Sigma}}$ and $\widehat{\boldsymbol{\mathcal{K}}}$ are filtered in separate bases with independently calibrated cores, nothing guarantees that they correspond to the moments of a single probability distribution. A necessary condition for such joint distributional consistency (as distinct from asymptotic statistical consistency as $\Delta t\to\infty$) is that every centered portfolio return $R = \boldsymbol{w}^\top(\boldsymbol{x}-\boldsymbol{\mu})$ satisfies the Cauchy--Schwarz inequality $\mathbb{E}[R^4]\geq\mathbb{E}[R^2]^2$, or equivalently $\kappa\geq 1$. Sample estimators satisfy this bound by construction, whereas separately filtered estimators may violate it. In our experiments, the HOAO kurtosis estimate falls below one for 3.16\% of random long-only portfolios with 100-day IS windows and for 0.051\% with 500-day windows. For CV, the corresponding shares are 0.54\% and 0\% (no violations among 500,000 portfolios for the longer window, SI, Sec.~\ref{app:compatibility}).

We tested several reconstruction schemes that enforce joint consistency while preserving the filtered covariance (SI, Sec.~\ref{app:compatibility}). Although the consistent HOAO variants yield valid moment pairs and further lower its tensor-level cokurtosis MSE at $n=50$, their resulting fourth-order tensors are close to those of an elliptical distribution whose fourth moment is a monotonic function of its variance; consequently, in the variance--fourth-moment allocation problem introduced below (Eq.~\eqref{eq:allocation}), penalizing the fourth moment has almost no effect, and the optimal allocation stays close to the minimum-variance portfolio. Directly minimizing the estimated kurtosis subject to a variance bound brings no general advantage either, whether for the original estimator or for the consistent corrections (SI, Sec.~\ref{app:kurtosis}). Even if all the moments do not stem from the same distribution, the fourth-moment form in Eq.~\eqref{eq:m4even} is the expectation of a fourth power under a sign-symmetrized mixture (SI, Sec.~\ref{app:compatibility}) and is therefore convex in the portfolio weights $\boldsymbol{w}$, so that the allocation problem is well-posed and strictly convex. We report the original estimators in the main text, while noting that this choice rests on an exploratory comparison and that cross-moment distributional inconsistency remains a conceptual limitation, not a practical one. Controlled experiments on synthetic Gaussian and Student-$t$ populations with known analytical moments further support the estimation accuracy of the proposed filters and their consistent variants (SI, Sec.~\ref{app:synthetic}).

\section{Empirical Study}
\label{sec:realdataapp}

We now evaluate the proposed estimators on U.S. equity returns at the levels of tensor estimation and portfolio optimization. After detailing the dataset and experimental protocol (Sec.~\ref{sec:data_protocol}), we measure how accurately the filtered tensors approximate their realized OOS counterparts (Sec.~\ref{sec:tensor_accuracy}). We next examine whether tensor-level gains translate into more accurate forecasts of portfolio volatility and kurtosis (Sec.~\ref{sec:portfolio_moments}) and into lower-risk optimized portfolios (Sec.~\ref{sec:allocation}). Finally, Section~\ref{app:stability} verifies the temporal stability of our findings across market regimes. Throughout the analysis, the sample estimator (Sample) serves as the primary benchmark, while Sample$_{\mathrm{even}}$ isolates the contribution of the even projection.

\subsection{Data and protocol}
\label{sec:data_protocol}

We use daily total returns and historical Russell 1000 membership from Norgate Data \cite{norgateData}. On each investment date $t$, we start with the securities in the index at the close and apply price, trading-history and liquidity filters (SI, Sec.~\ref{app:data}). We require average daily traded value of at least USD~5 million over the trailing three calendar months, on every listing exchange. We exclude real estate investment trusts (REITs).

Each experiment randomly draws a investment date $t$ and $n$ stocks from the eligible universe on that date. The In-Sample (IS) estimation window consists of the $\Delta t_{\mathrm{IS}}$ daily returns ending on day $t$. The Out-of-Sample (OOS) evaluation window comprises 100 daily returns starting on day $t+2$, allowing for execution at the next day's close. We keep the stock selection fixed through the OOS window and keep returns even if an asset is removed during the OOS period. When an OOS return is missing, we record a zero for that interval. The snapshot does not establish terminal cash proceeds.

For each portfolio size and IS length, we initialize the HOAO core with 10,000 historical pairs. Each pair contains $\Delta t_{\mathrm{IS}}$ training returns and 100 testing returns. The initial investment dates start in 1995 and all initial testing windows end by December 2009. Every two years thereafter, the calibration pool is expanded with 2,000 new window pairs drawn from the preceding two years, all terminating strictly before the update date of the portfolio. Out-of-sample evaluations span investment dates from January 2010 to August 2025, so that all OOS windows end by December 2025. We run 5,000 experiments for each portfolio size $n\in\{5,10,20,30,40,50\}$ with $\Delta t_{\mathrm{IS}}=100$ days and 5,000 experiments with $\Delta t_{\mathrm{IS}}=500$ days at $n=50$. The CV estimator uses 10 random splits of the IS window, with 65\% of the observations in the training subsample and 35\% in the testing subsample. All estimators are evaluated on the exact same draws of dates and assets (SI, Sec.~\ref{app:protocol}).

\subsection{Tensor accuracy}
\label{sec:tensor_accuracy}

We first evaluate estimation accuracy directly at the tensor level. For each moment order $k\in\{2,4\}$ and estimator $m\in\{\mathrm{Sample},\mathrm{Sample}_{\mathrm{even}},\mathrm{CV},\mathrm{HOAO}\}$, we measure the relative squared Frobenius error
\begin{equation}
    e_m^{(k)}
    :=
    \frac{
        \left\|\boldsymbol{\mathcal{M}}_m^{(k)}-\boldsymbol{\mathcal{M}}_{\mathrm{OOS}}^{(k)}\right\|_F^2
    }{
        \left\|\boldsymbol{\mathcal{M}}_{\mathrm{OOS}}^{(k)}\right\|_F^2
    },
    \qquad
    k\in\{2,4\},
    \quad
    m\in\{\mathrm{Sample},\mathrm{Sample}_{\mathrm{even}},\mathrm{CV},\mathrm{HOAO}\},
    \label{eq:tensor-loss}
\end{equation}
where $\boldsymbol{\mathcal{M}}^{(2)}=\boldsymbol{C}$ is the correlation matrix and $\boldsymbol{\mathcal{M}}^{(4)}=\boldsymbol{\mathcal{K}}_z$ is the standardized cokurtosis tensor. The tensor $\boldsymbol{\mathcal{M}}_m^{(k)}$ is the estimate produced by $m$ from the IS window, and $\boldsymbol{\mathcal{M}}_{\mathrm{OOS}}^{(k)}$ is the sample tensor of Eq.~\eqref{eq:sample_estimators} computed on standardized OOS returns. We refer to the average of $e_m^{(k)}$ across experiments as the relative OOS mean squared error (MSE). Because $\boldsymbol{\mathcal{M}}_{\mathrm{OOS}}^{(k)}$ is itself a finite-sample realized tensor rather than the unobservable population moment, the relative OOS MSE measures forecast accuracy with respect to realized OOS co-moments. It therefore includes sampling noise from the OOS benchmark and cannot be directly interpreted as the estimation error relative to the true population moments (SI, Sec.~\ref{app:inference}). As a benchmark, predicting the zero tensor yields a relative MSE of exactly one; any value above one indicates that the estimator is, on average, farther from the realized OOS tensor than a trivial zero forecast.

\begin{figure}[htbp]
    \centering
    \includegraphics[width=0.95\linewidth]{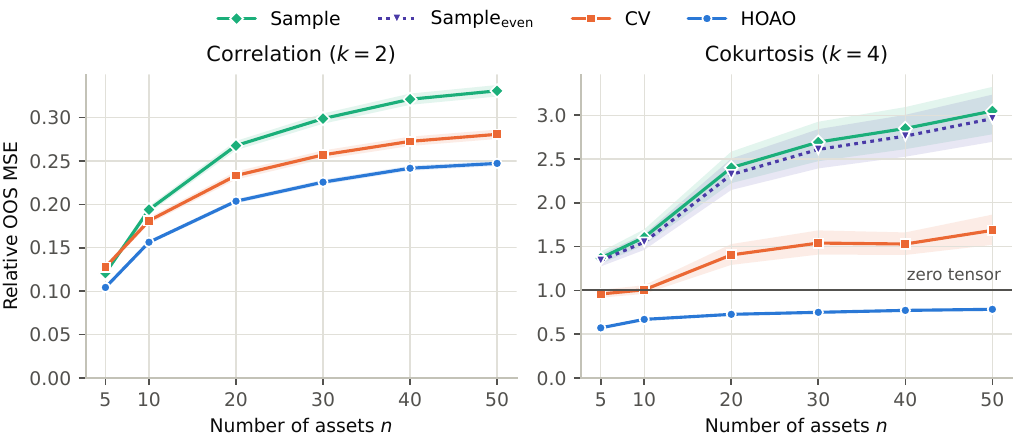}
    \caption{Relative OOS MSE of the correlation matrix (left) and of the standardized cokurtosis (right) against the number $n\in\{5,10,20,30,40,50\}$ of assets. $\Delta t_{\mathrm{IS}}=\Delta t_{\mathrm{OOS}}=100$ days; means over 5,000 experiments for each $n$. Bands are 95\% bootstrap intervals that treat experiments as independent (SI, Sec.~\ref{app:inference}). For $k=2$, Sample$_{\mathrm{even}}$ coincides with Sample. The horizontal line marks the relative MSE of the zero tensor.}
    \label{fig:accuracy_v_dim}
\end{figure}

Across the portfolio sizes shown in Fig.~\ref{fig:accuracy_v_dim}, HOAO achieves both the lowest correlation MSE and the lowest cokurtosis MSE among the four estimators. For cokurtosis, Sample exceeds the unit error of the zero tensor for every dimension, from 1.368 at $n=5$ to 3.048 at $n=50$. CV falls below one only at $n=5$ (0.958) and reaches 1.685 at $n=50$. HOAO's cokurtosis error stays below 1 (ranging from  0.572 for $n=5$ to 0.783 for $n=50$).

Comparing HOAO with the control Sample$_{\mathrm{even}}$ demonstrates that this improvement does not merely stem from zeroing out odd-multiplicity core entries. Applying $\Pi_{\mathrm{even}}$ without calibration barely reduces the cokurtosis MSE at $n=50$, from 3.048 to 2.961. Because Sample$_{\mathrm{even}}$, CV, and HOAO share the same orthogonal basis $\widehat{\boldsymbol{U}}$, the component of the OOS tensor orthogonal to the even subspace contributes identically to their errors and cancels out in pairwise MSE differences. The drop from 2.961 to 0.783 is therefore attributable to the out-of-sample calibration of the even core coefficients, which accounts for 96.2\% of HOAO's total error reduction relative to Sample, and 93.6\% of CV's reduction (SI, Table~\ref{tab:ablation}).

\subsection{Portfolio moment accuracy}
\label{sec:portfolio_moments}

While Frobenius norms weight all tensor entries equally, portfolio risk management ultimately depends on univariate portfolio moments, which contract these entries with portfolio weights. We therefore evaluate how accurately the estimators forecast the OOS volatility and kurtosis of random portfolios. In each experiment with $n=50$, we draw 100 long-only weight vectors from a uniform Dirichlet distribution on the simplex. Denoting by $\widehat\theta_{m,t,p}$ the IS forecast of $\theta\in\{\sigma,\kappa\}$ produced by estimator $m$ for portfolio $p$ in experiment $t$, and by $\theta_{\mathrm{OOS},t,p}$ its realized OOS counterpart, we compute the relative squared error
\begin{equation}
    e_{m,t,p}^{(\theta)}
    :=
    \left(
        \frac{
            \widehat\theta_{m,t,p}
            -\theta_{\mathrm{OOS},t,p}
        }{
            \theta_{\mathrm{OOS},t,p}
        }
    \right)^2,
    \label{eq:portfolio-loss}
\end{equation}
whose average across portfolios and experiments defines the relative OOS MSE of the portfolio moment.

\begin{table}[htbp]
\centering
\setlength{\tabcolsep}{7pt}
\begin{tabular}{llcccc}
\toprule
$\Delta t_{\mathrm{IS}}$ & Relative MSE & Sample & Sample$_{\mathrm{even}}$ & CV & HOAO \\
\midrule
\multirow{4}{*}{100} & Correlation & 0.331 & 0.331 & 0.281 & \textbf{0.247}$^{\dagger}$ \\
 & Cokurtosis & 3.048 & 2.961 & 1.685$^{\dagger}$ & \textbf{0.783}$^{\dagger}$ \\
 & Volatility & 0.202 & 0.202 & 0.190 & \textbf{0.101}$^{\dagger}$ \\
 & Kurtosis & 0.339$^{\dagger}$ & 0.335$^{\dagger}$ & 0.259$^{\dagger}$ & \textbf{0.148}$^{\dagger}$ \\
\midrule
\multirow{4}{*}{500} & Correlation & 0.296 & 0.296 & 0.279 & \textbf{0.229}$^{\dagger}$ \\
 & Cokurtosis & 8.157 & 8.051 & 6.087 & \textbf{0.715}$^{\dagger}$ \\
 & Volatility & 0.310 & 0.310 & 0.300 & \textbf{0.145}$^{\dagger}$ \\
 & Kurtosis & 2.146 & 2.117 & 1.671 & \textbf{0.124}$^{\dagger}$ \\
 
\bottomrule
\end{tabular}
\caption{Relative OOS MSE for $n=50$ assets of the correlation matrix and the cokurtosis (Eq.~\eqref{eq:tensor-loss}), and of the volatility and the kurtosis of 100 random long-only portfolios per experiment (Eq.~\eqref{eq:portfolio-loss}). 5,000 experiments for each IS length; $\Delta t_{\mathrm{OOS}}=100$ days. Sample$_{\mathrm{even}}$ coincides with Sample for the correlation and the volatility. Bold values are the smallest MSE of each row, and daggers mark the estimators in the 95\% model confidence set among the four estimators (SI, Sec.~\ref{app:inference}).}
\label{tab:portfolio-moment-errors}
\end{table}

Table~\ref{tab:portfolio-moment-errors} reports the portfolio-level and tensor-level MSEs at $n=50$ for IS windows of $\Delta t_{\mathrm{IS}}=100$ and $500$ days. For 100-day windows, the tensor-level gains carry over directly to portfolio moments. HOAO reduces the volatility MSE by 50\% relative to Sample and 47\% relative to CV. Its portfolio kurtosis MSE is 56\% lower than Sample's and 43\% lower than CV's. The uncalibrated projection reduces Sample's portfolio kurtosis MSE by 1.3\%.

Extending the IS estimation window to 500 days degrades the performance of sample-based estimators with respect to fourth-order quantities. The cokurtosis MSE of Sample rises from 3.048 to 8.157, and its portfolio kurtosis MSE from 0.339 to 2.146. Longer estimation windows are more likely to span extreme market events (such as the March 2020 crash), which can have a disproportionate influence on sample fourth moments, making them less representative of the subsequent 100-day OOS period. Because HOAO is calibrated on historical train/test window pairs matching the target window lengths, it dampens such transient extremes, and its errors remain stable and even improve slightly at $\Delta t_{\mathrm{IS}}=500$ days.

The daggers in Table~\ref{tab:portfolio-moment-errors} indicate the 95\% Model Confidence Set (MCS) \cite{MCS},  which identifies models that cannot be statistically distinguished from the best-performing model. To account for dependence caused by overlapping OOS windows, we compute the MCS using a circular block bootstrap with blocks of 252 consecutive trading days (approximately one trading year; SI, Sec.~\ref{app:inference}). HOAO belongs to the set for every metric and both IS lengths. At 100 days, CV joins HOAO for cokurtosis, and the set retains all four estimators for portfolio kurtosis despite their different mean errors. At 500 days, only HOAO remains for each metric.

\subsection{Portfolio allocation and out-of-sample efficient frontiers}
\label{sec:allocation}

\begin{figure}[htbp]
    \centering
    \includegraphics[width=0.95\linewidth]{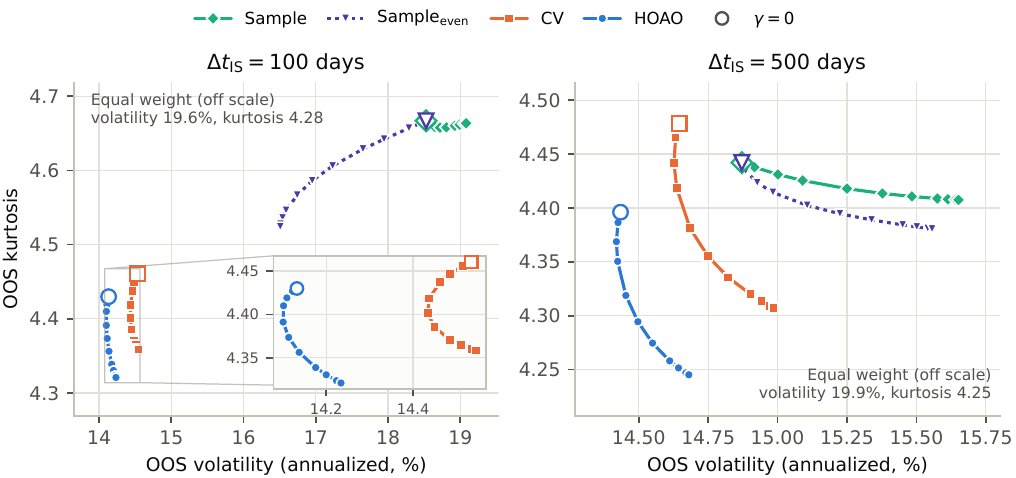}
    \caption{OOS efficient frontiers of the portfolios that solve Eq.~\eqref{eq:allocation} for $\gamma$ between 0 and 100, with $n=50$ assets and IS windows of 100 (left) and 500 days (right). Volatility is annualized; kurtosis is the mean OOS kurtosis. Each point is one value of $\gamma$, and open markers show the minimum-variance portfolios ($\gamma=0$). The equally weighted portfolio lies outside the plotted range; its values are given in a corner of each panel. The inset zooms on the CV and HOAO frontiers. 5,000 experiments for each IS length.}
    \label{fig:frontier}
\end{figure}

Accurate co-moment estimates are particularly important in portfolio selection with higher moments \cite{jondeau2006optimal,martellini2010improved,harvey2010portfolio}, because constrained optimizers over-allocate to directions where sampling noise leads risk to be underestimated, thereby amplifying in-sample estimation error \cite{michaud1989markowitz}. To test our estimators in an allocation setting, we construct portfolios that minimize a weighted sum of estimated variance and the fourth central moment, and we compare their realized OOS volatility and kurtosis across estimators. In each experiment with $n=50$, we solve
\begin{equation}
\label{eq:allocation}
\min_{\boldsymbol{w}}\;\frac{\hat\sigma^2(\boldsymbol{w})}{s^2}+\gamma\,\frac{\hat m_4(\boldsymbol{w})}{s^4}
\qquad\text{subject to}\qquad
\sum_{i=1}^n w_i=1,
\end{equation}
where $\hat m_4(\boldsymbol{w})=\hat\kappa(\boldsymbol{w})\,\hat\sigma^4(\boldsymbol{w})$ is the estimated fourth central moment of the portfolio, and $s^2$ denotes the IS variance of the equally weighted portfolio, which normalizes the penalty parameter $\gamma$ across dates and volatility regimes. Setting $\gamma=0$ yields the global minimum-variance portfolio, while increasing $\gamma$ places greater weight on fourth-moment reduction. We allow short sales and omit expected return forecasts to focus purely on risk control. Varying $\gamma$ from 0 to 100 and averaging the realized OOS volatility and kurtosis across experiments traces out the OOS variance--kurtosis efficient frontier of each estimator (Fig.~\ref{fig:frontier}).

For 100-day IS windows (Fig.~\ref{fig:frontier}, left), HOAO yields lower OOS volatility and kurtosis than CV, Sample and Sample$_{\mathrm{even}}$ for each of the 12 values of $\gamma$. One sees that using sample estimates only reduces marginally the OOS kurtosis when $\gamma$ increases, which is due to estimation noise. While HOAO and CV are sentitive to $\gamma$  roughly in the same way, HOAO leads to systematically lower kurtosis and volatility.

To assess the statistical significance of the differences shown in Fig.~\ref{fig:frontier}, we compute paired 95\% confidence intervals for the mean OOS variance and kurtosis differences between HOAO and each benchmark, using calendar-block bootstrap resampling (SI, Sec.~\ref{app:inference}). HOAO has lower OOS variance than Sample and Sample$_{\mathrm{even}}$ at all 12 penalties and both IS lengths, with every interval excluding zero. Against CV, the variance intervals exclude zero at all 12 penalties for 100 days and at the five penalties $\gamma\geq2$ for 500 days. HOAO's mean OOS kurtosis is lower than CV's at every penalty for both IS lengths. All 12 kurtosis intervals include zero at 100 days, while five of 12 exclude it at 500 days ($\gamma\leq0.2$). Against Sample and Sample$_{\mathrm{even}}$ the kurtosis intervals exclude zero at all 12 penalties at 100 days, and at the six penalties $\gamma\geq1$ at 500 days.

When $\gamma$ is large, the optimizer also tends to probe directions where the moments filtered separately violate the consistency bound discussed in Sec.~\ref{sec:compatibility}. With 100-day windows, the in-sample estimated kurtosis of optimized HOAO portfolios falls below one in 12.3\% of experiments at $\gamma=10$ and 13.2\% at $\gamma=100$, compared to 3.16\% for random long-only portfolios (SI, Table~\ref{tab:violations}). While the consistent variants eliminate such violations by construction, their near-ellipticity makes the fourth-moment penalty largely ineffectual, so that the OOS kurtosis stays essentially unchanged or slightly higher as $\gamma$ increases (SI, Fig.~\ref{fig:compatible_frontier} and Sec.~\ref{app:compatibility}).

As a further robustness check, the improvements in cokurtosis estimation are not driven solely by extreme market periods. Excluding 2025, the mean cokurtosis MSE with 100-day IS windows remains lower for HOAO (0.77) than for Sample (1.78). HOAO also achieves lower annual cokurtosis MSE than Sample in all 16 years, and than CV in 13 years with 100-day IS windows and all 16 years with 500-day windows (SI, Sec.~\ref{app:stability}).

\section{Conclusion}
\label{sec:conclusion}

Spectral covariance cleaning extends naturally to even-order co-moment tensors. By decomposing standardized co-moment tensors via a symmetric HOSVD, projecting the core onto even-multiplicity index tuples, and calibrating the remaining coefficients either by random cross-validation (CV) or across historical train/test windows (HOAO), we obtain tractable estimators whose portfolio moments and gradients can be evaluated in $O(n^2)$ operations. Across historical Russell 1000 constituents from 2010 to 2025, HOAO has the lowest mean correlation and cokurtosis errors among the four main estimators. An ablation against the uncalibrated projection Sample$_{\mathrm{even}}$ confirms that the cokurtosis gains largely stem from calibrating the core coefficients on historical data rather than from merely discarding odd-multiplicity entries. These tensor-level improvements translate directly into more accurate portfolio volatility and kurtosis forecasts, into variance--fourth-moment efficient frontiers with lower realized OOS volatility and kurtosis, and into more stable performance in extreme market regimes which distort long-window sample estimators.

Two main limitations point to avenues for future research. First, because singular vectors are defined only up to a sign, basis-agnostic core averaging annihilates odd-order co-moments; filtering coskewness therefore requires estimators that explicitly account for factor orientation. Second, filtering covariance and cokurtosis in separate bases does not guarantee that the resulting estimates are jointly consistent as the moments of a single distribution. While the consistent HOAO reconstructions explored in the SI enforce $\kappa\geq 1$ and further lower the tensor-level cokurtosis MSE, they bring the fourth-order structure close to an elliptical tensor, so that the fourth-moment penalty has almost no effect in portfolio optimization. Designing jointly consistent filters that preserve directional tail anisotropy, and integrating filtered higher moments into semi-parametric quantile expansions such as Cornish--Fisher for Value-at-Risk and Expected Shortfall \cite{CornishFisher1937,Withers1984,MaillardCornishFisher,AmedeeComputation}, are promising directions for future work.

\section*{Acknowledgments}

P.R. acknowledges funding from ANRT, under the CIFRE contract nr.~2025/0279.

This work used HPC resources from the ``Mésocentre'' computing center of CentraleSupélec and École Normale Supérieure Paris-Saclay supported by CNRS and Région Île-de-France.

\bibliographystyle{unsrt}  
\bibliography{references}

@article{MaillardCornishFisher,
author = {Maillard, Didier},
year = {2012},
month = {02},
pages = {},
title = {A User’s Guide to the Cornish Fisher Expansion},
journal = {SSRN Electronic Journal},
doi = {10.2139/ssrn.1997178}
}

@article{AmedeeComputation,
author = {Amédée-Manesme, Charles-Olivier and Barthélémy, Fabrice and Maillard, Didier},
year = {2019},
month = {10},
pages = {},
title = {Computation of the corrected {C}ornish–{F}isher expansion using the response surface methodology: application to {VaR} and {CVaR}},
volume = {281},
journal = {Annals of Operations Research},
doi = {10.1007/s10479-018-2792-4}
}

@article{harvey2010portfolio,
  title={Portfolio selection with higher moments},
  author={Harvey, Campbell R and Liechty, John C and Liechty, Merrill W and M{\"u}ller, Peter},
  journal={Quantitative Finance},
  volume={10},
  number={5},
  pages={469--485},
  year={2010},
  publisher={Taylor \& Francis}
}

@article{martellini2010improved,
  title={Improved estimates of higher-order comoments and implications for portfolio selection},
  author={Martellini, Lionel and Ziemann, Volker},
  journal={The Review of Financial Studies},
  volume={23},
  number={4},
  pages={1467--1502},
  year={2010},
  publisher={Oxford University Press}
}

@article{ledoit2004well,
  title={A well-conditioned estimator for large-dimensional covariance matrices},
  author={Ledoit, Olivier and Wolf, Michael},
  journal={Journal of multivariate analysis},
  volume={88},
  number={2},
  pages={365--411},
  year={2004},
  publisher={Elsevier}
}

@article{stuaricua2005nonstationarities,
  title={Nonstationarities in stock returns},
  author={St{\u{a}}ric{\u{a}}, C{\u{a}}t{\u{a}}lin and Granger, Clive},
  journal={Review of economics and statistics},
  volume={87},
  number={3},
  pages={503--522},
  year={2005},
  publisher={MIT Press 238 Main St., Suite 500, Cambridge, MA 02142-1046, USA journals~…}
}

@article{laloux1999noise,
  title={Noise dressing of financial correlation matrices},
  author={Laloux, Laurent and Cizeau, Pierre and Bouchaud, Jean-Philippe and Potters, Marc},
  journal={Physical review letters},
  volume={83},
  number={7},
  pages={1467},
  year={1999},
  publisher={APS}
}

@article{kondor2007noise,
  title={Noise sensitivity of portfolio selection under various risk measures},
  author={Kondor, Imre and Pafka, Szil{\'a}rd and Nagy, G{\'a}bor},
  journal={Journal of Banking \& Finance},
  volume={31},
  number={5},
  pages={1545--1573},
  year={2007},
  publisher={Elsevier}
}

@article{bun2017cleaning,
  title={Cleaning large correlation matrices: tools from random matrix theory},
  author={Bun, Jo{\"e}l and Bouchaud, Jean-Philippe and Potters, Marc},
  journal={Physics Reports},
  volume={666},
  pages={1--109},
  year={2017},
  publisher={Elsevier}
}

@article{ledoit2011eigenvectors,
  title={Eigenvectors of some large sample covariance matrix ensembles},
  author={Ledoit, Olivier and P{\'e}ch{\'e}, Sandrine},
  journal={Probability Theory and Related Fields},
  volume={151},
  number={1},
  pages={233--264},
  year={2011},
  publisher={Springer}
}

@article{ledoit2003improved,
  title={Improved estimation of the covariance matrix of stock returns with an application to portfolio selection},
  author={Ledoit, Olivier and Wolf, Michael},
  journal={Journal of empirical finance},
  volume={10},
  number={5},
  pages={603--621},
  year={2003},
  publisher={Elsevier}
}

@article{ledoit2012nonlinear,
  title={Nonlinear shrinkage estimation of large-dimensional covariance matrices},
  author={Ledoit, Olivier and Wolf, Michael},
  journal={The Annals of Statistics},
  volume={40},
  number={2},
  pages={1024--1060},
  year={2012}
}

@article{bongiorno2023filtering,
  title={Filtering time-dependent covariance matrices using time-independent eigenvalues},
  author={Bongiorno, Christian and Challet, Damien and Loeper, Gr{\'e}goire},
  journal={Journal of Statistical Mechanics: Theory and Experiment},
  volume={2023},
  number={2},
  pages={023402},
  year={2023},
  publisher={IOP Publishing}
}

@article{bongiorno2024averageoracle,
  title={Covariance matrix filtering and portfolio optimisation: the average oracle vs non-linear shrinkage and all the variants of {DCC-NLS}},
  author={Bongiorno, Christian and Challet, Damien},
  journal={Quantitative Finance},
  volume={24},
  number={9},
  pages={1227--1234},
  year={2024},
  publisher={Taylor \& Francis}
}

@article{de2000multilinear,
  title={A multilinear singular value decomposition},
  author={De Lathauwer, Lieven and De Moor, Bart and Vandewalle, Joos},
  journal={SIAM journal on Matrix Analysis and Applications},
  volume={21},
  number={4},
  pages={1253--1278},
  year={2000},
  publisher={SIAM}
}

@article{kolda2009tensor,
  title={Tensor decompositions and applications},
  author={Kolda, Tamara G and Bader, Brett W},
  journal={SIAM review},
  volume={51},
  number={3},
  pages={455--500},
  year={2009},
  publisher={SIAM}
}

@article{cichocki2015tensor,
  title={Tensor decompositions for signal processing applications: From two-way to multiway component analysis},
  author={Cichocki, Andrzej and Mandic, Danilo and De Lathauwer, Lieven and Zhou, Guoxu and Zhao, Qibin and Caiafa, Cesar and Phan, Huy Anh},
  journal={IEEE signal processing magazine},
  volume={32},
  number={2},
  pages={145--163},
  year={2015},
  publisher={IEEE}
}

@article{michaud1989markowitz,
  title={The Markowitz optimization enigma: Is ‘optimized’optimal?},
  author={Michaud, Richard O},
  journal={Financial analysts journal},
  volume={45},
  number={1},
  pages={31--42},
  year={1989},
  publisher={Taylor \& Francis}
}

@article{burda2022cleaning,
  title={Cleaning large-dimensional covariance matrices for correlated samples},
  author={Burda, Zdzislaw and Jarosz, Andrzej},
  journal={Physical Review E},
  volume={105},
  number={3},
  pages={034136},
  year={2022},
  publisher={APS}
}

@article{bartz2016cross,
  title={Cross-validation based nonlinear shrinkage},
  author={Bartz, Daniel},
  journal={arXiv preprint arXiv:1611.00798},
  year={2016}
}

@article{MCS,
author = {Hansen, Peter R. and Lunde, Asger and Nason, James M.},
title = {The Model Confidence Set},
journal = {Econometrica},
volume = {79},
number = {2},
pages = {453-497},
doi = {https://doi.org/10.3982/ECTA5771},
year = {2011}
}

@article{joanes1998comparing,
  title={Comparing measures of sample skewness and kurtosis},
  author={Joanes, Derrick N and Gill, Christine A},
  journal={Journal of the Royal Statistical Society: Series D (The Statistician)},
  volume={47},
  number={1},
  pages={183--189},
  year={1998},
  publisher={Wiley Online Library}
}

@article{cont2013running,
  title={Running for the exit: distressed selling and endogenous correlation in financial markets},
  author={Cont, Rama and Wagalath, Lakshithe},
  journal={Mathematical Finance: An International Journal of Mathematics, Statistics and Financial Economics},
  volume={23},
  number={4},
  pages={718--741},
  year={2013},
  publisher={Wiley Online Library}
}

@article{CornishFisher1937,
  author  = {Cornish, E. A. and Fisher, R. A.},
  title   = {Moments and Cumulants in the Specification of Distributions},
  journal = {Review of the International Statistical Institute},
  volume  = {5},
  number  = {4},
  pages   = {307--322},
  year    = {1937}
}

@article{Withers1984,
  author  = {Withers, Christopher S.},
  title   = {Asymptotic Expansions for Distributions and Quantiles with Power Series Cumulants},
  journal = {Journal of the Royal Statistical Society: Series B (Methodological)},
  volume  = {46},
  number  = {3},
  pages   = {389--396},
  year    = {1984}
}

@article{anandkumar2014tensor,
  title={Tensor decompositions for learning latent variable models.},
  author={Anandkumar, Animashree and Ge, Rong and Hsu, Daniel J and Kakade, Sham M and Telgarsky, Matus and others},
  journal={J. Mach. Learn. Res.},
  volume={15},
  number={1},
  pages={2773--2832},
  year={2014}
}

@article{afendras2020limiting,
  title={On the limiting distribution of sample central moments},
  author={Afendras, Georgios and Papadatos, Nickos and Piperigou, Violetta E},
  journal={Annals of the Institute of Statistical Mathematics},
  volume={72},
  number={2},
  pages={399--425},
  year={2020},
  publisher={Springer}
}

@incollection{politis1992circular,
  title={A circular block-resampling procedure for stationary data},
  author={Politis, Dimitris N. and Romano, Joseph P.},
  booktitle={Exploring the Limits of Bootstrap},
  editor={LePage, Raoul and Billard, Lynne},
  pages={263--270},
  year={1992},
  publisher={Wiley},
  address={New York}
}

@article{patton2011volatility,
  title={Volatility forecast comparison using imperfect volatility proxies},
  author={Patton, Andrew J.},
  journal={Journal of Econometrics},
  volume={160},
  number={1},
  pages={246--256},
  year={2011},
  publisher={Elsevier}
}

@misc{norgateData,
  author = {{Norgate Data}},
  title = {{U.S. equities and historical index constituents}},
  year = {2026},
  howpublished = {\href{https://norgatedata.com/data-content-tables.php}{Data content tables}
                  and \href{https://norgatedata.com/data-package-faq.php}{data definitions}},
  note = {Data snapshot of October 30, 2026}
}

@misc{nasdaqLiquidity,
  author = {{Nasdaq}},
  title = {{Nasdaq-100 Index Product Guide}},
  howpublished = {\href{https://www.nasdaq.com/docs/nasdaq-100-index-product-guide}{Index product guide}},
  note = {Eligibility criterion for average daily traded value. Accessed November 2, 2026}
}

@article{jondeau2006optimal,
  title={Optimal portfolio allocation under higher moments},
  author={Jondeau, Eric and Rockinger, Michael},
  journal={European Financial Management},
  volume={12},
  number={1},
  pages={29--55},
  year={2006},
  doi={10.1111/j.1354-7798.2006.00309.x},
  publisher={Wiley}
}

@article{pearson1916mathematical,
  title={Mathematical contributions to the theory of evolution. {XIX}. {S}econd supplement to a memoir on skew variation},
  author={Pearson, Karl},
  journal={Philosophical Transactions of the Royal Society of London. Series A},
  volume={216},
  pages={429--457},
  year={1916},
  doi={10.1098/rsta.1916.0009}
}

@book{boyd2004convex,
  title={Convex Optimization},
  author={Boyd, Stephen and Vandenberghe, Lieven},
  year={2004},
  publisher={Cambridge University Press},
  address={Cambridge}
}

@book{mccullagh1987tensor,
  title={Tensor Methods in Statistics},
  author={McCullagh, Peter},
  year={1987},
  publisher={Chapman and Hall},
  address={London}
}

@book{brillinger2001time,
  title={Time Series: Data Analysis and Theory},
  author={Brillinger, David R.},
  year={2001},
  publisher={SIAM},
  address={Philadelphia}
}

@book{horn2013matrix,
  title={Matrix Analysis},
  author={Horn, Roger A. and Johnson, Charles R.},
  edition={2nd},
  year={2013},
  publisher={Cambridge University Press},
  address={Cambridge}
}

@book{fang1990symmetric,
  title={Symmetric Multivariate and Related Distributions},
  author={Fang, Kai-Tai and Kotz, Samuel and Ng, Kai Wang},
  year={1990},
  publisher={Chapman and Hall},
  address={London}
}

@article{mardia1970measures,
  title={Measures of multivariate skewness and kurtosis with applications},
  author={Mardia, K. V.},
  journal={Biometrika},
  volume={57},
  number={3},
  pages={519--530},
  year={1970},
  doi={10.1093/biomet/57.3.519}
}

@article{berkane1986moments,
  title={Moments of elliptically distributed random variates},
  author={Berkane, Maia and Bentler, Peter M.},
  journal={Statistics \& Probability Letters},
  volume={4},
  number={6},
  pages={333--335},
  year={1986},
  doi={10.1016/0167-7152(86)90054-4}
}

@article{jammalamadaka2021multivariate,
  title={On multivariate skewness and kurtosis},
  author={Jammalamadaka, S. Rao and Taufer, Emanuele and Terdik, Gy{\"o}rgy H.},
  journal={Sankhya A},
  volume={83},
  number={2},
  pages={607--644},
  year={2021},
  doi={10.1007/s13171-020-00211-6}
}

@article{tofallis2015better,
  title={A better measure of relative prediction accuracy for model selection and model estimation},
  author={Tofallis, Chris},
  journal={Journal of the Operational Research Society},
  volume={66},
  number={8},
  pages={1352--1362},
  year={2015},
  doi={10.1057/jors.2014.103}
}

@book{hardy1952inequalities,
  title={Inequalities},
  author={Hardy, G. H. and Littlewood, J. E. and P{\'o}lya, G.},
  edition={2nd},
  year={1952},
  publisher={Cambridge University Press},
  address={Cambridge}
}

\clearpage
\appendix
\renewcommand{\thesection}{S\arabic{section}}
\renewcommand{\theHsection}{S\arabic{section}}
\setcounter{figure}{0}
\renewcommand{\thefigure}{S\arabic{figure}}
\renewcommand{\theHfigure}{S\arabic{figure}}
\setcounter{table}{0}
\renewcommand{\thetable}{S\arabic{table}}
\renewcommand{\theHtable}{S\arabic{table}}
\setcounter{equation}{0}
\renewcommand{\theequation}{S\arabic{equation}}
\renewcommand{\theHequation}{S\arabic{equation}}

\begin{center}
{\Large\bfseries Supplementary Information}
\end{center}

\section{Tensor algebra}
\subsection{General definitions and symmetric HOSVD}
\label{sec:appmodematrices}
The main text presents the framework for $k\in\{2,4\}$. This section gives the definitions at any order $k$, on which the proofs of the next sections rely.

For a random vector $\boldsymbol{x}\in\mathbb{R}^n$ with mean $\boldsymbol{\mu}$ and finite moments of order $k$, the central co-moment tensor of order $k$ is
\begin{equation}
\label{eq:comoment_order_k}
\boldsymbol{\mathcal{M}}^{(k)} := \mathbb{E}\!\left[(\boldsymbol{x} - \boldsymbol{\mu})^{\otimes k}\right]
\in (\mathbb{R}^n)^{\otimes k},
\qquad
\mathcal{M}^{(k)}_{i_1,\dots,i_k}
= \mathbb{E}\!\left[(x_{i_1}-\mu_{i_1})\cdots(x_{i_k}-\mu_{i_k})\right],
\end{equation}
for all $(i_1,\dots,i_k)\in[\![1,n]\!]^k$. It has $n^k$ entries, with $\boldsymbol{\Sigma}=\boldsymbol{\mathcal{M}}^{(2)}$, $\boldsymbol{\mathcal{S}}=\boldsymbol{\mathcal{M}}^{(3)}$, and $\boldsymbol{\mathcal{K}}=\boldsymbol{\mathcal{M}}^{(4)}$. Its sample version $\widehat{\boldsymbol{\mathcal{M}}}^{(k)}$ replaces the expectation by the average over the observations, centered at the sample mean. A tensor $\boldsymbol{\mathcal{T}}\in\mathbb{R}^{n\times\cdots\times n}$ of order $k$ is called supersymmetric if its entries are invariant under any permutation of its indices,
\begin{equation}
\label{eq:supersymmetry}
\mathcal{T}_{i_{\sigma(1)},\dots,i_{\sigma(k)}}
=
\mathcal{T}_{i_1,\dots,i_k},
\qquad
\forall~(i_1,\dots,i_k)\in[\![1,n]\!]^k,~
\forall~\sigma\text{ permutation of }\{1,\dots,k\}.
\end{equation}
The entries of $\boldsymbol{\mathcal{M}}^{(k)}$ are expectations of products of centered components, which do not depend on the order of the factors, so that $\boldsymbol{\mathcal{M}}^{(k)}$ and its sample version are supersymmetric.

For a tensor $\boldsymbol{\mathcal{T}} \in \mathbb{R}^{n_1 \times \cdots \times n_k}$, the mode-$r$ unfolding (or matricization) takes mode $r$ as the row index and merges all other modes into a single column index (see Fig.~\ref{fig:mode_2_unfolding} of the main text),
\begin{equation}
\label{eq:unfolding}
\mathrm{unfold}_r\boldsymbol{\mathcal{T}} \equiv \boldsymbol{T}_{(r)} \in \mathbb{R}^{n_r \times p_r}, \quad p_r:=  \prod_{j \neq r}^k n_j .
\end{equation}
The mode-$r$ product of $\boldsymbol{\mathcal{T}}$ by a matrix $\boldsymbol{A}\in \mathbb{R}^{m \times n_r}$ is the tensor $\boldsymbol{\mathcal{T}} \times_r \boldsymbol{A}\in \mathbb{R}^{n_1 \times \cdots \times n_{r-1} \times m \times n_{r+1} \times \cdots \times n_k}$ with entries
\begin{equation}
\label{eq:mode_product}
(\boldsymbol{\mathcal{T}} \times_r \boldsymbol{A})_{i_1 \cdots i_{r-1} j_r i_{r+1} \cdots i_k}
:= \sum_{i_r=1}^{n_r} \mathcal{T}_{i_1 \cdots i_r\cdots i_k} \, A_{j_r i_r}.
\end{equation}
This is the matrix multiplication on the mode-$r$ unfolding,
\begin{equation}
\label{eq:mode_product_unfolding}
\mathrm{unfold}_r
\left(
\boldsymbol{\mathcal{T}}\times_r \boldsymbol{A}
\right)
=
\boldsymbol{A}\,
\mathrm{unfold}_r
\boldsymbol{\mathcal{T}}.
\end{equation}

Let $\boldsymbol{\mathcal{M}}\in\mathbb{R}^{n\times\cdots\times n}$ be an order-$k$ supersymmetric tensor, such as $\boldsymbol{\mathcal{M}}^{(k)}$. By supersymmetry, all of its mode-$r$ unfoldings are identical, that is, $\boldsymbol{M}_{(r)} = \boldsymbol{M}_{(1)}$ for all $r \in [\![1,k]\!]$. In practice, however, computing separate SVDs across modes may yield distinct left singular vectors in the presence of degenerate singular values or sign ambiguities. Instead of estimating $k$ potentially different mode matrices, we enforce a single shared orthogonal basis $\boldsymbol{U}$ extracted from the mode-$1$ unfolding $\boldsymbol{M}_{(1)} \in \mathbb{R}^{n \times n^{k-1}}$ of $\boldsymbol{\mathcal{M}}$. Its matrix SVD is
\begin{equation}
\boldsymbol{M}_{(1)} = \boldsymbol{U} \boldsymbol{\Lambda} \boldsymbol{V}^\top.
\end{equation}
We then define the core tensor by contracting all $k$ modes of $\boldsymbol{\mathcal{M}}$ with $\boldsymbol{U}^\top$:
\begin{equation}
\boldsymbol{\mathcal{G}} := \boldsymbol{\mathcal{M}}\times_1 \boldsymbol{U}^\top \times_2 \boldsymbol{U}^\top \cdots \times_k \boldsymbol{U}^\top,
\end{equation}
where $\boldsymbol{\mathcal{G}}$ belongs to the all-orthogonal set
\begin{equation}
\label{eq:coreconstraint}
\boldsymbol{\mathcal{G}}^{(k)} \in \mathbb{A}_k := \left\{ \boldsymbol{\mathcal{A}} \in \mathbb{R}^{n \times \cdots \times n} ~\middle|~ \forall~ r \leq k, ~ \forall ~\alpha, ~\beta \leq n, ~ \left\langle \boldsymbol{\mathcal{A}}_{i_r = \alpha}, \boldsymbol{\mathcal{A}}_{i_r = \beta} \right\rangle = 0 ~ \textrm{if} ~ \alpha \neq \beta \right\},
\end{equation}
and $\langle \bullet, \bullet \rangle$ denotes the canonical Frobenius inner product on the tensor space, $\langle\boldsymbol{\mathcal{A}},\boldsymbol{\mathcal{B}} \rangle := \sum_{i_1, \ldots, i_k} \mathcal{A}_{i_1 \ldots i_k} \mathcal{B}_{i_1 \ldots i_k}$. Equivalently, slices along any given mode are mutually orthogonal. Any candidate core tensor must satisfy these constraints to qualify as a valid HOSVD core.

In component form, the adjoint relation reads
\begin{equation}
\mathcal{G}_{a_1\ldots a_k} = \sum_{i_1,\ldots,i_k=1}^n
U_{i_1 a_1}\cdots U_{i_k a_k}\,\mathcal{M}_{i_1\ldots i_k}.
\end{equation}
Since $\boldsymbol{U}$ is orthogonal, this construction is the symmetric HOSVD of order $k$,
\begin{equation}
\label{eq:symhosvd_core_k}
\boldsymbol{\mathcal{M}}
= \boldsymbol{\mathcal{G}}\times_1 \boldsymbol{U} \times_2 \boldsymbol{U} \cdots \times_k \boldsymbol{U},
\qquad
\boldsymbol{\mathcal{G}} \in \mathbb{A}_k,
\quad
\boldsymbol{U} \in O(n),
\end{equation}
which is Eq.~\eqref{eq:symhosvd_core} for $k=4$. The equality of the mode matrices is enforced by construction through a single unfolding, rather than emerging from $k$ separate SVDs as in the general HOSVD algorithm of Ref.~\cite{de2000multilinear}. This reduces memory requirements (only one orthogonal matrix is stored per co-moment tensor) without any loss in reconstruction accuracy. Note that when applied to a symmetric matrix $\boldsymbol{\Sigma}$ with eigenvalue decomposition $\boldsymbol{\Sigma} = \boldsymbol{V} \boldsymbol{\Lambda} \boldsymbol{V}^\top$, this procedure reduces exactly to standard spectral diagonalization. Since $\boldsymbol{\Sigma} = \boldsymbol{\Sigma}_{(1)}$, the left singular vectors coincide with the eigenvectors $\boldsymbol{V}$, and the resulting second-order core is $\boldsymbol{G} = \boldsymbol{\Sigma} \times_1 \boldsymbol{V}^\top \times_2 \boldsymbol{V}^\top = \boldsymbol{V}^\top \boldsymbol{\Sigma} \boldsymbol{V} = \boldsymbol{\Lambda}$.

\subsection{Proof of the unconstrained oracle core}
\label{app:proofoptimality}
This section proves for any order $k\ge 2$ the explicit solution of the unconstrained oracle problem stated for $k=4$ in Sec.~\ref{sec:oracle_comoment}. Let $\boldsymbol{\mathcal{M}}\in\mathbb{R}^{n\times\cdots\times n}$ be any tensor, for instance $\boldsymbol{\mathcal{M}}^{(k)}$. Fix an orthogonal mode matrix $\boldsymbol{U}\in O(n)$ and define the multilinear reconstruction map for this matrix (Eq.~\eqref{eq:oracle_reconstruction_map} is the case $k=4$ and $\boldsymbol{U}=\widehat{\boldsymbol{U}}$)
\begin{equation}
\label{eq:reconstruction_map}
\widehat{\boldsymbol{\mathcal{M}}}(\boldsymbol{\mathcal{G}})
\;:=\;
\boldsymbol{\mathcal{G}}\times_1 \boldsymbol{U}  \ldots \times_k \boldsymbol{U},
\qquad
\boldsymbol{\mathcal{G}}\in\mathbb{R}^{n\times\cdots\times n}.
\end{equation}
We equip the tensor space $(\mathbb{R}^{n})^{\otimes k}$ with the canonical Frobenius inner product and norm:
\[
\langle \boldsymbol{\mathcal{A}},\boldsymbol{\mathcal{B}}\rangle
\;:=\;
\sum_{i_1,\dots,i_k} \mathcal{A}_{i_1\dots i_k}\,\mathcal{B}_{i_1\dots i_k},
\qquad
\|\boldsymbol{\mathcal{A}}\|_F^2:=\langle \boldsymbol{\mathcal{A}},\boldsymbol{\mathcal{A}}\rangle.
\]

For any core tensor $\boldsymbol{\mathcal{G}}$, the map in Eq.~\eqref{eq:reconstruction_map} is an isometry with respect to the Frobenius norm:
\begin{equation}
\label{eq:isometry_identity}
\|\widehat{\boldsymbol{\mathcal{M}}}(\boldsymbol{\mathcal{G}})\|_F
=
\|\boldsymbol{\mathcal{G}}\|_F.
\end{equation}
Indeed, expanding the multilinear rotation entrywise gives
\[
\widehat{\mathcal{M}}(\mathcal{G})_{j_1\dots j_k}
=
\sum_{a_1=1}^{n}\cdots\sum_{a_k=1}^{n}
U_{j_1 a_1}\cdots U_{j_k a_k}\,\mathcal{G}_{a_1\dots a_k},
\]
so that
\begin{align*}
\|\widehat{\boldsymbol{\mathcal{M}}}(\boldsymbol{\mathcal{G}})\|_F^2
&=
\sum_{j_1,\dots,j_k}
\widehat{\mathcal{M}}(\mathcal{G})_{j_1\dots j_k}\;
\widehat{\mathcal{M}}(\mathcal{G})_{j_1\dots j_k} \\
&=
\sum_{j_1,\dots,j_k}
\sum_{a_1,\dots,a_k}
\sum_{b_1,\dots,b_k}
\left(\prod_{m=1}^k U_{j_m a_m}\,U_{j_m b_m}\right)
\mathcal{G}_{a_1\dots a_k}\,\mathcal{G}_{b_1\dots b_k}.
\end{align*}
Factoring the sums over $j_1,\dots,j_k$ across modes yields
\[
\|\widehat{\boldsymbol{\mathcal{M}}}(\boldsymbol{\mathcal{G}})\|_F^2
=
\sum_{a_1,\dots,a_k}
\sum_{b_1,\dots,b_k}
\left(\prod_{m=1}^k \sum_{j_m} U_{j_m a_m}U_{j_m b_m}\right)
\mathcal{G}_{a_1\dots a_k}\,\mathcal{G}_{b_1\dots b_k}.
\]
Because $\boldsymbol{U}\in O(n)$ satisfies $\sum_{j_m} U_{j_m a_m}U_{j_m b_m}=\delta_{a_m b_m}$, this expression simplifies to
\[
\|\widehat{\boldsymbol{\mathcal{M}}}(\boldsymbol{\mathcal{G}})\|_F^2
=
\sum_{a_1,\dots,a_k}\mathcal{G}_{a_1\dots a_k}^2
=
\|\boldsymbol{\mathcal{G}}\|_F^2.
\]

Next, we define the adjoint map
\begin{equation}
\label{eq:adjoint_map}
\widehat{\boldsymbol{\mathcal{M}}}^{\,*}(\boldsymbol{\mathcal{S}})
\;:=\;
\boldsymbol{\mathcal{S}}\times_1 \boldsymbol{U}^\top \cdots \times_k \boldsymbol{U}^\top,
\qquad
\boldsymbol{\mathcal{S}}\in\mathbb{R}^{n\times\cdots\times n}.
\end{equation}
For all $\boldsymbol{\mathcal{G}},\boldsymbol{\mathcal{S}}\in\mathbb{R}^{n\times\cdots\times n}$, one has
\begin{equation}
\label{eq:adjoint_identity}
\langle \widehat{\boldsymbol{\mathcal{M}}}(\boldsymbol{\mathcal{G}}),\,\boldsymbol{\mathcal{S}}\rangle
\;=\;
\langle \boldsymbol{\mathcal{G}},\,\widehat{\boldsymbol{\mathcal{M}}}^{\,*}(\boldsymbol{\mathcal{S}})\rangle.
\end{equation}
Indeed, expanding $\widehat{\boldsymbol{\mathcal{M}}}(\boldsymbol{\mathcal{G}})$ entrywise gives
\begin{align*}
\langle \widehat{\boldsymbol{\mathcal{M}}}(\boldsymbol{\mathcal{G}}),\,\boldsymbol{\mathcal{S}}\rangle
&=
\sum_{j_1,\dots,j_k}
\widehat{\mathcal{M}}(\mathcal{G})_{j_1\dots j_k}\;\mathcal{S}_{j_1\dots j_k} \\
&=
\sum_{j_1,\dots,j_k}
\sum_{a_1,\dots,a_k}
\left(\prod_{m=1}^k U_{j_m a_m}\right)
\mathcal{G}_{a_1\dots a_k}\;\mathcal{S}_{j_1\dots j_k}.
\end{align*}
Interchanging the two sums reveals the adjoint map acting on $\boldsymbol{\mathcal{S}}$:
\begin{equation}
\langle \widehat{\boldsymbol{\mathcal{M}}}(\boldsymbol{\mathcal{G}}),\,\boldsymbol{\mathcal{S}}\rangle
=
\sum_{a_1,\dots,a_k}
\mathcal{G}_{a_1\dots a_k}\;
\underbrace{\sum_{j_1,\dots,j_k}
\left(\prod_{m=1}^k U_{j_m a_m}\right)\mathcal{S}_{j_1\dots j_k}}_{=\;(\widehat{\mathcal{M}}^{\,*}(\mathcal{S}))_{a_1\dots a_k}}
=
\langle \boldsymbol{\mathcal{G}},\,\widehat{\boldsymbol{\mathcal{M}}}^{\,*}(\boldsymbol{\mathcal{S}})\rangle, 
\end{equation}
which establishes Eq.~\eqref{eq:adjoint_identity}.

\subsubsection*{Optimality of $\widetilde{\boldsymbol{\mathcal{G}}}$}
Consider the unconstrained least-squares problem
\begin{equation}
\label{eq:oracle_problem}
\widetilde{\boldsymbol{\mathcal{G}}}
:=
\arg\min_{\boldsymbol{\mathcal{G}}\in\mathbb{R}^{n\times\cdots\times n}}
\|
\widehat{\boldsymbol{\mathcal{M}}}(\boldsymbol{\mathcal{G}}) - \boldsymbol{\mathcal{M}}
\|_F^2.
\end{equation}
Expanding the squared Frobenius norm via the inner product and applying Eqs.~\eqref{eq:isometry_identity} and~\eqref{eq:adjoint_identity}, we obtain
\begin{align*}
\|
\widehat{\boldsymbol{\mathcal{M}}}(\boldsymbol{\mathcal{G}}) - \boldsymbol{\mathcal{M}}
\|_F^2
&=
\|\widehat{\boldsymbol{\mathcal{M}}}(\boldsymbol{\mathcal{G}})\|_F^2
-2\langle \widehat{\boldsymbol{\mathcal{M}}}(\boldsymbol{\mathcal{G}}),\,\boldsymbol{\mathcal{M}}\rangle
+\|\boldsymbol{\mathcal{M}}\|_F^2 \\
&=
\|\boldsymbol{\mathcal{G}}\|_F^2
-2\langle \boldsymbol{\mathcal{G}},\,\widehat{\boldsymbol{\mathcal{M}}}^{\,*}(\boldsymbol{\mathcal{M}})\rangle
+\|\boldsymbol{\mathcal{M}}\|_F^2 \\
&=
\|\boldsymbol{\mathcal{G}}-\widehat{\boldsymbol{\mathcal{M}}}^{\,*}(\boldsymbol{\mathcal{M}})\|_F^2
+\|\boldsymbol{\mathcal{M}}\|_F^2
-\|\widehat{\boldsymbol{\mathcal{M}}}^{\,*}(\boldsymbol{\mathcal{M}})\|_F^2.
\end{align*}
Since the last two terms do not depend on $\boldsymbol{\mathcal{G}}$, the unique minimizer is $\widetilde{\boldsymbol{\mathcal{G}}}=\widehat{\boldsymbol{\mathcal{M}}}^{\,*}(\boldsymbol{\mathcal{M}})=\boldsymbol{\mathcal{M}}\times_1 \boldsymbol{U}^\top \cdots \times_k \boldsymbol{U}^\top$, which is Eq.~\eqref{eq:unconstrained_oracle} for $k=4$ (with zero residual $\|\widehat{\boldsymbol{\mathcal{M}}}(\widetilde{\boldsymbol{\mathcal{G}}}) - \boldsymbol{\mathcal{M}}\|_F = 0$ because $\widehat{\boldsymbol{\mathcal{M}}}$ is invertible on $\mathbb{R}^{n\times\cdots\times n}$).

Under the assumption of rotational invariance, this expression also defines the optimal rotationally invariant estimator (RIE) for an order-$k$ co-moment tensor (in the sense of fixing the empirical mode basis $\boldsymbol{U}$ and selecting the core tensor that minimizes the Frobenius distance to a target tensor), although the algebraic optimality of this solution holds without requiring rotational equivariance.

\subsection{Even-pairing projection and all-orthogonality}
\label{sec:appevenprojection}

This section proves, at any order $k$, two statements of Sec.~\ref{sec:oracle_comoment}. The average over random signs cancels every core entry with an odd index multiplicity (Eq.~\eqref{eq:sign_expectation} for $k=4$), and any tensor supported on even-multiplicity tuples belongs to $\mathbb{A}_k$ (Eq.~\eqref{eq:even_projection_allorthogonal} for $k=4$). At order $k$, the projector $\Pi_{\mathrm{even}}$ keeps the entries of a core whose index tuple has an even multiplicity for every index and sets the others to zero, which is Eq.~\eqref{eq:even_projection} for $k=4$.

Let $\boldsymbol{\mathcal{H}}_b$ denote a fold-level reference core prior to resolving the sign ambiguity of the singular vectors, and let $\boldsymbol{S}_b = \mathrm{diag}(\eta_{b,1},\ldots,\eta_{b,n})$ be the diagonal matrix encoding the (uniformly) random signs of the singular vectors in fold $b$. Replacing the multilinear basis $\boldsymbol{U}_b$ with $\boldsymbol{U}_b \boldsymbol{S}_b$ transforms the corresponding core tensor into
\begin{equation}
\boldsymbol{\mathcal{G}}_b
=
\boldsymbol{\mathcal{H}}_b \times_1 \boldsymbol{S}_b \cdots \times_k \boldsymbol{S}_b.
\end{equation}
Entrywise, this transformation reads
\begin{equation}
(\mathcal{G}_b)_{i_1,\dots,i_k}
=
\left(\prod_{\ell=1}^k \eta_{b,i_\ell}\right)
(\mathcal{H}_b)_{i_1,\dots,i_k}.
\label{eq:appendix_signed_coefficient}
\end{equation}

Let $m_a(i_1,\dots,i_k)$ denote the multiplicity of index $a\in[\![1,n]\!]$ in the tuple $(i_1,\dots,i_k)$, so that
\begin{equation}
\prod_{\ell=1}^k \eta_{b,i_\ell}
=
\prod_{a=1}^n \eta_{b,a}^{\,m_a(i_1,\dots,i_k)}.
\end{equation}
If at least one multiplicity $m_a(i_1,\dots,i_k)$ is odd, then under independent symmetric signs $\eta_{b,a}\sim\mathcal{U}(\{-1,1\})$,
\begin{equation}
\label{eq:sign_average}
\mathbb{E}\!\left[(\mathcal{G}_b)_{i_1,\dots,i_k}\,\middle|\,\boldsymbol{\mathcal{H}}_b\right]
=
(\mathcal{H}_b)_{i_1,\dots,i_k}
\prod_{a=1}^n
\mathbb{E}\!\left[\eta_{b,a}^{\,m_a(i_1,\dots,i_k)}\right]
=
0.
\end{equation}
Consequently, when averaging entrywise across folds or calibration windows, every core entry with at least one odd index multiplicity vanishes asymptotically. In particular, if the tensor order $k$ is odd, the sum of multiplicities
\begin{equation}
\sum_{a=1}^n m_a(i_1,\dots,i_k)=k
\end{equation}
is odd, which implies that every index tuple $(i_1,\dots,i_k)$ contains at least one index with odd multiplicity; thus, the entire averaged odd-order core vanishes asymptotically.

It remains to show that any tensor $\boldsymbol{\mathcal{T}}$ supported exclusively on even-multiplicity index tuples belongs to $\mathbb{A}_k$. Fix a mode $r\leq k$ and two distinct indices $\alpha\neq\beta$, and consider the inner product of the two mode-$r$ slices:
\begin{equation}
\left\langle \boldsymbol{\mathcal{T}}_{i_r=\alpha},\boldsymbol{\mathcal{T}}_{i_r=\beta}\right\rangle
=
\sum_{i_1,\dots,i_{r-1},\,i_{r+1},\dots,i_k}
\mathcal{T}_{i_1,\dots,i_{r-1},\alpha,i_{r+1},\dots,i_k}\,
\mathcal{T}_{i_1,\dots,i_{r-1},\beta,i_{r+1},\dots,i_k}.
\label{eq:appendix_scalar_product_slices}
\end{equation}
For a summand in Eq.~\eqref{eq:appendix_scalar_product_slices} to be non-zero, both index tuples
\begin{equation}
(i_1,\dots,i_{r-1},\alpha,i_{r+1},\dots,i_k)
\qquad\text{and}\qquad
(i_1,\dots,i_{r-1},\beta,i_{r+1},\dots,i_k)
\end{equation}
would need to have exclusively even multiplicities. This is impossible, because replacing $\alpha$ with $\beta$ in the $r$-th position changes the parity of the multiplicities of both $\alpha$ and $\beta$. Every summand therefore vanishes, yielding
\begin{equation}
\left\langle \boldsymbol{\mathcal{T}}_{i_r=\alpha},\boldsymbol{\mathcal{T}}_{i_r=\beta}\right\rangle=0
\qquad
\forall~ r\leq k,\ \forall~ \alpha\neq\beta,
\end{equation}
which proves that $\boldsymbol{\mathcal{T}}\in \mathbb{A}_k$.

\subsection{Even cores of order four}
\label{app:evencore}

This section gives the computations behind the $O(n^2)$ formulas of Sec.~\ref{sec:oracle_comoment}. An even core of order four has $O(n^2)$ distinct parameters rather than $O(n^4)$, so that tensor norms, portfolio moments, and gradients can be computed directly in $O(n^2)$ operations without forming the full tensor. Throughout, $\boldsymbol{B}$ is the matrix of Eq.~\eqref{eq:B_matrix_def}. For CV and HOAO it is the average of the matrices of Eq.~\eqref{eq:Bempirical} across cross-validation folds or historical calibration pairs, and its entries are non-negative by construction.

\paragraph{Portfolio fourth moment and gradient.}

Let $\boldsymbol{v}=\widehat{\boldsymbol{U}}^\top
\widehat{\boldsymbol{D}}\boldsymbol{w}$.
By Eq.~\eqref{eq:adjoint_identity},
\[
\begin{aligned}
\widehat m_4(\boldsymbol{w})
&=
\left\langle
\mathcal{G}\times_1\widehat{\boldsymbol{U}}
\cdots\times_4\widehat{\boldsymbol{U}},
(\widehat{\boldsymbol{D}}\boldsymbol{w})^{\otimes 4}
\right\rangle \\
&=\left\langle\mathcal{G},\boldsymbol{v}^{\otimes 4}\right\rangle
=\sum_{a,b,c,d}\mathcal{G}_{abcd}\,v_av_bv_cv_d.
\end{aligned}
\]
For an even core, supersymmetry and $B_{ij}=\mathcal{G}_{iijj}$ give
\[
\widehat m_4(\boldsymbol{w})
=\sum_i B_{ii}v_i^4
+\frac{4!}{2!\,2!}\sum_{i<j}B_{ij}v_i^2v_j^2,
\]
which is Eq.~\eqref{eq:m4even}.

Differentiating its second form with respect to $v_i$ and using the symmetry of $\boldsymbol{B}$ gives
\begin{equation}
\frac{\partial\hat m_4}{\partial v_i}
=12\,v_i\sum_{j=1}^n B_{ij}v_j^2-8B_{ii}v_i^3,
\end{equation}
that is, $\nabla_{\boldsymbol{v}}\hat m_4=12\,\boldsymbol{v}\circ\boldsymbol{B}(\boldsymbol{v}\circ\boldsymbol{v})-8\,\mathrm{diag}(\boldsymbol{B})\circ\boldsymbol{v}\circ\boldsymbol{v}\circ\boldsymbol{v}$. Since $\boldsymbol{v}=\widehat{\boldsymbol{U}}^\top\widehat{\boldsymbol{D}}\boldsymbol{w}$, the chain rule gives $\nabla_{\boldsymbol{w}}=\widehat{\boldsymbol{D}}\widehat{\boldsymbol{U}}\nabla_{\boldsymbol{v}}$ and
\begin{equation}
\nabla_{\boldsymbol{w}}\hat m_4
=\widehat{\boldsymbol{D}}\widehat{\boldsymbol{U}}\left[12\,\boldsymbol{v}\circ\boldsymbol{B}(\boldsymbol{v}\circ\boldsymbol{v})-8\,\mathrm{diag}(\boldsymbol{B})\circ\boldsymbol{v}\circ\boldsymbol{v}\circ\boldsymbol{v}\right].
\end{equation}
Once $\boldsymbol{v}$ is formed, the fourth moment and its gradient require only $O(n^2)$ operations, dominated by the product $\boldsymbol{B}(\boldsymbol{v}\circ\boldsymbol{v})$.

\paragraph{Frobenius norm and OOS inner product.} Since $\widehat{\boldsymbol{U}}$ is orthogonal, the reconstruction map is an isometry (Eq.~\eqref{eq:isometry_identity}), so that the squared Frobenius norm of the reconstructed tensor equals that of the core. Counting the even tuples as above gives
\begin{equation}
\|\boldsymbol{\mathcal{G}}_{\mathrm{even}}\|_F^2
=\sum_{i=1}^n B_{ii}^2+6\sum_{i<j}B_{ij}^2
=3\sum_{i,j=1}^n B_{ij}^2-2\sum_{i=1}^n B_{ii}^2 .
\end{equation}
Let $\boldsymbol{z}'_1,\ldots,\boldsymbol{z}'_{\Delta t_{\mathrm{OOS}}}$ be the standardized OOS observations and $\boldsymbol{y}'_t=\widehat{\boldsymbol{U}}^\top\boldsymbol{z}'_t$. The Frobenius inner product between the reconstructed tensor and the sample fourth-moment tensor of these observations equals, by the adjoint identity, the OOS time average of $3\sum_{i,j}B_{ij}(y'_{it})^2(y'_{jt})^2-2\sum_iB_{ii}(y'_{it})^4$. The squared norm of the OOS tensor is $\Delta t_{\mathrm{OOS}}^{-2}\sum_{s,t}(\boldsymbol{z}_s'^\top\boldsymbol{z}'_t)^4$. The tensor losses in Eq.~\eqref{eq:tensor-loss} can therefore be evaluated exactly without ever allocating an $n^4$ array in memory.

\paragraph{Computation of the basis.} The basis $\widehat{\boldsymbol{U}}$ consists of the left singular vectors of the mode-1 unfolding $\widehat{\boldsymbol{K}}_{(1)}\in\mathbb{R}^{n\times n^3}$ of the sample fourth-moment tensor of the standardized observations (Sec.~\ref{sec:appmodematrices}), that is, the eigenvectors of $\widehat{\boldsymbol{K}}_{(1)}\widehat{\boldsymbol{K}}_{(1)}^\top$. Denote by $\boldsymbol{Z}\in\mathbb{R}^{T\times n}$ the matrix of standardized observations, with $T=\Delta t$. Then $\widehat{\boldsymbol{K}}_{(1)}=\frac{1}{T}\boldsymbol{Z}^\top\boldsymbol{W}$, where the $t$-th row of $\boldsymbol{W}\in\mathbb{R}^{T\times n^3}$ is $\boldsymbol{z}_t^{\otimes 3}$, and $(\boldsymbol{W}\boldsymbol{W}^\top)_{st}=(\boldsymbol{z}_s^\top\boldsymbol{z}_t)^3$. The $n\times n$ Gram matrix is therefore
\begin{equation}
\widehat{\boldsymbol{K}}_{(1)}\widehat{\boldsymbol{K}}_{(1)}^\top
=\frac{1}{T^2}\,\boldsymbol{Z}^\top\left[(\boldsymbol{Z}\boldsymbol{Z}^\top)^{\circ 3}\right]\boldsymbol{Z},
\end{equation}
where $(\cdot)^{\circ 3}$ denotes the elementwise cube. Forming and diagonalizing this matrix costs $O(T^2n+Tn^2+n^3)$ operations instead of $O(Tn^4)$. Because the even projection depends on the specific basis chosen within any degenerate eigenspace, whenever two eigenvalues are nearly tied (relative gap below $100\,n$ times machine precision), we fall back to the SVD of the dense unfolding.

\section{Data}
\label{app:data}

Our source is Norgate Data's U.S. equity price history, including delisted securities, and its historical Russell 1000 constituent series \cite{norgateData}. We use the data snapshot of October~30, 2026. The panel runs from January~2, 1990 to December~31, 2025 and contains 3,557 distinct securities with at least one index membership during that period.

\paragraph{Returns.} We compute daily close-to-close total returns from consecutive adjusted closes, including dividends and distributions. We retain prices and returns before a security first enters the index and after it leaves. Data before 1995 provide history for screening and estimation.

\paragraph{Universe.} At the close of trading day $t$, we start from the effective Russell 1000 constituents on that date. Historical membership supplies the large-capitalization universe. Each share class retains its own permanent security identifier. We apply the following screens in order.
\begin{itemize}
    \item \emph{Long history.} An active session requires a finite total return, a positive raw closing price and positive share volume. We require activity on at least 95\% of sessions in every rolling 252-session subwindow of the last 1,199 sessions, plus an active observation before that window.
    \item \emph{Recent activity and price.} The security must have been active on each of the last 20 sessions, including $t$. Its unadjusted closing price on $t$ must be between USD~10 and USD~2,000, inclusive.
    \item \emph{Stale prices.} We screen the log of sample return volatility first over five sessions and then over 20. With at least 100 candidates of positive finite volatility, we retain values at or above the first quartile minus 1.5 times the interquartile range. We recompute the quartiles among the survivors for the second screen.
    \item \emph{Dollar turnover.} Following the Nasdaq-100 traded-value criterion \cite{nasdaqLiquidity}, we require mean daily traded value of at least USD~5 million over the three calendar months ending on $t$, including $t$. We average over all trading sessions and count missing or inconsistent turnover as zero. This rule applies on every listing exchange.
    \item \emph{Near duplicates.} We consider candidates in descending order of their trailing 20-session median dollar turnover, breaking ties by ascending security identifier. We remove a security if its signed Pearson return correlation with an earlier survivor is at least 0.95 over the last 1,199 sessions, using at least 60 common observations.
    \item \emph{Security type.} We exclude real estate investment trusts (REITs).
\end{itemize}
The eligible pool averaged 772 securities per session in 2010 and 824 in 2025. Its size varies with index membership and the screens. The procedure uses no historical shares-outstanding threshold or additional capitalization ranking.

\paragraph{Timing.} The IS estimation window terminates on the investment date $t$, whose closing data also define the eligible universe. A portfolio formed after the close of day $t$ is assumed to be executed at the close of day $t+1$; accordingly, the return on day $t+1$ is excluded from both windows, and the OOS evaluation window begins on day $t+2$.

\paragraph{Exits and missing returns.} For each experiment, we draw $n$ eligible securities uniformly without replacement, requiring a complete IS return history and strictly positive IS variance. We keep the same securities throughout the OOS window, including those that leave the index. Each OOS interval without two adjacent valid closes receives a zero return for scoring. This includes the first quote after a gap. Observed returns resume with the next pair of consecutive closes. At $n=50$, 70,236 of 25 million security-day returns were missing with 100-day IS windows (0.281\%), and 71,511 of 25 million with 500-day windows (0.286\%). Of these, 69,663 and 71,010, respectively, followed the last quote of a security classified as delisted. The snapshot supplies neither dated trading suspensions nor verified terminal settlement amounts. A constant OOS return series is retained as a zero standardized component.

\section{Experimental protocol}
\label{app:protocol}

This section provides the implementation details required to reproduce the empirical results in Sec.~\ref{sec:realdataapp}. It covers the calibration and updating of HOAO, the cross-validation scheme, the sampling of experiments and test portfolios, and the numerical solver for portfolio optimization.

\paragraph{Calibration of HOAO.} A calibration pair consists of a training window of $\Delta t_{\mathrm{IS}}$ daily returns ending on a randomly drawn investment date $\tau$, paired with a testing window of the next 100 daily returns starting on date $\tau+2$, for $n$ stocks drawn uniformly at random from the eligible universe on date $\tau$. Each training and testing window is centered and standardized using its own sample mean and volatility vector. Each calibration pair then provides a realized oracle core via Eq.~\eqref{eq:hoao_oracle_core_tau} for $k=4$ (by projecting the standardized testing tensor onto the training basis), and the realized variances of the standardized testing window along the training eigenvectors for $k=2$. Initial calibration uses 10,000 pairs with investment dates starting in January 1995 and testing windows ending on or before December~31, 2009. On January~1 of 2012, 2014, \ldots, 2024, the calibration pool is augmented with 2,000 new pairs whose investment dates fall within the preceding two calendar years and whose testing windows end strictly prior to the update date. Realized cores are accumulated as running sums and counts, so that the HOAO core used for an evaluation experiment at investment date $t$ is the cumulative average over all calibration pairs available prior to $t$. Calibration is performed separately for each portfolio size $n$, IS window length $\Delta t_{\mathrm{IS}}$ and specification of the turnover screen.

\paragraph{Cross-validation.} Given the standardized IS return matrix, CV generates 10 random permutations of the IS dates. For each permutation, the first 65\% of observations form the training subsample and the remaining 35\% form the testing subsample. Both subsamples are re-centered, and the testing co-moment tensor is projected onto the training HOSVD basis and restricted to the even subspace via Eq.~\eqref{eq:cv_test_core}; the resulting ten cores are then averaged.

\paragraph{Experiments.} We run 5,000 experiments with $\Delta t_{\mathrm{IS}}=100$ for each $n\in\{5,10,20,30,40,50\}$, and 5,000 experiments with $\Delta t_{\mathrm{IS}}=500$ at $n=50$. The OOS window length is fixed at $\Delta t_{\mathrm{OOS}}=100$ days throughout. Investment dates are drawn uniformly at random from all trading days between January 2010 and August 2025, so that every OOS window ends on or before December~31, 2025. Because investment dates are sampled densely over the 16-year period, evaluation windows overlap in calendar time. All estimators, random portfolio evaluations, and optimal allocations are computed on the exact same experimental draws.

\paragraph{Losses.} In Eq.~\eqref{eq:tensor-loss}, the target OOS tensors are the sample correlation matrix and the sample fourth-moment tensor of OOS returns, standardized using the OOS sample means and volatilities. For the portfolio moment losses in Eq.~\eqref{eq:portfolio-loss}, each experiment draws 100 long-only weight vectors from a uniform Dirichlet distribution on the simplex; all estimators are evaluated on the same 100 portfolios using Eqs.~\eqref{eq:destandardization} and~\eqref{eq:m4even}.

\paragraph{Allocation.} Problem~\eqref{eq:allocation} is solved on the $(n-1)$-dimensional affine space that satisfies the budget constraint, starting from equal weights, for $\gamma\in\{0, 0.01, 0.03, 0.1, 0.2, 0.5, 1, 2, 5, 10, 30, 100\}$. For $\gamma=0$, the solution follows from a linear system. Otherwise, we use BFGS with analytic gradients, followed by Newton steps with the analytic Hessian. A solution is accepted when the largest component of the projected gradient is below $10^{-6}$. All optimizations satisfy this criterion.
Table~\ref{tab:allocation} reports OOS volatility, OOS kurtosis and gross leverage $\sum_i|w_i|$ for a selection of penalties.

\begin{table}[htbp]
\centering
\setlength{\tabcolsep}{7pt}
\begin{tabular}{llcccccccccccc}
\toprule
& & \multicolumn{3}{c}{Sample} & \multicolumn{3}{c}{Sample$_{\mathrm{even}}$} & \multicolumn{3}{c}{CV} & \multicolumn{3}{c}{HOAO} \\
\cmidrule(lr){3-5}\cmidrule(lr){6-8}\cmidrule(lr){9-11}\cmidrule(lr){12-14}
$\Delta t_{\mathrm{IS}}$ & $\gamma$ & vol. & $\kappa$ & lev. & vol. & $\kappa$ & lev. & vol. & $\kappa$ & lev. & vol. & $\kappa$ & lev. \\
\midrule
\multirow{5}{*}{100} & 0 & 18.53 & 4.67 & 3.76 & 18.53 & 4.67 & 3.76 & 14.53 & 4.46 & 1.86 & 14.13 & 4.43 & 1.66 \\
 & 0.1 & 18.56 & 4.66 & 3.77 & 17.94 & 4.64 & 3.54 & 14.49 & 4.45 & 1.84 & 14.11 & 4.42 & 1.65 \\
 & 1 & 18.72 & 4.66 & 3.83 & 16.95 & 4.59 & 3.14 & 14.43 & 4.40 & 1.77 & 14.11 & 4.37 & 1.61 \\
 & 10 & 18.99 & 4.66 & 3.92 & 16.54 & 4.54 & 2.93 & 14.51 & 4.36 & 1.73 & 14.20 & 4.33 & 1.58 \\
 & 100 & 19.08 & 4.66 & 3.95 & 16.51 & 4.52 & 2.91 & 14.55 & 4.36 & 1.73 & 14.23 & 4.32 & 1.57 \\
\cmidrule(l){2-14}
 & EW & \multicolumn{12}{l}{volatility 19.63, kurtosis 4.28, leverage 1.00} \\
\midrule
\multirow{5}{*}{500} & 0 & 14.87 & 4.44 & 2.34 & 14.87 & 4.44 & 2.34 & 14.65 & 4.48 & 1.99 & 14.43 & 4.40 & 1.68 \\
 & 0.1 & 15.00 & 4.43 & 2.43 & 14.93 & 4.42 & 2.36 & 14.63 & 4.44 & 1.98 & 14.42 & 4.37 & 1.68 \\
 & 1 & 15.38 & 4.41 & 2.63 & 15.22 & 4.39 & 2.44 & 14.75 & 4.36 & 1.97 & 14.50 & 4.29 & 1.70 \\
 & 10 & 15.61 & 4.41 & 2.73 & 15.50 & 4.38 & 2.49 & 14.94 & 4.31 & 1.98 & 14.64 & 4.25 & 1.72 \\
 & 100 & 15.65 & 4.41 & 2.75 & 15.56 & 4.38 & 2.50 & 14.99 & 4.31 & 1.98 & 14.68 & 4.25 & 1.73 \\
\cmidrule(l){2-14}
 & EW & \multicolumn{12}{l}{volatility 19.90, kurtosis 4.25, leverage 1.00} \\
 
\bottomrule
\end{tabular}
\caption{Optimized portfolios of Eq.~\eqref{eq:allocation} with $n=50$, for selected penalties $\gamma$. The table reports the annualized OOS volatility (in \%), the mean OOS kurtosis $\kappa$ and the mean gross leverage. EW denotes the equally weighted portfolio. 5,000 experiments for each IS length.}
\label{tab:allocation}
\end{table}

\section{Statistical significance}
\label{app:inference}

This section explains how we assess the statistical significance of performance differences across estimators. Testing is non-trivial because experiments have overlapping time windows and the 100 random portfolios within a given experiment share the same co-moment estimates.

To account for serial dependence across overlapping windows, we use the Model Confidence Set (MCS) procedure of Ref.~\cite{MCS}, which constructs a subset of candidate estimators that contains the best-performing estimator(s) in expected loss with 95\% confidence (inclusion means that an estimator cannot be rejected as significantly inferior given the available sample, rather than proving exact equivalence). Specifically, for each relative squared-error loss, we first average the losses of all experiments sharing the same investment date. We then resample the trading calendar in circular blocks of 252 sessions \cite{politis1992circular} over 1,999 bootstrap replications, keeping all estimators paired on the exact same resampled calendar dates in every replication, and apply the range statistic and sequential elimination rule of Ref.~\cite{MCS} at the 95\% confidence level. Circular blocks of 126 and 504 sessions serve as sensitivity checks. Because the MCS operates on date-averaged losses, its sample means differ slightly from the experiment-level averages reported in the main text.

By contrast, the 95\% confidence bands in Figs.~\ref{fig:accuracy_v_dim} and~\ref{fig:compatible_accuracy} are computed via a bootstrap over experiments (19,999 draws) that resamples experiments with replacement while keeping all estimators paired on the same resampled experiments in each draw. These bands summarize cross-experiment variability conditional on the sample, but they treat overlapping windows as independent and do not account for dependence between them. Table~\ref{tab:mcs} reports the MCS $p$-values for $n=50$, among the four main estimators and after adding the three consistent variants. HOAO has $p$-value one among the four main estimators for every metric, IS length and block length. With 126-session blocks, CV also enters the set for correlation at 100 days. With 504-session blocks at 500 days, CV and Sample$_{\mathrm{even}}$ enter for cokurtosis, and CV enters for portfolio kurtosis. Among all seven estimators, HOAO$^{\mathrm{cr}}$ has the smallest date-averaged cokurtosis MSE at both IS lengths and the original HOAO lies outside this set. For portfolio kurtosis at 500 days, the 252-session set among all seven retains HOAO, Sample, Sample$_{\mathrm{even}}$, CV, and CV$^{\mathrm{c}}$.

All losses compare estimates with noisy OOS sample moments rather than with the unobservable true moments. For variance forecasts, the mean squared error ranks forecasts as the true variance would when the proxy is conditionally unbiased \cite{patton2011volatility}. This property does not extend to relative losses or to fourth moments, so that our rankings should be read as comparisons with realized moments.

\begin{table}[htbp]
\centering
\setlength{\tabcolsep}{7pt}
\begin{tabular}{llccccccc}
\toprule
$\Delta t_{\mathrm{IS}}$ & Relative MSE & Sample & Sample$_{\mathrm{even}}$ & CV & HOAO & CV$^{\mathrm{c}}$ & HOAO$^{\mathrm{c}}$ & HOAO$^{\mathrm{cr}}$ \\
\midrule
 & & \multicolumn{4}{c}{Four estimators} \\
\cmidrule(lr){3-6}
\multirow{4}{*}{100} & Correlation & 0.001 & 0.001 & 0.026 & 1$^{\dagger}$ \\
 & Cokurtosis & 0.001 & 0.029 & 0.073$^{\dagger}$ & 1$^{\dagger}$ \\
 & Volatility & 0.011 & 0.011 & 0.011 & 1$^{\dagger}$ \\
 & Kurtosis & 0.142$^{\dagger}$ & 0.142$^{\dagger}$ & 0.142$^{\dagger}$ & 1$^{\dagger}$ \\
\cmidrule{1-6}
\multirow{4}{*}{500} & Correlation & 0.001 & 0.001 & 0.015 & 1$^{\dagger}$ \\
 & Cokurtosis & 0.020 & 0.043 & 0.043 & 1$^{\dagger}$ \\
 & Volatility & 0.008 & 0.008 & 0.008 & 1$^{\dagger}$ \\
 & Kurtosis & 0.038 & 0.038 & 0.038 & 1$^{\dagger}$ \\
\midrule
 & & \multicolumn{7}{c}{All estimators} \\
\cmidrule(lr){3-9}
\multirow{4}{*}{100} & Correlation & 0.001 & 0.001 & 0.026 & 1$^{\dagger}$ & 0.026 & 1$^{\dagger}$ & 1$^{\dagger}$ \\
 & Cokurtosis & 0.001 & 0.038 & 0.011 & 0.004 & 0.066$^{\dagger}$ & 0.038 & 1$^{\dagger}$ \\
 & Volatility & 0.011 & 0.011 & 0.011 & 1$^{\dagger}$ & 0.011 & 1$^{\dagger}$ & 1$^{\dagger}$ \\
 & Kurtosis & 0.226$^{\dagger}$ & 0.226$^{\dagger}$ & 0.226$^{\dagger}$ & 0.492$^{\dagger}$ & 0.226$^{\dagger}$ & 0.001 & 1$^{\dagger}$ \\
\cmidrule{1-9}
\multirow{4}{*}{500} & Correlation & 0.001 & 0.001 & 0.015 & 1$^{\dagger}$ & 0.015 & 1$^{\dagger}$ & 1$^{\dagger}$ \\
 & Cokurtosis & 0.033 & 0.043 & 0.043 & 0.001 & 0.043 & 0.034 & 1$^{\dagger}$ \\
 & Volatility & 0.008 & 0.008 & 0.008 & 1$^{\dagger}$ & 0.008 & 1$^{\dagger}$ & 1$^{\dagger}$ \\
 & Kurtosis & 0.052$^{\dagger}$ & 0.052$^{\dagger}$ & 0.052$^{\dagger}$ & 1$^{\dagger}$ & 0.052$^{\dagger}$ & 0.001 & 0.003 \\
 
\bottomrule
\end{tabular}
\caption{MCS $p$-values for $n=50$, with calendar blocks of 252 sessions. Daggers mark the estimators in the 95\% model confidence set. ``Four estimators'' is the set of the main text, and ``All estimators'' adds the consistent variants of Sec.~\ref{app:compatibility}.}
\label{tab:mcs}
\end{table}

\section{Joint distributional consistency of second and fourth moments}
\label{app:compatibility}

Section~\ref{sec:compatibility} noted that filtering covariance and cokurtosis in separate bases with independently calibrated cores can produce moment pairs $(\widehat{\boldsymbol{\Sigma}}, \widehat{\boldsymbol{\mathcal{K}}})$ that no probability distribution can generate. Here we derive the underlying mixture representation of an even core, show how the mismatch between the filtered covariance $\widehat{\boldsymbol{\Sigma}}$ and the mixture's own covariance $\boldsymbol{\Omega}$ causes violations of the Cauchy--Schwarz bound while preserving the convexity of the allocation problem (Sec.~\ref{app:cs_violations}), define the consistent reconstructions (Sec.~\ref{app:consistent_variants}) together with the elliptical control and anisotropy metric (Sec.~\ref{app:anisotropy}), and report their empirical performance (Sec.~\ref{app:compatibility_results}).

\subsection{Sign-symmetrized mixture, companion covariance, and Cauchy--Schwarz violations}
\label{app:cs_violations}

For any random vector $\boldsymbol{x}$ with mean $\boldsymbol{\mu}$ and finite fourth moments, applying the Cauchy--Schwarz inequality to the squared centered portfolio return $R^2 = [\boldsymbol{w}^\top(\boldsymbol{x}-\boldsymbol{\mu})]^2$ yields
\begin{equation}
\label{eq:cs_portfolio_bound}
\mathbb{E}\!\left[R^4\right] \geq \left(\mathbb{E}\!\left[R^2\right]\right)^2,
\qquad \text{or equivalently} \qquad
\kappa(\boldsymbol{w}) := \frac{m_4(\boldsymbol{w})}{\sigma^4(\boldsymbol{w})} \geq 1,
\qquad \forall\,\boldsymbol{w}\neq\boldsymbol{0}.
\end{equation}
It is also implied by Pearson's inequality, which bounds the kurtosis below by one plus the squared skewness~\cite{pearson1916mathematical}. A covariance estimate $\widehat{\boldsymbol{\Sigma}}$ and a cokurtosis estimate $\widehat{\boldsymbol{\mathcal{K}}}$ are jointly consistent as the moments of a single distribution only if $\hat{\kappa}(\boldsymbol{w}) = \hat{m}_4(\boldsymbol{w})/\hat{\sigma}^4(\boldsymbol{w}) \geq 1$ holds for every portfolio $\boldsymbol{w}$.

Although $\widehat{\boldsymbol{\Sigma}}$ and $\widehat{\boldsymbol{\mathcal{K}}}$ may violate this bound when paired together, any fourth-moment tensor reconstructed from an even core ($\mathrm{Sample}_{\mathrm{even}}$, CV, or HOAO) is by itself the exact fourth-moment tensor of a well-defined probability distribution. Consider a calibration pair $\tau$ (or a cross-validation fold $b$, or the full IS window for $\mathrm{Sample}_{\mathrm{even}}$), take a standardized observation rotated into the fourth-order HOSVD basis, $\boldsymbol{y} = \widehat{\boldsymbol{U}}_{\mathrm{train}}^\top\boldsymbol{z}$, and multiply each coordinate by an independent Rademacher sign $\eta_i \sim \mathcal{U}(\{-1,+1\})$. Because the sign-symmetrized vector $\tilde{\boldsymbol{y}} = \boldsymbol{\eta}\circ\boldsymbol{y}$ is invariant in distribution under coordinate reflections ($\tilde{y}_i \mapsto -\tilde{y}_i$), all of its odd-multiplicity moments vanish, while its even-multiplicity fourth moments $\mathbb{E}[\tilde{y}_i^2\tilde{y}_j^2] = \mathbb{E}[y_i^2 y_j^2]$ coincide with the matrix $\widehat{\boldsymbol{B}}$ of Eq.~\eqref{eq:Bempirical}. Forming an equally weighted mixture of these sign-symmetrized laws over observations and calibration pairs (or folds) defines a centered latent random vector $\tilde{\boldsymbol{y}}$ whose fourth-moment tensor is the calibrated even core $\widehat{\boldsymbol{\mathcal{G}}}$.

Mapping $\tilde{\boldsymbol{y}}$ into return space via the current IS fourth-order basis $\widehat{\boldsymbol{U}}$ and volatility matrix $\widehat{\boldsymbol{D}}$ defines the random vector $\tilde{\boldsymbol{x}} = \widehat{\boldsymbol{D}}\widehat{\boldsymbol{U}}\tilde{\boldsymbol{y}}$, whose fourth-moment tensor is $\widehat{\boldsymbol{\mathcal{K}}}$. Because the random signs make the coordinates of $\tilde{\boldsymbol{y}}$ mutually uncorrelated ($\mathbb{E}[\tilde{y}_i\tilde{y}_j] = \mathbb{E}[\eta_i\eta_j]\mathbb{E}[y_i y_j] = 0$ for $i\neq j$) with variances $a_i := \mathbb{E}[\tilde{y}_i^2] = \mathbb{E}[y_i^2]$, the covariance matrix of $\tilde{\boldsymbol{x}}$ is the companion covariance
\begin{equation}
\label{eq:companion_covariance}
\boldsymbol{\Omega} := \mathbb{E}\!\left[\tilde{\boldsymbol{x}}\tilde{\boldsymbol{x}}^\top\right] = \widehat{\boldsymbol{D}}\,\widehat{\boldsymbol{U}}\,\mathrm{diag}(\boldsymbol{a})\,\widehat{\boldsymbol{U}}^\top\widehat{\boldsymbol{D}}.
\end{equation}
This representation has two immediate consequences. First, the allocation problem in Eq.~\eqref{eq:allocation} remains well-posed and strictly convex even when $\boldsymbol{\Omega}\neq\widehat{\boldsymbol{\Sigma}}$. Indeed, $\hat{m}_4(\boldsymbol{w}) = \mathbb{E}[(\boldsymbol{w}^\top\tilde{\boldsymbol{x}})^4]$ is the expectation of the fourth power of a linear form and is therefore convex in $\boldsymbol{w}$~\cite[Sec.~3.2]{boyd2004convex}; since the filtered covariance $\widehat{\boldsymbol{\Sigma}}$ is positive definite, the objective of Eq.~\eqref{eq:allocation} is strictly convex on the budget hyperplane $\boldsymbol{1}^\top\boldsymbol{w}=1$ for every $\gamma\geq 0$, and any stationary point is its unique global minimum~\cite[Sec.~4.2]{boyd2004convex}.

Second, applying Cauchy--Schwarz to the scalar random variable $\boldsymbol{w}^\top\tilde{\boldsymbol{x}}$ shows that $\hat{m}_4(\boldsymbol{w})$ always satisfies the lower bound with respect to its own companion variance $\boldsymbol{w}^\top\boldsymbol{\Omega}\boldsymbol{w}$:
\begin{equation}
\label{eq:cs_omega}
\hat{m}_4(\boldsymbol{w}) = \mathbb{E}\!\left[(\boldsymbol{w}^\top\tilde{\boldsymbol{x}})^4\right] \geq \left(\mathbb{E}\!\left[(\boldsymbol{w}^\top\tilde{\boldsymbol{x}})^2\right]\right)^2 = (\boldsymbol{w}^\top\boldsymbol{\Omega}\boldsymbol{w})^2,
\qquad \forall\,\boldsymbol{w}\in\mathbb{R}^n.
\end{equation}
In the estimated kurtosis $\hat{\kappa}(\boldsymbol{w}) = \hat{m}_4(\boldsymbol{w}) / (\boldsymbol{w}^\top\widehat{\boldsymbol{\Sigma}}\boldsymbol{w})^2$, however, the denominator uses the filtered covariance $\widehat{\boldsymbol{\Sigma}}$ rather than $\boldsymbol{\Omega}$. A violation $\hat{\kappa}(\boldsymbol{w}) < 1$ can therefore occur if and only if
\begin{equation}
\label{eq:violation_condition}
\boldsymbol{w}^\top\widehat{\boldsymbol{\Sigma}}\boldsymbol{w} > \sqrt{\hat{m}_4(\boldsymbol{w})} \geq \boldsymbol{w}^\top\boldsymbol{\Omega}\boldsymbol{w},
\end{equation}
which in particular requires directions along which $\widehat{\boldsymbol{\Sigma}}$ assigns a larger variance than the companion covariance $\boldsymbol{\Omega}$.

Equation~\eqref{eq:violation_condition} explains both why the even projection alone can break the bound in $\mathrm{Sample}_{\mathrm{even}}$ and why violations are more frequent for CV and HOAO. In $\mathrm{Sample}_{\mathrm{even}}$, both $\widehat{\boldsymbol{\Sigma}}$ and $\boldsymbol{\Omega}$ are computed on the same IS window without shrinkage, so $a_i = (\widehat{\boldsymbol{U}}^\top\widehat{\boldsymbol{C}}_{\mathrm{IS}}\widehat{\boldsymbol{U}})_{ii}$. However, $\widehat{\boldsymbol{U}}$ is the eigenbasis of $\widehat{\boldsymbol{K}}_{(1)}\widehat{\boldsymbol{K}}_{(1)}^\top$ rather than the eigenbasis of the sample correlation matrix $\widehat{\boldsymbol{C}}_{\mathrm{IS}}$. In the basis $\widehat{\boldsymbol{U}}$, the matrix $\widehat{\boldsymbol{U}}^\top\widehat{\boldsymbol{C}}_{\mathrm{IS}}\widehat{\boldsymbol{U}}$ has non-zero off-diagonal entries $\mathbb{E}[y_i y_j]$ ($i\neq j$), which sign symmetrization ($\Pi_{\mathrm{even}}$) sets to zero in $\boldsymbol{\Omega}$. Along portfolio vectors $\boldsymbol{w}$ where the discarded off-diagonal covariance terms sum positively, $\boldsymbol{w}^\top\widehat{\boldsymbol{\Sigma}}\boldsymbol{w}$ exceeds $\boldsymbol{w}^\top\boldsymbol{\Omega}\boldsymbol{w}$ and can surpass $\sqrt{\hat{m}_4(\boldsymbol{w})}$.

In CV and HOAO, $\widehat{\boldsymbol{\Sigma}}$ and $\boldsymbol{\Omega}$ are also calibrated separately. For HOAO, the entries of $\widehat{\boldsymbol{B}}_{\mathrm{HOAO}}$ (and the companion variances $a_i$) are fixed historical averages indexed by the ordering $i\in\{1,\ldots,n\}$ of the training fourth-order singular values. Whenever a current IS fourth-order basis vector $\widehat{\boldsymbol{u}}_i$ has a small fourth-order singular value, that is, a large index $i$ in the ordering (and thus typically a small calibrated $B_{\mathrm{HOAO},ii}$ and $a_i$) but projects strongly onto high-variance eigenvectors of the current correlation matrix $\widehat{\boldsymbol{C}}$, $\boldsymbol{w}^\top\widehat{\boldsymbol{\Sigma}}\boldsymbol{w}$ exceeds $\boldsymbol{w}^\top\boldsymbol{\Omega}\boldsymbol{w}$ along $\boldsymbol{w}\propto\widehat{\boldsymbol{D}}^{-1}\widehat{\boldsymbol{u}}_i$.

Table~\ref{tab:violations} reports the percentage of portfolios with $\hat{\kappa}(\boldsymbol{w}) < 1$ in our experiments at $n=50$. The standard sample estimator never violates the bound. Filtered estimators violate it rarely on random long-only portfolios ($0.54\%$ for CV and $3.16\%$ for HOAO at $\Delta t_{\mathrm{IS}}=100$; $0.00\%$ and $0.05\%$ at $\Delta t_{\mathrm{IS}}=500$), but much more often on portfolios that solve Eq.~\eqref{eq:allocation} for large penalties $\gamma$ (up to $10.32\%$ for CV and $13.18\%$ for HOAO at $\gamma=100$ with $\Delta t_{\mathrm{IS}}=100$). Minimizing $\hat{\sigma}^2(\boldsymbol{w})/s^2 + \gamma\hat{m}_4(\boldsymbol{w})/s^4$ at large $\gamma$ actively searches for directions where $\hat{m}_4(\boldsymbol{w})$ is small relative to $\hat{\sigma}^4(\boldsymbol{w})$, and therefore tends to select directions satisfying Eq.~\eqref{eq:violation_condition}.

\begin{table}[htbp]
\centering
\setlength{\tabcolsep}{7pt}
\begin{tabular}{llccccccc}
\toprule
$\Delta t_{\mathrm{IS}}$ & Portfolios & Sample & Sample$_{\mathrm{even}}$ & CV & HOAO & CV$^{\mathrm{c}}$ & HOAO$^{\mathrm{c}}$ & HOAO$^{\mathrm{cr}}$ \\
\midrule
\multirow{5}{*}{100} & Random portfolios & 0.00 & 0.00 & 0.54 & 3.16 & 0.00 & 0.00 & 0.00 \\
 & Optimized, $\gamma = 0.1$ & 0.00 & 0.00 & 0.02 & 0.06 & 0.00 & 0.00 & 0.00 \\
 & Optimized, $\gamma = 1$ & 0.00 & 0.00 & 0.62 & 5.64 & 0.00 & 0.00 & 0.00 \\
 & Optimized, $\gamma = 10$ & 0.00 & 0.30 & 7.50 & 12.30 & 0.00 & 0.00 & 0.00 \\
 & Optimized, $\gamma = 100$ & 0.00 & 1.22 & 10.32 & 13.18 & 0.00 & 0.00 & 0.00 \\
\midrule
\multirow{5}{*}{500} & Random portfolios & 0.00 & 0.00 & 0.00 & 0.05 & 0.00 & 0.00 & 0.00 \\
 & Optimized, $\gamma = 0.1$ & 0.00 & 0.00 & 0.00 & 0.18 & 0.00 & 0.00 & 0.00 \\
 & Optimized, $\gamma = 1$ & 0.00 & 0.00 & 0.00 & 1.06 & 0.00 & 0.00 & 0.00 \\
 & Optimized, $\gamma = 10$ & 0.00 & 0.00 & 0.32 & 2.36 & 0.00 & 0.00 & 0.00 \\
 & Optimized, $\gamma = 100$ & 0.00 & 0.00 & 0.60 & 2.58 & 0.00 & 0.00 & 0.00 \\
 
\bottomrule
\end{tabular}
\caption{Percentage of portfolios with estimated kurtosis below one, $n=50$. Random portfolios are the 100 long-only portfolios of each of the 5,000 experiments; optimized portfolios are the solutions of Eq.~\eqref{eq:allocation}, one per experiment.}
\label{tab:violations}
\end{table}

\subsection{Consistent covariance-preserving reconstructions}
\label{app:consistent_variants}

To guarantee $\hat{\kappa}(\boldsymbol{w})\geq 1$ for every portfolio while leaving the filtered covariance $\widehat{\boldsymbol{\Sigma}}$ (and thus the $\gamma=0$ minimum-variance portfolio) unchanged, we modify the fourth-moment estimator so that it represents the fourth-moment tensor of a random vector whose covariance matrix equals $\widehat{\boldsymbol{\Sigma}}$.

The first pair of estimators, CV$^{\mathrm{c}}$ and HOAO$^{\mathrm{c}}$, applies a linear transport map to the mixture vector $\tilde{\boldsymbol{x}}$. We store the companion variance vector $\boldsymbol{a}$ ($a_i = \mathbb{E}[y_i^2]$) alongside the core matrix $\boldsymbol{B}$ ($B_{ij} = \mathbb{E}[y_i^2 y_j^2]$), averaged over the exact same calibration pairs (or folds), bases, and weights. In each evaluation experiment, we transform $\tilde{\boldsymbol{x}}$ by
\begin{equation}
\label{eq:linear_transport_A}
\boldsymbol{A} = \widehat{\boldsymbol{\Sigma}}^{1/2}\,\boldsymbol{\Omega}^{-1/2},
\end{equation}
where $(\cdot)^{1/2}$ denotes the principal symmetric square root~\cite[Sec.~7.2]{horn2013matrix} (singular matrices were not regularized and never occurred in our experiments). The transformed random vector $\tilde{\boldsymbol{x}}^{\mathrm{c}} = \boldsymbol{A}\tilde{\boldsymbol{x}}$ has covariance $\boldsymbol{A}\boldsymbol{\Omega}\boldsymbol{A}^\top = \widehat{\boldsymbol{\Sigma}}$ and, by the transformation rule of moment tensors under linear maps~\cite[Ch.~2]{mccullagh1987tensor}, fourth-moment tensor
\begin{equation}
\label{eq:Kc_reconstruction}
\widehat{\boldsymbol{\mathcal{K}}}^{\mathrm{c}} = \widehat{\boldsymbol{\mathcal{K}}} \times_1 \boldsymbol{A} \times_2 \boldsymbol{A} \times_3 \boldsymbol{A} \times_4 \boldsymbol{A}.
\end{equation}
Because $\widehat{\boldsymbol{\Sigma}}$ and $\widehat{\boldsymbol{\mathcal{K}}}^{\mathrm{c}}$ are the second and fourth central moments of the same random vector $\tilde{\boldsymbol{x}}^{\mathrm{c}}$, $\hat{\kappa}(\boldsymbol{w})\geq 1$ holds for every $\boldsymbol{w}$. Since the effective mode matrix $\boldsymbol{A}\widehat{\boldsymbol{D}}\widehat{\boldsymbol{U}}$ is no longer orthogonal, Frobenius norms and inner products are evaluated via the general multilinear expansion rather than Eq.~\eqref{eq:isometry_identity}.

In HOAO$^{\mathrm{c}}$, the linear map $\boldsymbol{A}$ alters the geometry in which the core is applied at evaluation time, whereas the core $\widehat{\boldsymbol{B}}_{\mathrm{HOAO}}$ was calibrated in the original orthogonal basis $\widehat{\boldsymbol{U}}_{\mathrm{train}}$. The recalibrated variant HOAO$^{\mathrm{cr}}$ re-estimates the core directly in the transformed geometry. At each calibration update date, the HOAO correlation estimator and the pair $(\widehat{\boldsymbol{B}}_{\mathrm{HOAO}}, \boldsymbol{a})$ of HOAO$^{\mathrm{c}}$ are first computed on all available calibration pairs and frozen. For each calibration pair $\tau$, denoting by $\widehat{\boldsymbol{C}}_{\mathrm{train}(\tau)}$ the HOAO-filtered training correlation matrix and by $\boldsymbol{\Omega}_{z,\mathrm{train}(\tau)} = \widehat{\boldsymbol{U}}_{\mathrm{train}(\tau)}\,\mathrm{diag}(\boldsymbol{a})\,\widehat{\boldsymbol{U}}_{\mathrm{train}(\tau)}^\top$ the standardized training companion covariance, the linear decoder
\begin{equation}
\label{eq:decoder_H}
\boldsymbol{H}_\tau = \widehat{\boldsymbol{C}}_{\mathrm{train}(\tau)}^{1/2}\,\boldsymbol{\Omega}_{z,\mathrm{train}(\tau)}^{-1/2}\,\widehat{\boldsymbol{U}}_{\mathrm{train}(\tau)}\,\mathrm{diag}(\boldsymbol{a})^{1/2}
\end{equation}
maps latent coordinates with unit covariance $\boldsymbol{I}_n$ to standardized returns with covariance $\boldsymbol{H}_\tau\boldsymbol{H}_\tau^\top = \widehat{\boldsymbol{C}}_{\mathrm{train}(\tau)}$. On the subsequent standardized testing returns $\boldsymbol{z}_t$, we solve $\boldsymbol{H}_\tau\boldsymbol{y}_t = \boldsymbol{z}_t$ and accumulate $\tilde{B}_{ij} = \mathbb{E}[y_i^2 y_j^2]$ and $\tilde{a}_i = \mathbb{E}[y_i^2]$ across the same calibration pairs. After aggregation, we normalize the new core by its companion second moments via $B_{\mathrm{HOAO}^{\mathrm{cr}},ij} = \tilde{B}_{ij}/(\tilde{a}_i\tilde{a}_j)$, so that the latent mixture has unit covariance $\boldsymbol{I}_n$ and decoding by $\widehat{\boldsymbol{D}}\boldsymbol{H}$ on the evaluation window yields a distribution with covariance $\widehat{\boldsymbol{\Sigma}}$.

We also tested three additional constructions (HOAO$^{\mathrm{g}}$, HOAO$^{\mathrm{pca}}$, and HOAO$^{\mathrm{pre}}$) that preserve the filtered covariance $\widehat{\boldsymbol{\Sigma}}$ without applying the non-orthogonal transport map $\boldsymbol{A}$. HOAO$^{\mathrm{g}}$ scales the mixture variable $\tilde{\boldsymbol{x}}$ by $\sqrt{\alpha}$ and adds an independent Gaussian vector $\boldsymbol{\xi}\sim\mathcal{N}(\boldsymbol{0},\widehat{\boldsymbol{\Sigma}}-\alpha\boldsymbol{\Omega})$, where $\alpha = \min\!\big(1, \lambda_{\min}(\boldsymbol{\Omega}^{-1/2}\widehat{\boldsymbol{\Sigma}}\boldsymbol{\Omega}^{-1/2})\big) \leq 1$ is the largest value in $[0,1]$ for which $\widehat{\boldsymbol{\Sigma}}-\alpha\boldsymbol{\Omega}$ is positive semidefinite~\cite[Secs.~4.2 and~4.5]{horn2013matrix}. The sum $\sqrt{\alpha}\tilde{\boldsymbol{x}}+\boldsymbol{\xi}$ has covariance $\widehat{\boldsymbol{\Sigma}}$; by additivity of fourth cumulants for independent summands~\cite[Sec.~2.3]{brillinger2001time}, its fourth central moment is that of a Gaussian vector with covariance $\widehat{\boldsymbol{\Sigma}}$ plus $\alpha^2$ times the fourth-moment excess of $\tilde{\boldsymbol{x}}$ over a Gaussian vector with covariance $\boldsymbol{\Omega}$:
\begin{equation}
\label{eq:hoaog_m4}
\hat{m}_4^{\mathrm{g}}(\boldsymbol{w}) = 3\big(\boldsymbol{w}^\top\widehat{\boldsymbol{\Sigma}}\boldsymbol{w}\big)^2 + \alpha^2\Big[\hat{m}_4(\boldsymbol{w}) - 3\big(\boldsymbol{w}^\top\boldsymbol{\Omega}\boldsymbol{w}\big)^2\Big].
\end{equation}
This construction avoids rotating the fourth-order basis $\widehat{\boldsymbol{U}}$, but retains only a fraction $\alpha^2$ of the non-Gaussian fourth-moment excess (the median of $\alpha^2$ across experiments is $0.27$ with 100-day windows and $0.29$ with 500-day windows).

HOAO$^{\mathrm{pca}}$ and HOAO$^{\mathrm{pre}}$ estimate both the second and fourth moments in the same basis, namely the eigenvector basis of the sample correlation matrix, so that the filtered correlation matrix (before the restoration of its unit diagonal) and the standardized companion covariance share the same eigenvectors. They keep the first principal component separate and average the fourth-order core of the remaining $n-1$ components over all orthogonal rotations in $O(n-1)$. Invariance under $O(n-1)$ reduces the $n\times n$ even-core matrix $\boldsymbol{B}$ to three scalar coefficients~\cite[Ch.~2]{fang1990symmetric}, namely $B_{11}$ for the first component, $B_{1i}=B_{i1}=b_{\mathrm{cross}}$ ($i\geq 2$) for its interaction with the remaining components, and $B_{ij} = b_{\mathrm{bulk}}(1+2\delta_{ij})$ ($i,j\geq 2$) among the remaining $n-1$ components. HOAO$^{\mathrm{pca}}$ normalizes the pooled core by the pooled companion variances after aggregating across calibration pairs (as in HOAO$^{\mathrm{cr}}$), whereas HOAO$^{\mathrm{pre}}$ normalizes each calibration pair to unit covariance prior to aggregation.

All three constructions use the exact same experiments, calibration pairs, and filtered covariance $\widehat{\boldsymbol{\Sigma}}$ as HOAO. In the same experiments, we also evaluated variants using a sample-whitened basis, a full $n\times n$ even core in the correlation eigenbasis without $O(n-1)$ averaging, pre-aggregation normalization applied to HOAO$^{\mathrm{cr}}$, and the companion covariance $\boldsymbol{\Omega}$ paired directly with the original HOAO tensor; we omit them here and provide their aggregates with the paper data.

\subsection{Elliptical control and directional anisotropy}
\label{app:anisotropy}

For each fourth-moment tensor $\widehat{\boldsymbol{\mathcal{K}}}$ paired with the filtered covariance $\widehat{\boldsymbol{\Sigma}}$, we construct an elliptical control tensor with the same covariance $\widehat{\boldsymbol{\Sigma}}$ by whitening the return distribution via $\widehat{\boldsymbol{\Sigma}}^{-1/2}$, averaging its whitened fourth-moment tensor $\boldsymbol{\mathcal{K}}_{\mathrm{white}} = \widehat{\boldsymbol{\mathcal{K}}}\times_1\widehat{\boldsymbol{\Sigma}}^{-1/2}\times_2\widehat{\boldsymbol{\Sigma}}^{-1/2}\times_3\widehat{\boldsymbol{\Sigma}}^{-1/2}\times_4\widehat{\boldsymbol{\Sigma}}^{-1/2}$ uniformly over all rotations in $O(n)$, and mapping the result back via $\widehat{\boldsymbol{\Sigma}}^{1/2}$. Averaging over $O(n)$ preserves the mean fourth power of the Euclidean norm of the whitened vector (Mardia's multivariate kurtosis~\cite{mardia1970measures}) and yields the isotropic whitened tensor~\cite[Ch.~2]{fang1990symmetric}
\begin{equation}
\label{eq:elliptical_white_tensor}
\mathcal{K}_{\mathrm{white},ijkl}^{\mathrm{ell}} = \frac{\kappa_0}{3}\big(\delta_{ij}\delta_{kl} + \delta_{ik}\delta_{jl} + \delta_{il}\delta_{jk}\big),
\qquad
\kappa_0 = \frac{3}{n(n+2)}\sum_{i,j=1}^n \mathcal{K}_{\mathrm{white},iijj}.
\end{equation}
Under the resulting elliptical fourth-moment tensor, every portfolio $\boldsymbol{w}$ has the same kurtosis $\hat{\kappa}(\boldsymbol{w}) = \kappa_0$~\cite{berkane1986moments,jammalamadaka2021multivariate}, so that $\hat{m}_4(\boldsymbol{w}) = \kappa_0\,\hat{\sigma}^4(\boldsymbol{w})$. The objective of Eq.~\eqref{eq:allocation} is then strictly increasing in $\hat{\sigma}^2(\boldsymbol{w})$, and the minimum-variance portfolio ($\gamma=0$) solves Eq.~\eqref{eq:allocation} for every penalty $\gamma\geq 0$.

We measure how far a candidate tensor $\widehat{\boldsymbol{\mathcal{K}}}$ departs from this elliptical case by its anisotropy, defined as the relative Frobenius distance between its whitened fourth-moment tensor and the nearest elliptical tensor:
\begin{equation}
\label{eq:anisotropy_def}
\mathrm{Anisotropy}(\widehat{\boldsymbol{\mathcal{K}}},\widehat{\boldsymbol{\Sigma}}) := \frac{\big\|\boldsymbol{\mathcal{K}}_{\mathrm{white}} - \boldsymbol{\mathcal{K}}_{\mathrm{white}}^{\mathrm{ell}}\big\|_F}{\big\|\boldsymbol{\mathcal{K}}_{\mathrm{white}}\big\|_F}.
\end{equation}
The whitening in Eq.~\eqref{eq:anisotropy_def} uses the filtered covariance $\widehat{\boldsymbol{\Sigma}}$, which for the original HOAO estimator is not the covariance $\boldsymbol{\Omega}$ of its own mixture; for such estimators the control is the algebraic isotropic projection, which need not be the fourth-moment tensor of a distribution with covariance $\widehat{\boldsymbol{\Sigma}}$. Distributional consistency does not imply zero anisotropy, because valid non-elliptical distributions can be strongly directional. That the consistent estimators below have small anisotropy is an empirical property of the historically averaged cores, and does not show that the remaining anisotropy cannot be predicted. In Tables~\ref{tab:corrections} and~\ref{tab:corrections_allocation}, the column ``Elliptical'' is the elliptical control of HOAO$^{\mathrm{cr}}$.

\subsection{Empirical results for the consistent estimators}
\label{app:compatibility_results}

\begin{figure}[htbp]
    \centering
    \includegraphics[width=0.5\linewidth]{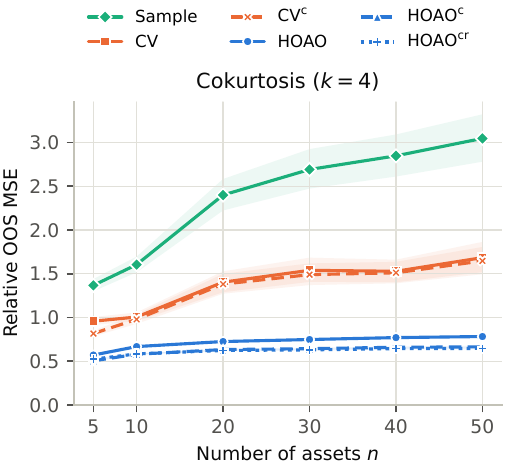}
    \caption{Relative OOS MSE of the cokurtosis for the original and consistent estimators against the number of assets, with $\Delta t_{\mathrm{IS}}=\Delta t_{\mathrm{OOS}}=100$ days. Bands are 95\% bootstrap intervals over experiments.}
    \label{fig:compatible_accuracy}
\end{figure}

\begin{figure}[htbp]
    \centering
    \includegraphics[width=0.95\linewidth]{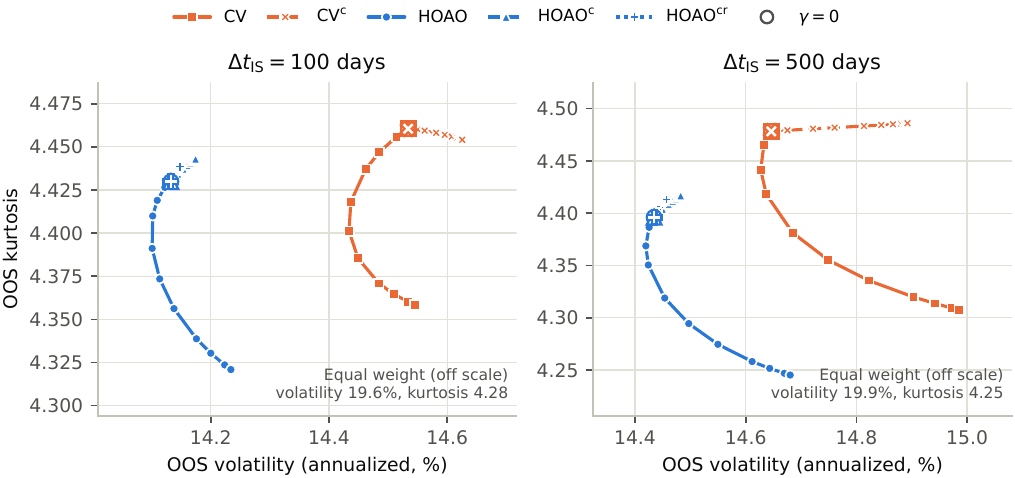}
    \caption{OOS efficient frontiers of the original and consistent estimators, as in Fig.~\ref{fig:frontier}.}
    \label{fig:compatible_frontier}
\end{figure}

By construction, all consistent variants share the correlation and portfolio volatility MSEs of their parent estimators (Table~\ref{tab:portfolio-moment-errors}). With 100-day IS windows, CV$^{\mathrm{c}}$, HOAO$^{\mathrm{c}}$, and HOAO$^{\mathrm{cr}}$ lower their parents' standardized cokurtosis MSE for every dimension $n\geq 5$ (Fig.~\ref{fig:compatible_accuracy}). At $n=50$ (Table~\ref{tab:compatible}), the cokurtosis MSE with 100-day windows falls from $0.783$ for HOAO to $0.665$ for HOAO$^{\mathrm{c}}$ and $0.647$ for HOAO$^{\mathrm{cr}}$; with 500-day windows, the corresponding errors are $0.715$, $0.680$, and $0.672$, whereas the transport map raises the error of CV from $6.087$ to $6.270$. Among the estimators in Table~\ref{tab:compatible}, HOAO$^{\mathrm{cr}}$ has the lowest portfolio kurtosis MSE with 100-day windows ($0.134$) and the original HOAO has the lowest with 500-day windows ($0.124$). In portfolio allocation, each consistent variant shares its parent's minimum-variance portfolio at $\gamma=0$, but increasing the fourth-moment penalty $\gamma$ changes the consistent allocations very little (Fig.~\ref{fig:compatible_frontier}). From $\gamma=0$ to $\gamma=100$ with 100-day windows, mean OOS kurtosis moves from $4.43$ to $4.44$ for both HOAO$^{\mathrm{c}}$ and HOAO$^{\mathrm{cr}}$, and to $4.32$ for HOAO, while staying near $4.46$ for CV$^{\mathrm{c}}$. Enforcing consistency prevents the optimizer from exploiting directions with spuriously low fourth moments, but does not by itself improve the OOS portfolio kurtosis.

\begin{table}[htbp]
\centering
\setlength{\tabcolsep}{7pt}
\begin{tabular}{llccccc}
\toprule
$\Delta t_{\mathrm{IS}}$ & Relative MSE & CV & CV$^{\mathrm{c}}$ & HOAO & HOAO$^{\mathrm{c}}$ & HOAO$^{\mathrm{cr}}$ \\
\midrule
\multirow{2}{*}{100} & Cokurtosis & 1.685 & 1.647 & 0.783 & 0.665 & \textbf{0.647} \\
 & Portfolio kurtosis & 0.259 & 0.298 & 0.148 & 0.172 & \textbf{0.134} \\
\midrule
\multirow{2}{*}{500} & Cokurtosis & 6.087 & 6.270 & 0.715 & 0.680 & \textbf{0.672} \\
 & Portfolio kurtosis & 1.671 & 2.116 & \textbf{0.124} & 0.166 & 0.151 \\
 
\bottomrule
\end{tabular}
\caption{Relative OOS MSE of the original and consistent estimators for $n=50$, for the cokurtosis (Eq.~\eqref{eq:tensor-loss}) and for the portfolio kurtosis (Eq.~\eqref{eq:portfolio-loss}). For the correlation and the volatility, the consistent variants have the MSE of their parent estimators (Table~\ref{tab:portfolio-moment-errors}). Bold values are the smallest MSE of each row.}
\label{tab:compatible}
\end{table}

Tables~\ref{tab:corrections} and~\ref{tab:corrections_allocation} compare all five consistent HOAO estimators and the elliptical control with the original HOAO for $n=50$. All five corrections eliminate the violations ($\hat{\kappa}<1$) and lower the cokurtosis MSE of HOAO from $0.783$ to between $0.623$ and $0.665$ with 100-day windows, and from $0.715$ to between $0.617$ and $0.680$ with 500-day windows. At the portfolio kurtosis level, the ranking depends on the loss function. Minimizing the relative squared error $\mathbb{E}[(c/\kappa - 1)^2]$ over a constant forecast $c$ gives $c^\star = \mathbb{E}[\kappa^{-1}]/\mathbb{E}[\kappa^{-2}] < \mathbb{E}[\kappa]$ by the Cauchy--Schwarz and Jensen inequalities~\cite{hardy1952inequalities}, so the relative MSE penalizes overestimation more heavily than underestimation and favors downward-shrunk forecasts, as for percentage errors~\cite{tofallis2015better}. Accordingly, HOAO$^{\mathrm{g}}$, which damps the fourth-moment excess by $\alpha^2$, has the smallest relative kurtosis MSE ($0.076$ at 100 days and $0.074$ at 500 days). By contrast, the mean squared logarithmic error $\mathbb{E}[(\log\hat{\kappa}-\log\kappa)^2]$ is minimized by the geometric mean $\exp(\mathbb{E}[\log\kappa])$~\cite[Sec.~3.1]{tofallis2015better} and ranks the estimators differently. With 100-day windows, it equals $0.360$ for HOAO and $0.204$ for Sample, versus $0.115$ to $0.149$ for the consistent corrections; with 500-day windows, it equals $0.140$ for HOAO and $0.510$ for Sample, versus $0.112$ to $0.137$ for the corrections.

The anisotropy values in Table~\ref{tab:corrections} explain the allocation results in Table~\ref{tab:corrections_allocation}. All five consistent HOAO estimators have anisotropy between $0.010$ and $0.048$, compared with $0.351$ (100 days) and $0.328$ (500 days) for the original HOAO whitened by $\widehat{\boldsymbol{\Sigma}}$. Because whitening $\tilde{\boldsymbol{x}}^{\mathrm{c}} = \boldsymbol{A}\tilde{\boldsymbol{x}}$ by $\widehat{\boldsymbol{\Sigma}}^{-1/2}$ gives $\widehat{\boldsymbol{\Sigma}}^{-1/2}\boldsymbol{A}\tilde{\boldsymbol{x}} = \boldsymbol{\Omega}^{-1/2}\tilde{\boldsymbol{x}}$, a linear transport map preserves the whitened law; the anisotropy of HOAO$^{\mathrm{c}}$ ($0.042$ at 100 days and $0.048$ at 500 days) is therefore identical to the anisotropy of the original HOAO mixture measured with respect to its own companion covariance $\boldsymbol{\Omega}$. Between $\gamma=0$ and $\gamma=100$, the consistent estimators change portfolio weights by only $0.4\%$ to $7.3\%$ of the budget, against $28.9\%$ and $32.3\%$ for HOAO (while the elliptical control gives identical weights at every penalty). At $\gamma=100$, the consistent estimators therefore stay close to the minimum-variance portfolio and achieve slightly lower OOS volatility than HOAO at both IS lengths, but higher OOS fourth moments and kurtosis. At 100 days, the fourth root of the annualized OOS fourth moment is $32.25\%$--$32.41\%$ for the consistent estimators versus $31.95\%$ for HOAO; at 500 days, the corresponding values are $31.87\%$--$31.99\%$ versus $31.38\%$ (Table~\ref{tab:corrections_allocation}).

For HOAO$^{\mathrm{g}}$, HOAO$^{\mathrm{pca}}$, and HOAO$^{\mathrm{pre}}$, we also compared the optimized portfolio at $\gamma=100$ with the portfolio obtained from each correction's own elliptical control. At $\gamma=100$, the paired 95\% calendar-block bootstrap intervals include zero for every difference in OOS variance, fourth moment, and kurtosis at both IS window lengths except for a $0.08\%$ increase in kurtosis for HOAO$^{\mathrm{pre}}$ at 500 days. The largest point difference in the OOS fourth moment is a reduction of $1.16\%$ at 500 days.

\begin{table}[htbp]
\centering
\setlength{\tabcolsep}{7pt}
\begin{tabular}{llccccccc}
\toprule
$\Delta t_{\mathrm{IS}}$ & Measure & HOAO & HOAO$^{\mathrm{c}}$ & HOAO$^{\mathrm{cr}}$ & HOAO$^{\mathrm{g}}$ & HOAO$^{\mathrm{pca}}$ & HOAO$^{\mathrm{pre}}$ & Elliptical \\
\midrule
\multirow{5}{*}{100} & Cokurtosis MSE & 0.783 & 0.665 & 0.647 & \textbf{0.623} & 0.659 & 0.631 & 0.632 \\
 & Kurtosis MSE & 0.148 & 0.172 & 0.134 & \textbf{0.076} & 0.151 & 0.085 & 0.081 \\
 & Kurtosis log error & 0.360 & 0.141 & 0.128 & 0.149 & 0.133 & \textbf{0.115} & 0.114 \\
 & Violations (\%) & 3.16 & 0.00 & 0.00 & 0.00 & 0.00 & 0.00 & 0.00 \\
 & Anisotropy & 0.351 & 0.042 & 0.032 & 0.031 & 0.025 & 0.010 & 0.000 \\
\midrule
\multirow{5}{*}{500} & Cokurtosis MSE & 0.715 & 0.680 & 0.672 & \textbf{0.617} & 0.677 & 0.628 & 0.628 \\
 & Kurtosis MSE & 0.124 & 0.166 & 0.151 & \textbf{0.074} & 0.158 & 0.085 & 0.082 \\
 & Kurtosis log error & 0.140 & 0.136 & 0.131 & 0.137 & 0.133 & \textbf{0.112} & 0.110 \\
 & Violations (\%) & 0.05 & 0.00 & 0.00 & 0.00 & 0.00 & 0.00 & 0.00 \\
 & Anisotropy & 0.328 & 0.048 & 0.046 & 0.035 & 0.030 & 0.013 & 0.000 \\
 
\bottomrule
\end{tabular}
\caption{Original estimator, consistent variants, further corrections and elliptical control for $n=50$. Cokurtosis MSE and Kurtosis MSE are the relative OOS MSE of the cokurtosis and of the portfolio kurtosis, as in Table~\ref{tab:compatible}. Kurtosis log error is the mean squared difference between the logarithms of the estimated and of the OOS portfolio kurtosis. The portfolios are the 100 random long-only portfolios of each of the 5,000 experiments. Violations is the percentage of these portfolios with estimated kurtosis below one, and Anisotropy is defined in the text. Bold values are the smallest of each of the first three rows among the estimators. The elliptical control is that of HOAO$^{\mathrm{cr}}$.}
\label{tab:corrections}
\end{table}

\begin{table}[htbp]
\centering
\setlength{\tabcolsep}{5.5pt}
\begin{tabular}{llccccccc}
\toprule
$\Delta t_{\mathrm{IS}}$ & Measure & HOAO & HOAO$^{\mathrm{c}}$ & HOAO$^{\mathrm{cr}}$ & HOAO$^{\mathrm{g}}$ & HOAO$^{\mathrm{pca}}$ & HOAO$^{\mathrm{pre}}$ & Elliptical \\
\midrule
\multirow{4}{*}{100} & OOS volatility (\%) & 14.23 & 14.17 & 14.15 & 14.13 & 14.14 & 14.13 & 14.13 \\
 & OOS fourth moment (\%) & 31.95 & 32.41 & 32.33 & 32.25 & 32.27 & 32.27 & 32.27 \\
 & OOS kurtosis & 4.32 & 4.44 & 4.44 & 4.43 & 4.43 & 4.43 & 4.43 \\
 & Weight change (\%) & 28.9 & 4.9 & 2.7 & 2.6 & 1.3 & 0.4 & 0.0 \\
\midrule
\multirow{4}{*}{500} & OOS volatility (\%) & 14.68 & 14.48 & 14.46 & 14.44 & 14.44 & 14.43 & 14.43 \\
 & OOS fourth moment (\%) & 31.38 & 31.99 & 31.97 & 31.87 & 31.93 & 31.95 & 31.97 \\
 & OOS kurtosis & 4.25 & 4.42 & 4.41 & 4.39 & 4.41 & 4.40 & 4.40 \\
 & Weight change (\%) & 32.3 & 7.3 & 4.9 & 3.8 & 2.3 & 0.6 & 0.0 \\
 
\bottomrule
\end{tabular}
\caption{Optimized portfolios of Eq.~\eqref{eq:allocation} at $\gamma=100$ for the estimators of Table~\ref{tab:corrections}, $n=50$. OOS fourth moment is the fourth root of $252^2$ times the mean daily OOS fourth moment, in percent, so that it has the scale of a volatility. Weight change is half the sum of the absolute differences between the weights at $\gamma=100$ and at $\gamma=0$, in percent of the budget. The penalty $\gamma=100$ is the largest of the grid of Sec.~\ref{app:protocol} and was not selected with OOS data. At $\gamma=0$, all estimators give the minimum-variance portfolio of the filtered covariance, with a volatility of 14.13\% and a kurtosis of 4.43 with 100-day windows, and of 14.43\% and 4.40 with 500-day windows. 5,000 experiments for each IS length.}
\label{tab:corrections_allocation}
\end{table}

\section{Minimizing the kurtosis directly}
\label{app:kurtosis}

The allocation objective of Eq.~\eqref{eq:allocation} penalizes the fourth central moment $\hat{m}_4(\boldsymbol{w}) = \hat{\kappa}(\boldsymbol{w})\,\hat{\sigma}^4(\boldsymbol{w})$, which depends on both the scale and the shape of the portfolio return distribution. A natural alternative is to minimize the scale-invariant estimated kurtosis $\hat{\kappa}(\boldsymbol{w})$ directly, subject to an upper bound on the estimated variance. Because a lower kurtosis does not imply a lower variance or fourth moment, this formulation can reduce kurtosis at the cost of a higher variance. We test this problem using the exact same experiments, estimators, and filtered covariance $\widehat{\boldsymbol{\Sigma}}$ as in Sec.~\ref{app:compatibility}.

For each experiment, we solve
\begin{equation}
\label{eq:direct_kurtosis_min}
\min_{\boldsymbol{1}^\top\boldsymbol{w}=1}\;\hat{\kappa}(\boldsymbol{w})
\qquad\text{subject to}\qquad
\hat{\sigma}^2(\boldsymbol{w})\leq r\,\hat{\sigma}^2_{\mathrm{GMV}},
\end{equation}
where $\hat{\sigma}_{\mathrm{GMV}}^2$ is the estimated variance of the minimum-variance portfolio, and $r\in\{1, 1.0025, 1.005, 1.01, 1.02, 1.05, 1.1, 1.2, 1.5, 2, 3, 5\}$ bounds the ratio of the portfolio's IS estimated variance to $\hat{\sigma}_{\mathrm{GMV}}^2$. We apply Eq.~\eqref{eq:direct_kurtosis_min} to HOAO (whose estimated kurtosis can fall below one) and to the consistent single-basis estimators HOAO$^{\mathrm{pca}}$ and HOAO$^{\mathrm{pre}}$. For the last two, rotational invariance on the last $n-1$ principal components reduces Eq.~\eqref{eq:direct_kurtosis_min} to a one-variable problem whose solution is global. For HOAO, we use a multistart solver, and we do not know whether a guaranteed global solution would change the results.

\begin{table}[htbp]
\centering
\setlength{\tabcolsep}{7pt}
\begin{tabular}{lcccccc}
\toprule
 & \multicolumn{3}{c}{$\Delta t_{\mathrm{IS}}=100$} & \multicolumn{3}{c}{$\Delta t_{\mathrm{IS}}=500$} \\
\cmidrule(lr){2-4}\cmidrule(lr){5-7}
Variance ratio bound $r$ & Variance & Fourth moment & Kurtosis & Variance & Fourth moment & Kurtosis \\
\midrule
\multicolumn{7}{l}{Reference: penalty $\gamma=100$} \\
1 & $-1.75^{*}$ & $+1.96$ & $+2.24^{*}$ & $-3.74$ & $+6.94$ & $+3.63^{*}$ \\
1.0025 & $-1.97^{*}$ & $+1.24$ & $+2.05^{*}$ & $-3.73$ & $+5.20$ & $+3.12^{*}$ \\
1.005 & $-2.01^{*}$ & $+0.96$ & $+1.95^{*}$ & $-3.67$ & $+4.56$ & $+2.89^{*}$ \\
1.01 & $-2.03^{*}$ & $+0.60$ & $+1.80^{*}$ & $-3.51^{*}$ & $+3.73$ & $+2.55^{*}$ \\
1.02 & $-1.96^{*}$ & $+0.16$ & $+1.56^{*}$ & $-3.16^{*}$ & $+2.70$ & $+2.06^{*}$ \\
1.05 & $-1.46^{*}$ & $-0.41$ & $+1.02^{*}$ & $-1.96^{*}$ & $+1.19$ & $+1.06^{*}$ \\
1.1 & $-0.28$ & $-0.48$ & $+0.35$ & $+0.31$ & $+0.35$ & $+0.01$ \\
1.2 & $+2.64^{*}$ & $+0.77$ & $-0.60^{*}$ & $+5.43^{*}$ & $+1.10$ & $-1.20$ \\
1.5 & $+13.44^{*}$ & $+9.09^{*}$ & $-2.05^{*}$ & $+22.62^{*}$ & $+12.23^{*}$ & $-2.11$ \\
2 & $+33.30^{*}$ & $+29.83^{*}$ & $-3.02^{*}$ & $+53.34^{*}$ & $+46.58^{*}$ & $-1.76$ \\
3 & $+71.33^{*}$ & $+84.61^{*}$ & $-3.62^{*}$ & $+114.60^{*}$ & $+148.76^{*}$ & $-1.37$ \\
5 & $+136.25^{*}$ & $+227.89^{*}$ & $-4.25^{*}$ & $+226.34^{*}$ & $+441.97^{*}$ & $-1.17$ \\
\midrule
\multicolumn{7}{l}{Reference: penalized portfolio under the same ceiling} \\
1 & $0.00$ & $0.00$ & $0.00$ & $0.00$ & $0.00$ & $0.00$ \\
1.0025 & $-0.06$ & $-0.25$ & $-0.07^{*}$ & $-0.01$ & $-0.68$ & $-0.21^{*}$ \\
1.005 & $-0.06$ & $-0.29^{*}$ & $-0.08^{*}$ & $+0.01$ & $-0.76$ & $-0.19^{*}$ \\
1.01 & $-0.02$ & $-0.33$ & $-0.12^{*}$ & $+0.11$ & $-0.67$ & $-0.26^{*}$ \\
1.02 & $+0.03$ & $-0.25$ & $-0.17^{*}$ & $+0.21$ & $-0.71$ & $-0.33^{*}$ \\
1.05 & $+0.13$ & $-0.26$ & $-0.20^{*}$ & $+0.34$ & $-0.50$ & $-0.28^{*}$ \\
1.1 & $+0.74^{*}$ & $+0.29^{*}$ & $-0.25^{*}$ & $+1.16^{*}$ & $+0.10$ & $-0.33$ \\
1.2 & $+3.25^{*}$ & $+1.45^{*}$ & $-0.75^{*}$ & $+5.52^{*}$ & $+1.16$ & $-1.19$ \\
1.5 & $+13.56^{*}$ & $+9.14^{*}$ & $-2.08^{*}$ & $+22.62^{*}$ & $+12.23^{*}$ & $-2.11$ \\
2 & $+33.31^{*}$ & $+29.84^{*}$ & $-3.02^{*}$ & $+53.34^{*}$ & $+46.58^{*}$ & $-1.76$ \\
3 & $+71.33^{*}$ & $+84.61^{*}$ & $-3.62^{*}$ & $+114.60^{*}$ & $+148.76^{*}$ & $-1.37$ \\
5 & $+136.25^{*}$ & $+227.89^{*}$ & $-4.25^{*}$ & $+226.34^{*}$ & $+441.97^{*}$ & $-1.17$ \\
 
\bottomrule
\end{tabular}
\caption{Minimization of the estimated kurtosis of HOAO under a bound $r$ on the estimated variance, in units of the minimum-variance portfolio, $n=50$. The reference is the portfolio of Eq.~\eqref{eq:allocation} at $\gamma=100$ in the same experiment (first panel), or the penalized portfolio with the lowest estimated kurtosis that respects the same estimated variance bound (second panel). Entries are the percentage changes of the OOS variance, fourth moment and kurtosis. Means are taken first over the experiments of each investment date and then over dates, and 5,000 experiments are run for each IS length. An asterisk marks a 95\% calendar-block bootstrap interval, with blocks of 252 sessions and 1,999 draws, that excludes zero.}
\label{tab:kurtosis_objective}
\end{table}

Table~\ref{tab:kurtosis_objective} compares the HOAO solutions of Eq.~\eqref{eq:direct_kurtosis_min} with two reference portfolios from the same experiment, namely the portfolio of Eq.~\eqref{eq:allocation} at the fixed penalty $\gamma=100$ (first panel), and the portfolio from the penalty grid of Eq.~\eqref{eq:allocation} that has the lowest IS estimated kurtosis among those satisfying the same IS variance bound $\hat{\sigma}^2(\boldsymbol{w})\leq r\,\hat{\sigma}_{\mathrm{GMV}}^2$ (second panel). The second reference tests whether direct kurtosis minimization improves upon selecting a feasible portfolio from the fourth-moment penalty grid using only in-sample information.

Relative to the fixed reference at $\gamma=100$, tight variance bounds ($r\leq 1.05$) reduce OOS variance and raise OOS kurtosis with 100-day windows. At $r=1.1$, all three 95\% confidence intervals include zero. At $r=1.5$, direct minimization lowers OOS kurtosis by $2.05\%$ and raises OOS variance by $13.44\%$, with both intervals excluding zero. With 500-day windows, the point estimates of OOS kurtosis fall for $r\geq 1.2$, but every interval for these reductions includes zero, while OOS variance rises by $5.43\%$ at $r=1.2$ and $22.62\%$ at $r=1.5$.

Relative to the penalized portfolio chosen under the same variance ceiling $r$, direct minimization at $r=1.1$ reduces OOS kurtosis by $0.25\%$ and raises OOS variance by $0.74\%$ and the OOS fourth moment by $0.29\%$ with 100-day windows (all three 252-session intervals excluding zero). With 500-day windows, the changes in kurtosis and variance are $-0.33\%$ and $+1.16\%$ and only the variance interval excludes zero. At the tighter bound $r=1.02$, the OOS kurtosis reductions of $0.17\%$ (100 days) and $0.33\%$ (500 days) have intervals below zero.

For HOAO$^{\mathrm{pca}}$ and HOAO$^{\mathrm{pre}}$, direct minimization at $r=1.1$ raises OOS kurtosis relative to the fixed $\gamma=100$ penalty by $0.43\%$--$0.46\%$ at 100 days and $0.85\%$--$0.93\%$ at 500 days. Only the increase for HOAO$^{\mathrm{pre}}$ at 500 days has a 95\% interval excluding zero; its OOS variance is $0.07\%$ higher on average, with an interval spanning zero.

Overall, direct kurtosis minimization yields modest reductions in OOS kurtosis accompanied by higher OOS variance, while pointwise confidence intervals containing zero (which are not adjusted for multiple comparisons across bounds) leave the corresponding differences inconclusive rather than establishing equivalence. These results support retaining the fourth-moment penalty in the main text, although the relative merits of the two objectives depend on the desired variance--kurtosis trade-off.

\section{Additional results}
\label{app:additional}

This section decomposes the cokurtosis MSE reduction into projection and calibration components (Sec.~\ref{app:ablation}), examines error variation across investment years (Sec.~\ref{app:stability}), tests sensitivity to the dollar-turnover screen (Sec.~\ref{app:turnover}), and reports controlled experiments on synthetic data (Sec.~\ref{app:synthetic}).

\subsection{Decomposition of the cokurtosis gain}
\label{app:ablation}

Let $\bar{e}_m$ denote the mean relative OOS cokurtosis error $e_m^{(4)}$ of estimator $m$. The MSE reduction of a filtered estimator $m\in\{\mathrm{CV},\mathrm{HOAO}\}$ over Sample splits additively into the gain from the even projection $\Pi_{\mathrm{even}}$ alone and the gain from recalibrating the even-core coefficients:
\begin{equation}
\label{eq:ablation_split}
\bar{e}_{\mathrm{Sample}}-\bar{e}_m
=\big(\bar{e}_{\mathrm{Sample}}-\bar{e}_{\mathrm{Sample}_{\mathrm{even}}}\big)
+\big(\bar{e}_{\mathrm{Sample}_{\mathrm{even}}}-\bar{e}_m\big).
\end{equation}
Because $\mathrm{Sample}_{\mathrm{even}}$, CV, and HOAO share the same orthogonal basis $\widehat{\boldsymbol{U}}$ (so that Eq.~\eqref{eq:isometry_identity} holds) and are all supported on the even subspace, the component of the OOS tensor orthogonal to the even subspace contributes identically to their three MSEs and cancels in the second term. That second term therefore measures the pure effect of out-of-sample core calibration, whereas the first term can take either sign. Table~\ref{tab:ablation} reports this decomposition for $n=50$.

\begin{table}[htbp]
\centering
\setlength{\tabcolsep}{7pt}
\begin{tabular}{llcccccc}
\toprule
$\Delta t_{\mathrm{IS}}$ & $m$ & $\bar e_{\mathrm{Sample}}$ & $\bar e_{\mathrm{Sample}_{\mathrm{even}}}$ & $\bar e_m$ & Projection gain & Coefficient gain & Coefficient share (\%) \\
\midrule
\multirow{2}{*}{100} & CV & 3.048 & 2.961 & 1.685 & 0.087 & 1.276 & 93.6 \\
 & HOAO & 3.048 & 2.961 & 0.783 & 0.087 & 2.177 & 96.2 \\
\midrule
\multirow{2}{*}{500} & CV & 8.157 & 8.051 & 6.087 & 0.106 & 1.963 & 94.9 \\
 & HOAO & 8.157 & 8.051 & 0.715 & 0.106 & 7.335 & 98.6 \\
 
\bottomrule
\end{tabular}
\caption{Decomposition of the gain in cokurtosis MSE over Sample, $n=50$.}
\label{tab:ablation}
\end{table}

\subsection{Stability over time}
\label{app:stability}

\begin{figure}[htbp]
    \centering
    \includegraphics[width=0.95\linewidth]{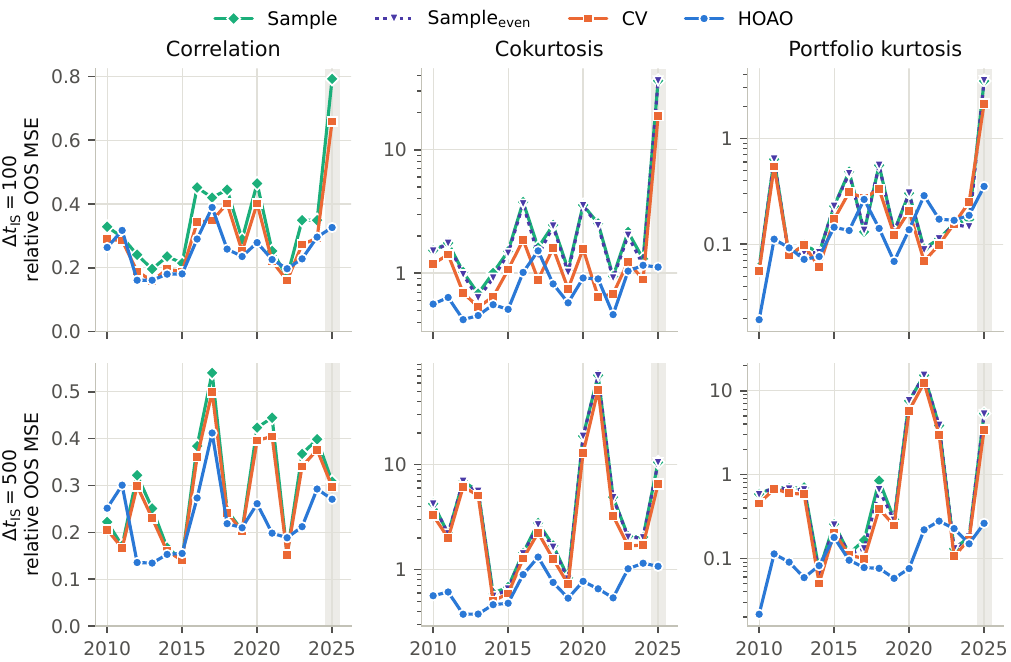}
    \caption{Relative OOS MSE by year of the investment date, $n=50$, with IS windows of 100 (top) and 500 days (bottom). From left to right, the panels show the relative OOS MSE of the correlation, of the cokurtosis and of the portfolio kurtosis, the last two on a logarithmic scale. The shaded band marks 2025, whose investment dates end in August.}
    \label{fig:annual}
\end{figure}

Figure~\ref{fig:annual} plots the relative OOS MSE grouped by the calendar year of the investment date. With 100-day IS windows, Sample's cokurtosis MSE reaches $36.30$ in 2025. It averages $0.87$ for January--March investment dates and exceeds $50$ for both the April--June and July--August cohorts, whose IS windows include the large market moves of early April 2025. Excluding 2025 yields overall cokurtosis MSEs of $1.78$ for Sample, $1.03$ for CV, and $0.77$ for HOAO. With 500-day windows, Sample's annual cokurtosis MSE reaches $18.75$ in 2020 and $70.96$ in 2021. Across the 16 years, HOAO's annual cokurtosis MSE peaks at $1.52$ for 100-day windows and $1.32$ for 500-day windows; it lies below Sample's in all 16 years for both window lengths, and below CV's in all years except 2017, 2021, and 2024 at 100 days and in all 16 years at 500 days. HOAO's correlation MSE is below Sample's in 14 of 16 years at 100 days and 11 at 500 days; for portfolio kurtosis, the counts are 10 and 14.

\subsection{Sensitivity to the dollar-turnover screen}
\label{app:turnover}

We repeat the 5,000 experiments for each IS length at $n=50$ after removing the USD~5 million daily dollar-turnover threshold, keeping all other screens (including the REIT exclusion) and all investment dates fixed, and recalibrating the HOAO cores separately on this expanded universe. Pairing the random asset priorities across the two specifications yields identical holdings in $3,459$ of the $5,000$ draws with 100-day IS windows and in $3,503$ with 500-day windows.

\begin{table}[htbp]
\centering
\setlength{\tabcolsep}{7pt}
\begin{tabular}{llcccc}
\toprule
& & \multicolumn{2}{c}{Window-weighted gain (\%)} & \multicolumn{2}{c}{Equal-date difference (pp)} \\
\cmidrule(lr){3-4}\cmidrule(lr){5-6}
$\Delta t_{\mathrm{IS}}$ & Benchmark & Screen on & Screen off & On minus off & 95\% interval \\
\midrule
\multirow{3}{*}{100} & Sample & 74.30 & 74.33 & $-0.04$ & $[-0.50, +0.10]$ \\
 & Sample$_{\mathrm{even}}$ & 73.55 & 73.58 & $-0.04$ & $[-0.55, +0.11]$ \\
 & CV & 53.52 & 53.65 & $-0.17$ & $[-1.02, +0.08]$ \\
\midrule
\multirow{3}{*}{500} & Sample & 91.23 & 91.26 & $-0.03$ & $[-0.28, +0.01]$ \\
 & Sample$_{\mathrm{even}}$ & 91.11 & 91.14 & $-0.03$ & $[-0.29, +0.01]$ \\
 & CV & 88.25 & 88.29 & $-0.04$ & $[-0.37, +0.02]$ \\
 
\bottomrule
\end{tabular}
\caption{HOAO's reduction in standardized cokurtosis MSE relative to each benchmark, $n=50$. The first two columns of results report one minus the ratio of mean losses across windows, in percent. For the difference, we first average losses within each investment date, then recompute the reductions across dates. Intervals in percentage points (pp) use 1,999 paired circular bootstrap draws with 252-session blocks.}
\label{tab:turnover}
\end{table}

The largest change in the window-weighted cokurtosis MSE reduction is against CV at 100 days, from $53.52\%$ with the screen to $53.65\%$ without it (Table~\ref{tab:turnover}). All six paired 95\% confidence intervals for the difference include zero.

\subsection{Controlled synthetic experiments}
\label{app:synthetic}

To evaluate the estimators against exact population targets in a stationary setting, we consider two centered $n=10$ populations sharing the same positive-definite covariance matrix $\boldsymbol{\Sigma}=\boldsymbol{A}\boldsymbol{A}^\top$ (where $\boldsymbol{A}$ is a fixed loadings matrix combining four common factors and ten idiosyncratic shocks with heterogeneous scales), namely i.i.d.\ multivariate Gaussian returns and i.i.d.\ multivariate Student-$t$ returns with $\nu=10$ degrees of freedom and normalized covariance, whose population portfolio kurtosis equals $3$ and $4$, respectively, in every direction~\cite[Sec.~5]{jammalamadaka2021multivariate}. For each population, we calibrate HOAO (and its consistent variants) on 100 independent train/test pairs of 100 observations each, and evaluate all estimators over 100 independent replications of length $\Delta t_{\mathrm{IS}}=100$ (using 10 random 65\%/35\% splits for CV and 100 random long-only portfolios drawn from $\mathrm{Dirichlet}(1,\ldots,1)$ per replication). Losses are defined as in Eqs.~\eqref{eq:tensor-loss} and~\eqref{eq:portfolio-loss}, but replace the noisy OOS sample proxy with the exact analytical population moments.

Table~\ref{tab:synthetic} reports the mean relative MSE and its standard error across the 100 replication means (conditional on the saved calibration core). In both populations, HOAO achieves lower errors than Sample, Sample$_{\mathrm{even}}$, and CV across all four metrics. None of the 10,000 random portfolios per method and population produces an estimated kurtosis below one (which does not imply global consistency over all weight vectors or rule out the violations observed on equity data). Because both populations are elliptical with constant true kurtosis across directions, the consistent reconstructions (CV$^{\mathrm{c}}$, HOAO$^{\mathrm{c}}$, HOAO$^{\mathrm{cr}}$) further reduce standardized cokurtosis and portfolio kurtosis errors relative to their parent estimators, but this setting cannot test directional kurtosis reduction in portfolio optimization.

\begin{table}[htbp]
\centering
\setlength{\tabcolsep}{7pt}
\begin{tabular}{@{}lccccccc@{}}
\toprule
Relative MSE & Sample & Sample$_{\mathrm{even}}$ & CV & HOAO & CV$^{\mathrm{c}}$ & HOAO$^{\mathrm{c}}$ & HOAO$^{\mathrm{cr}}$ \\
\midrule
\multicolumn{8}{@{}l}{Gaussian} \\
Correlation & 0.076646 & 0.076646 & 0.056385 & \textbf{0.039038} & 0.056385 & \textbf{0.039038} & \textbf{0.039038} \\
 & (0.001585) & (0.001585) & (0.001909) & (0.001479) & (0.001909) & (0.001479) & (0.001479) \\
\addlinespace[2pt]
Cokurtosis & 0.503350 & 0.278383 & 0.169926 & 0.118705 & 0.142758 & 0.091105 & \textbf{0.091082} \\
 & (0.010448) & (0.008916) & (0.004679) & (0.004297) & (0.004372) & (0.003389) & (0.003384) \\
\addlinespace[2pt]
Volatility & 0.005005 & 0.005005 & 0.003907 & \textbf{0.003415} & 0.003907 & \textbf{0.003415} & \textbf{0.003415} \\
 & (0.000259) & (0.000259) & (0.000216) & (0.000199) & (0.000216) & (0.000199) & (0.000199) \\
\addlinespace[2pt]
Kurtosis & 0.020702 & 0.021163 & 0.018877 & 0.010565 & 0.002207 & 0.000070 & \textbf{0.000042} \\
 & (0.000987) & (0.001729) & (0.001294) & (0.000700) & (0.000292) & (0.000004) & (0.000002) \\
\midrule
\multicolumn{8}{@{}l}{Student-$t_{10}$} \\
Correlation & 0.094526 & 0.094526 & 0.071547 & \textbf{0.048699} & 0.071547 & \textbf{0.048699} & \textbf{0.048699} \\
 & (0.002303) & (0.002303) & (0.002492) & (0.001621) & (0.002492) & (0.001621) & (0.001621) \\
\addlinespace[2pt]
Cokurtosis & 1.135796 & 0.857488 & 0.289427 & 0.204620 & 0.205875 & \textbf{0.114824} & 0.115623 \\
 & (0.084148) & (0.082705) & (0.007716) & (0.004120) & (0.006277) & (0.003469) & (0.003389) \\
\addlinespace[2pt]
Volatility & 0.006943 & 0.006943 & 0.005886 & \textbf{0.005216} & 0.005886 & \textbf{0.005216} & \textbf{0.005216} \\
 & (0.000404) & (0.000404) & (0.000443) & (0.000418) & (0.000443) & (0.000418) & (0.000418) \\
\addlinespace[2pt]
Kurtosis & 0.066852 & 0.063714 & 0.045759 & 0.018059 & 0.014690 & \textbf{0.001094} & 0.001499 \\
 & (0.005853) & (0.005585) & (0.004624) & (0.001567) & (0.001193) & (0.000041) & (0.000049) \\
 
\bottomrule
\end{tabular}
\caption{Relative MSE against analytical population moments on controlled synthetic data ($n=10$, $\Delta t_{\mathrm{IS}}=100$). Entries report the mean across 100 independent replications (100 random long-only portfolios per replication for volatility and kurtosis), with standard errors across replication means in parentheses. Bold values are the smallest MSE of each row.}
\label{tab:synthetic}
\end{table}

\end{document}